\documentclass{amsart}

\usepackage{hyperref} %
\usepackage{amssymb}
\usepackage{color}
\usepackage{cases}
\usepackage{graphicx}

\newcommand{\be}{\begin{equation}}
\newcommand{\ee}{\end{equation}}
\newcommand{\e}{{\rm{e}}}
\newcommand{\R}{\mathbb{R}}
\newcommand{\Z}{\mathbb{Z}}
\newcommand{\C}{\mathbb{C}}
\newcommand{\mP}{\mathbb{P}}
\newcommand{\bx}{\pmb{x}} 
\newcommand{\bkappa}{\pmb{\kappa}} 
\newcommand{\bz}{\pmb{z}}
\newcommand{\by}{\pmb{y}}
\newcommand{\bolde}{\pmb{e}}
\newcommand{\mL}{\mathcal{L_\alpha^{-\alpha}}}
\newcommand{\mD}{\mathcal{D}}
\newcommand{\mK}{\mathcal{K}}

\theoremstyle{definition}

\theoremstyle{remark}

\begin{document}
\title[SHOULD I STAY OR SHOULD I GO?] 
{SHOULD I STAY OR SHOULD I GO?\\ 
ZERO-SIZE JUMPS IN RANDOM WALKS FOR L\'EVY FLIGHTS$\dagger$\footnote{$\dagger$ 
Paper published in
Fract. Calc. Appl. Anal. 24(1), 137--167 (2021)
}}

\author{Gianni Pagnini}
\address{BCAM - Basque Center for Applied Mathematics,
Alameda Mazarredo 14, 48009 Bilbao, Basque Country - Spain}
\address{Ikerbasque - Basque Foundation for Science,
Euskadi Plaza 5, 48009 Bilbao, Basque Country - Spain}
\email{gpagnini@bcamath.org}

\author{Silvia Vitali}
\address{BCAM - Basque Center for Applied Mathematics,
Alameda Mazarredo 14, 48009 Bilbao, Basque Country - Spain}
\email{svitali@bcamath.org}

\thanks{
We are very grateful to Professor Francesco Mainardi 
for an endless list of personal merits
and best practices that motivate us, and we would also like to acknowledge Enrico Scalas for
valuable remarks on a draft version of this paper. 
This research is supported by the Basque Government through the BERC 2018--2021 program
and also funded by the Spanish Ministry of Economy and Competitiveness MINECO via the
BCAM Severo Ochoa SEV-2017-0718 accreditation.The code is available at: 
\href{https://gitlab.bcamath.org/svitali/should-i-stay-or-should-i-go}
{https://gitlab.bcamath.org/svitali/should-i-stay-or-should-i-g}.
}



 \begin{abstract}

We study Markovian continuous-time random walk models for L\'evy flights
and we show an example in which the convergence to stable densities
is not guaranteed when jumps follow a bi-modal power-law distribution
that is equal to zero in zero.
The significance of this result is two-fold: $i)$ with regard to the probabilistic
derivation of the fractional diffusion equation
and also $ii)$ with regard to the concept of site fidelity
in the framework of L\'evy-like motion for wild animals.

 \end{abstract}

 \maketitle


\section{Introduction}\label{Sec:1}
\setcounter{section}{1}
\setcounter{equation}{0}\setcounter{theorem}{0}

This research is motivated by the fact that,
in the literature dedicated to random walks for anomalous diffusion,
the specific value of the
frequency of the jumps with zero-size is disregarded
as if it does not affect the motion of the walker,
e.g.,
\cite{montroll_etal-jmp-1965,
shlesinger_etal-jsp-1982,
bouchaud_etal-pr-1990,
weiss-1994,
scalas_etal-pa-2000,
metzler_etal-pr-2000,
barkai-cp-2002,
metzler_etal-jpa-2004,
scalas_etal-pre-2004,
meerschaert_etal-jap-2004,
fulger_etal-pre-2008,
germano_etal-pre-2009,
gorenflo_etal-jcam-2009,
klafter_sokolov-2011,
meerschaert_sikorskii-2012,
zaburdaev_etal-rmp-2015,
kutner_etal-epjb-2017}.
Actually, in the literature it is disregarded
if the walker does not move in the majority of the iterations
because the most frequent jump-size is zero
(i.e., the jump-size distribution is unimodal with mode located in zero)
or, in opposition, if the walker always moves because
the jumps with zero-size never occur
(i.e., the jump-size distribution is bi-modal and equal to zero in zero).
As a matter of fact, in the large-time limit,
this irrelevance holds true for random-walk models of the Brownian motion
when the corresponding jump-processes follow
a Gaussian law or a coin-flipping rule.
On the other side, anomalous diffusion is explained by L\'evy flights,
rather than by the Brownian motion,
and L\'evy flights are defined as Markovian random walks
that converge to stable densities
because of power-law distributed jumps
\cite{shlesinger_etal-jsp-1982,bouchaud_etal-pr-1990}.

\smallskip

Before starting, we declare that we are more confident in
using terminology, notions and notation adopted in physics.
Hence, we do not refer to the considered diffusion processes
as random-walk models
with Lebesgue measure when they satisfy the Central Limit Theorem
or as long-jump processes with Hausdorff measure when they satisfy the
generalised Central Limit Theorem in the sense of L\'evy,
but we term the processes according to the resulting
probability density function ($pdf$),
namely we call Brownian motion the processes whose
walker's distribution converges to a Gaussian law
and we term L\'evy flights the processes whose walker's distribution converges
to a stable law.

\smallskip

The evolution in time of L\'evy flights
emerges to be governed by a fractional diffusion equation
\cite{uchaikin-ijtp-2000,
metzler_etal-pr-2000,metzler_etal-jpa-2004,
zaburdaev_etal-rmp-2015}.
A number of properties of L\'evy flights has been studied, e.g.,
\cite{chechkin_etal-jpa-2003,
chechkin_etal-anotrans-2008,
dybiec_etal-pre-2017,
palyulin_etal-njp-2019,
padash_etal-jpa-2019,padash_etal-jpa-2020}.
However, in the probabilistic derivation of the fractional
diffusion equation, the distinctive singularity of the fractional Laplacian is obtained,
with a constant time-step, when the distribution of jumps
is bi-modal and equal to zero in zero \cite{valdinoci-bsema-2009},
see also Appendix B.
Hence, the frequency of jumps with zero-size
is expected to play a key role in modelling
fractional anomalous diffusion.

\smallskip

Here we analyse this literature inconsistency between
the apparent irrelevance of the frequency of zero-size jumps,
as promoted by random-walk models for L\'evy flights,
and the link between the jump distribution and
the distinctive singularity of the fractional Laplacian,
as established by probability arguments
for deriving the fractional diffusion equation \cite{valdinoci-bsema-2009}.
In particular, in this paper we provide an example to show that
it is not guaranteed that a Markovian continuous-time random walk (CTRW),
with jump-sizes uncoupled from the waiting-times and displaying power-law tails,
converges to a stable density when the jumps follow
a bi-modal distribution equal to zero in zero,
that is the one in agreement with the probabilistic derivation
\cite{valdinoci-bsema-2009}, and,
moreover, the resulting diffusive process can be non self-similar.

\smallskip

The consequence of this loss of self-similarity is the
emergence of a time-scale for realizing the large-time limit.
Such time-scale results to be dependent on the stability parameter
by spanning from zero to infinity.
Hence, the large-time limit could not be reached in real systems.

\smallskip

Even if this can be considered a second order effect,
in diffusion processes the maximum of the walker's distribution
stays located in the starting site at all elapsed times and,
in the large-time limit,
the scaling-law in time of the distribution around its maximum
is important for determining the properties of recurrence and transience
of the random walk \cite{affili_etal-matrix-2020}.
Therefore, attaining the large-time limit
together with the scaling-law in time of the
walker's distribution maximum have a fundamental role on determining
the suitability of the CTRW approach
for modelling L\'evy flights, because of a failing performance
or a compatible performance by a CTRW model
for reproducing recurrence and transience of
the many observed signatures of L\'evy flights.
Since L\'evy flights can be modelled, for example,
also through stochastic differential equations driven by L\'evy-noise
\cite{fogedby-pre-1994,metzler_etal-pr-2000,chechkin_etal-anotrans-2008}, through parametric subordination
\cite{gorenflo_etal-csf-2007,gorenflo_etal-epjst-2011,gorenflo_etal-klm-2012} or through other methods
\cite{gorenflo_etal-fcaa-1998,gorenflo_etal-pa-1999,
gorenflo_etal-zaa-1999,vitali_etal-jrsi-2018},
this result establishes a criterion for the selection
of proper modelling approaches for L\'evy flights.

\smallskip

In particular, in the spirit of P\'olya's theorem \cite{polya-1921,novak-amm-2014},
recurrence and transience are of paramount importance on the way home.
The motion of wild animals has been associated many times
to power-law distributions both in the view of the celebrated,
and criticised, L\'evy flights foraging hypothesis
\cite{fritz_etal-prslb-2003,humphries_etal-pnas-2012,
palyulin_etal-pnas-2014,zeng_etal-fcaa-2014,
benhamou_etal-jtb-2015,pyke-mee-2015,klages-2018},
and also in the view of the concept of {\it site fidelity}
\cite{giuggioli_etal-jmb-2012,gautestad_etal-me-2013}:
the recurrent visit of an animal to a previously occupied location
\cite{greenwood-ab-1980,benhamou-e-2007,
boyer_etal-prl-2014,berthelot_etal-bioRxiv-2020}.
The fact that the large-time limit for determining the recurrence or transience of the
process could not be realistically reached clashes against
the concept of site fidelity,
which is straightforwardly related to recurrence,
and this provides a further weakness
of the power-law hypothesis for animal behaviour.

\smallskip

To conclude, our result highlights the need to investigate more
deeply the role of zero-size jumps in random walks with
power-law distributed jumps.

\smallskip

In Section \ref{sec:main}, we call the attention to
the small wavelength expansion of the characteristic function
of jumps that are power-law distributed and we derive the conditions
for the loss of self-similarity in the resulting process.
In Section \ref{sec:discussion}, we discuss the significance
of the derived result both $i)$ in the framework of the probabilistic
derivation of the fractional diffusion equation,
as far as the relation between zero-size jumps and
the distinctive singularity of the fractional Laplacian is concerned,
and $ii)$ in the framework of animal behaviour, as far as
the concept of site fidelity and the L\'evy flights foraging hypothesis
are concerned, and we furtherly highlight the effect due to zero-size jumps for reaching the
large-time limit.
In Section \ref{sec:conclusion}, we provide summary and conclusions
in the perspective of future research.

\vspace*{-4pt} 
\section{Power-law tails, zero-size jumps and self-similarity} 

\label{sec:main}
\setcounter{section}{2}
\setcounter{equation}{0}\setcounter{theorem}{0}

We denote by $\rho(\bx;t)$ the distribution of the walker's displacement
$\bx$ at time $t$, with $\bx=(x_1, ..., x_N) \in \R^N$ and $t > 0$,
such that
\be
\int_{\R^N} \rho(\bx;t) \, d\bx = 1
\,\,\, {\rm and} \,\,\,
\rho(\bx;t) > 0 \,\,\,
{\rm for} \,\, {\rm all} \,\,\,
(\bx,t) \in \R^N \times (0,+\infty) \,.
\label{defrho}
\ee
Moreover, we assume as initial datum $\rho(\bx;0)=\delta(\bx)$.
In a CTRW model, the distribution $\rho(\bx;t)$ is governed by the Montroll--Weiss equation
\cite{montroll_etal-jmp-1965,shlesinger-epjb-2017} that,
in the Markovian case with jump-sizes uncoupled from waiting-times, reads
\vskip -10pt
\be
\widehat{\rho}(\bkappa;t)
=\e^{-(1-\widehat{\varphi}(\bkappa)) t/\tau} \,,
\quad \bkappa \in \R^N \,, \label{MWeq}
\ee
where $\widehat{\rho}(\bkappa;t)$ and
$\widehat{\varphi}(\bkappa)=\widehat{\varphi}(\ell\bkappa)$
are the characteristic functions
of $\rho(\bx;t)$ and of the jump $pdf$ $\varphi(\bx)=\varphi(\bx/\ell)/\ell^N$,
respectively, with $\ell$ as the length-unit of the jumps and
$\tau$ as the time-unit - and also the mean value -
of the waiting-times that are exponentially distributed.

\smallskip

Here, we are interested in establishing an observable
that allows for discriminating between the case {\it ``Should I stay?"}:
when the walker does not move in the majority of the iterations
because the most frequent jump-size is zero, i.e.,
$\varphi(\bx)$ is an unimodal jump $pdf$
such that $\varphi(0)=\sup\{\varphi(\bx) : \bx \in \R^N \}$;
and the opposite case {\it ``Should I go?"}:
when the walker always moves because the jumps with
zero-size never occur, i.e.,
$\varphi(\bx)$ is a bi-modal jump $pdf$ such that
$\varphi(0)=\inf\{\varphi(\bx) : \bx \in \R^N \}=0$.
We say that jump-sizes in the {\it ``Should I go?"} condition
follow a rule {\it \`a la} coin-flipping.

\smallskip

By applying in one single step the analog of
the Kramers--Moyal expansion and of the Pawula theorem,
whatever the jump $pdf$ $\varphi(\bx)$ is such that it holds
\cite{metzler_etal-pr-2000,
metzler_etal-jpa-2004,
meerschaert_sikorskii-2012,
zaburdaev_etal-rmp-2015,
kutner_etal-epjb-2017}
\be
\widehat{\varphi}(\bkappa) \simeq 1 - \ell^\alpha \, |\bkappa|^\alpha
+ o(|\bkappa|^\alpha) \,, \quad
\ell |\bkappa| \ll 1 \,,
\quad 0 < \alpha \le 2 \,,
\label{varphiexpansion}
\ee
then, if in the small wavelength expansion (\ref{varphiexpansion})
we set $\alpha=2$,
from equation (\ref{MWeq}) we obtain that $\rho(\bx;t)$
solves the evolution problem
\be
\left\{
\begin{array}{l}
\displaystyle{
\frac{\partial \rho}{\partial t} = \mathcal{D} \Delta \rho} \,,
\quad {\rm in} \,\, \R^N \times (0,+\infty) \,, \\ \\
\rho(\bx;0)=\delta(\bx) \,,
\end{array}
\right.
\label{DE}
\ee
where $\mathcal{D}=\ell^2/\tau$ is the diffusion coefficient
and, actually, $\rho(\bx;t)$ is a Gaussian density:
\vskip -10pt
\be
\rho(\bx;t)=\mathcal{G}(\bx;t)=
\frac{1}{(\mathcal{D} t)^{1/2}} \, \mathcal{G}\!\left(
\frac{\bx}{(\mathcal{D} t)^{1/2}};1\right)
=\frac{1}{(4 \pi \mathcal{D} t)^{N/2}} \,
\e^{-\frac{|\bx|^2}{4 \mathcal{D} t}}
\,,
\label{gausspdf}
\ee
so we say that this CTRW is a model for the Brownian motion,
with variance
\be
\sigma^2=\int_{\R^N} |\bx|^2 \rho(\bx;t) \, d\bx = 2 N \mathcal{D} t \,.
\ee

\smallskip

On the contrary, if in the small wavelength expansion (\ref{varphiexpansion})
we consider the interval $0 < \alpha < 2$,
from equation (\ref{MWeq}) we obtain that $\rho(\bx;t)$
solves the fractional evolution problem
\be
\left\{
\begin{array}{l}
\displaystyle{
\frac{\partial \rho}{\partial t}
+ \mathcal{D}_{\! \alpha} \, (-\Delta)^{\frac{\alpha}{2}} \rho} = 0 \,,
\quad 0 < \alpha < 2 \,, \quad {\rm in} \,\, \R^N \times (0,+\infty) \,, \\ \\
\rho(\bx;0)=\delta(\bx) \,,
\end{array}
\right.
\label{SFDE}
\ee
where $\mathcal{D}_{\! \alpha}=\ell^\alpha/\tau$
is the fractional diffusion coefficient, therefore $\mD_2=\mD$,
and $(-\Delta)^{\alpha/2}$, $\alpha \in (0,2)$,
is the fractional Laplacian
\cite{bucur_valdinoci-2016,kwasnicki-fcaa-2017,kwasnicki-hfca-2019}
such that, actually,
$\rho(\bx;t)$ is a L\'evy stable density
\cite{hanyga-prsla-2001}, i.e.,
\be
\rho(\bx;t)=\mathcal{L}_\alpha(\bx;t)=
\frac{1}{(\mD_{\!\alpha} \, t)^{N/\alpha}}
 \, \mathcal{L}_\alpha\!\left(\frac{\bx}{(\mD_{\!\alpha} \, t)^{1/\alpha}};1\right)
\,,
\ee
and we say that this CTRW is a model for L\'evy flights
\cite{zaburdaev_etal-rmp-2015}, with
fractional absolute moments \cite{metzler_etal-pr-2000,metzler_etal-cp-2002}
\be
\sigma^q=\int_{\R^N} |\bx|^q \rho(\bx;t) \, d\bx \propto
(\mD_{\!\alpha} t)^{q/\alpha} \,,
\quad 0 < q < \alpha < 2 \,.
\ee
Here we do not specify any particular definition of the fractional
Laplacian, because we consider only processes
in an unbounded domain and in this case there are at least ten
equivalent definitions \cite{kwasnicki-fcaa-2017}.
In bounded domains, the spectral representation results to be favourite,
with respect others, for diffusion problems because it is
based on the heat kernel, namely the Brownian motion
\cite{cusimano_etal-sjna-2018}.
Mathematical and physical interpretations of the fractional
Laplacian are provided by Hilfer
\cite{hilfer-fcaa-2015,hilfer-hfca-2019}
and a number of physical systems governed by space fractional
kinetics are reported, for example, by Uchaikin \& Sibatov
\cite{uchaikin_sibatov-2018}.
A noteworthy case of diffusion problem (\ref{SFDE}) is the
special case $\alpha=1$ that leads to the Cauchy (Lorentz) distribution
\cite{hanyga-prsla-2001,affili_etal-matrix-2020}
\vskip -12pt
\be
\rho(\bx;t)=\mathcal{L}_1(\bx;t) =
\frac{\mathcal{N}}{(\mD_1 t)^N}
\, \frac{1}{[1 +(|\bx|/(\mD_{1}t))^2]^{(N+1)/2}} \,, 
\label{cauchy}
\ee
where $\mathcal{N}$ is the normalization factor.

\smallskip

Whenever the derivation of CTRW models for the Brownian motion
and for L\'evy flights is strictly based on
the small wavelength expansion of the characteristic function
of jumps (\ref{varphiexpansion}),
the difference between
the conditions {\it ``Should I stay?"} and {\it ``Should I go?"}
is neglected because this limit provides the behaviour of the tails of the
resulting walker's distribution $\rho(\bx;t)$
and then, in this respect, the distribution of small jump-sizes is irrelevant.
Actually, the application of this method could mislead
to the {\it undeclared statement} - on the back of the mind -
that {\it the small wavelength expansion of the characteristic
function of the jump $pdf$ should be a series with alternating signs,
namely an alternating series,
like the Taylor expansion of a completely monotonic function is,
but this is not always true for stable densities}.
Here we investigate the effect of this non-alternation of signs
in the resulting random walk.

\begin{center} 
\begin{figure}
  \includegraphics[scale=0.14]{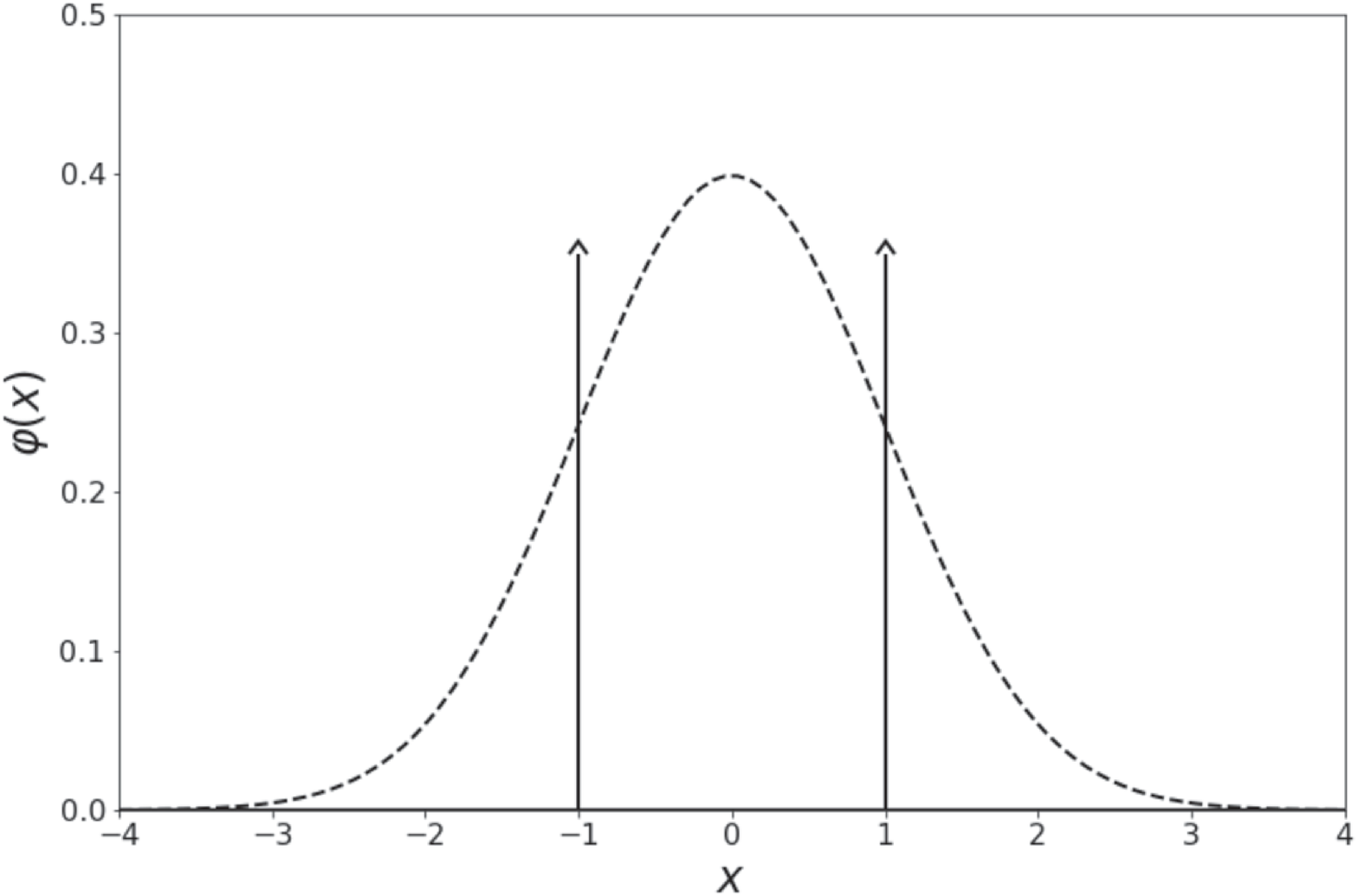}
  \includegraphics[scale=0.14]{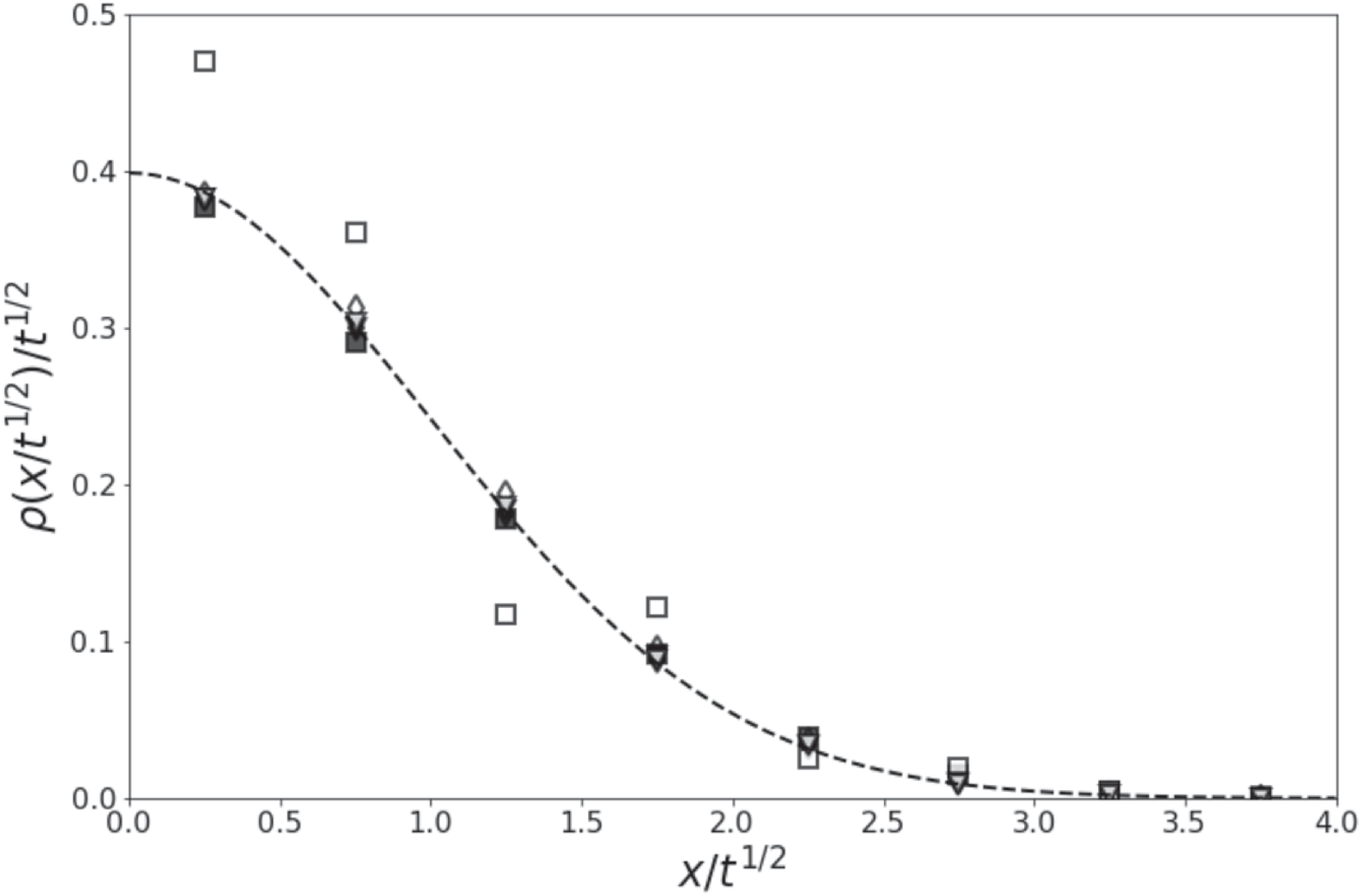}
\caption{
Left: plots of the one-dimensional ($N=1$)
jump $pdfs$ (\ref{gauss}) and (\ref{coinflip})
corresponding to the {\it Should I stay?} and {\it Should I go?} conditions,
respectively, for the generation of the Brownian motion from CTRW models.
Right: plots of the Gaussian walker's distribution $\rho(\bx;t)$ (\ref{gausspdf})
of the CTRW models for the Brownian motion
as generated by using the one-dimensional ($N=1$)
jump $pdfs$ (\ref{gauss}) (filled symbols) and (\ref{coinflip}) (empty symbols)
at $t= 10\tau \,, 100\tau \,, 1000\tau$ represented by
squares, diamons and triangles, respectively:
the short-time effects of the coin-flipping rule (\ref{coinflip})
is visible.}
\label{fig:BM}
\end{figure}
\end{center} 

In the case of the CTRW model for the Brownian motion,
this {\it undeclared statement} is true both in the
{\it ``Should I stay?"} and {\it ``Should I go?"} conditions,
see Figure \ref{fig:BM}, in fact, in the isotropic case,
it holds
\begin{subequations}
\be
\varphi(\bx)=
\frac{1}{(4 \pi \ell^2)^{N/2}}
\, \e^{-\frac{|\bx|^2}{4\ell^2}} \,,
\label{gauss}
\ee
\vspace*{-5pt}
\be
\widehat{\varphi}(\bkappa)= \e^{-\ell^2 |\bkappa|^2}
\simeq 1 - \ell^2 |\bkappa|^2 + \frac{\ell^4}{2}|\bkappa|^4
+ o(|\bkappa|^4) \,, \quad \ell |\bkappa| \ll 1 \,,
\ee
\end{subequations}
and also
\begin{subequations}
\be
\varphi(\bx)= \frac{1}{2} [
\delta(\bx - \sqrt{2}\ell \, \widehat{\bolde}) +
\delta(\bx + \sqrt{2}\ell \, \widehat{\bolde})] \,,
\label{coinflip}
\ee
where $\widehat{\bolde}$ is the basis vector
($\widehat{\bolde} \cdot \widehat{\bolde}=1$)
with isotropic symmetry so
$\bx=|\bx| \, \widehat{\bolde}= \sqrt{2}\ell \, \widehat{\bolde}$
and $\bkappa=|\bkappa| \, \widehat{\bolde}$ such that
\begin{eqnarray}
\widehat{\varphi}(\bkappa)
&=& \cos (\sqrt{2}\ell \, \bkappa \cdot \widehat{\bolde})
= \cos (\sqrt{2}\ell |\bkappa|
\, \widehat{\bolde} \cdot \widehat{\bolde}) \\
&\simeq& 1 - \ell^2 \, |\bkappa|^2 + \frac{\ell^4}{6} |\bkappa|^4
+ o(|\bkappa|^4) \,,
\quad \ell |\bkappa| \ll 1 \,. \nonumber
\label{coinflipseries}
\end{eqnarray}
\end{subequations}

\smallskip

Conversely, in the case of CTRW models for L\'evy flights,
although this {\it undeclared statement} is true in the
{\it ``Should I stay?"} condition
and in fact, with $0 < \alpha < 2$, it holds
\begin{subequations}
\be
\varphi(\bx)=
\frac{1}{\ell^N} \mathcal{L}_\alpha\left(\frac{\bx}{\ell}\right)
\sim \frac{1}{|\bx|^{\alpha+N}} \,, \quad
|\bx| \to + \infty \,,
\label{Ljump}
\ee
\be
\widehat{\varphi}(\bkappa)=\e^{-\ell^\alpha \, |\bkappa|^\alpha}
\simeq 1 - \ell^\alpha \, |\bkappa|^\alpha
+ \frac{\ell^{2\alpha}}{2} |\bkappa|^{2\alpha}
+ o(|\bkappa|^{2\alpha}) \,,
\quad \ell |\bkappa| \ll 1 \,,
\label{Lseries}
\ee
\end{subequations}
and $\rho(\bx;t)$ solves (\ref{SFDE}),
unfortunately,
the alternating sign expansion (\ref{Lseries}) is not always true in the
{\it ``Should I go?"} condition.

\smallskip

For mathematical convenience, we provide an example,
in the one-dimensional ($N=1$) case, of a jump $pdf$ whose
small wavelength expansion of the characteristic function
is not an alternating series.

\smallskip

In order to arrange the {\it ``Should I go?"} condition within
the framework of power-law distributed jumps, we consider
the one-sided (extremal) L\'evy densities
$\mathcal{L}_{\alpha}^{-\alpha}(x)$
\cite{penson_etal-prl-2010}, with $x \in \R$,
i.e.,
$\mathcal{L}_{\alpha}^{-\alpha}(x) > 0$ when $x > 0$ and
$\mathcal{L}_{\alpha}^{-\alpha}(x) = 0$ when $x \le 0$,
with $0 < \alpha < 1$.
Thus one-sided L\'evy densities can be used
for defining a jump rule {\it \`a la} coin-flipping
by taking into account also the remarkable
limit $\mathcal{L}_1^{-1}(x)=\delta(x-1)$.
The power-law of the tails of the jump $pdf$ $\varphi(x)$
is spanned inside the range of the stable parameter
$(0,1) \cup (1,2)$ as follows
\begin{subnumcases}{\varphi(x)=}
\displaystyle{
\frac{1}{2} \, \frac{1}{\sqrt{2} \, \ell}
\, \mL\!\left(\frac{|x|}{\sqrt{2} \ell} \right)
\sim \frac{1}{|x|^{\alpha+1}} \,, \quad
|x| \to + \infty} \,,
\label{jumppdf}
\\
\nonumber \\
\displaystyle{
\frac{1}{2} \, \frac{\alpha}{\Gamma(1/\alpha) |x|}
\, \mL\!\left(\frac{|x|}{\sqrt{2} \ell} \right)
\sim \frac{1}{|x|^{(\alpha+1)+1}} \,, \quad
|x| \to + \infty} \,.
\label{jumppdf2}
\end{subnumcases}
We observe that the Brownian coin-flipping rule (\ref{coinflip}) is recovered
from both the L\'evy coin-flipping rules
(\ref{jumppdf}) and (\ref{jumppdf2}) when $\alpha=1$,
while the special case of the Cauchy distribution (\ref{cauchy})
is not achievable.
A study of the considered L\'evy coin-flipping rules
(\ref{jumppdf}, \ref{jumppdf2}) is moved in Appendix A
where formulae useful for the following analysis are derived.

  \begin{center} 
\begin{figure}
  \includegraphics[scale=0.14]{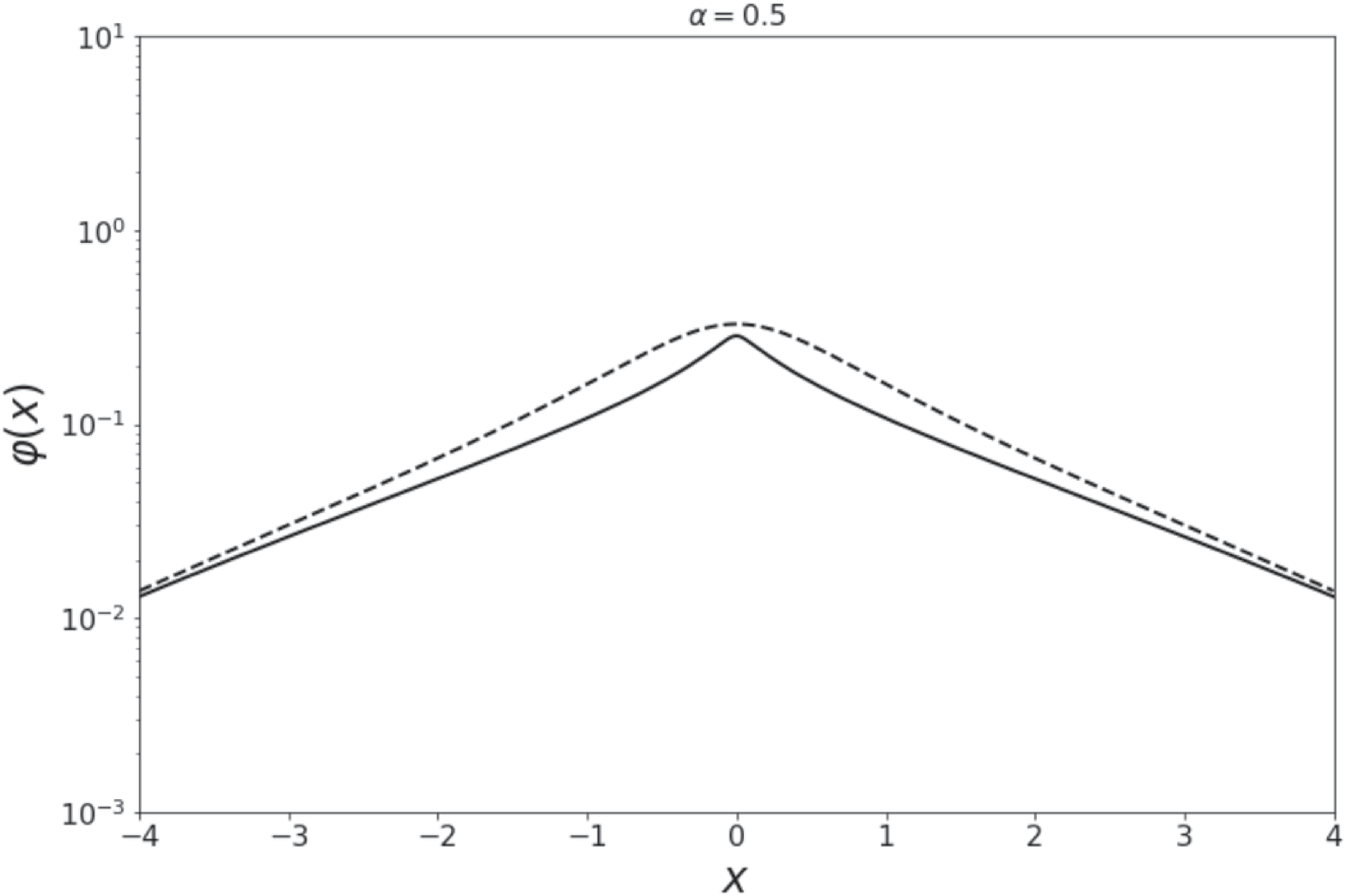}
  \includegraphics[scale=0.195]{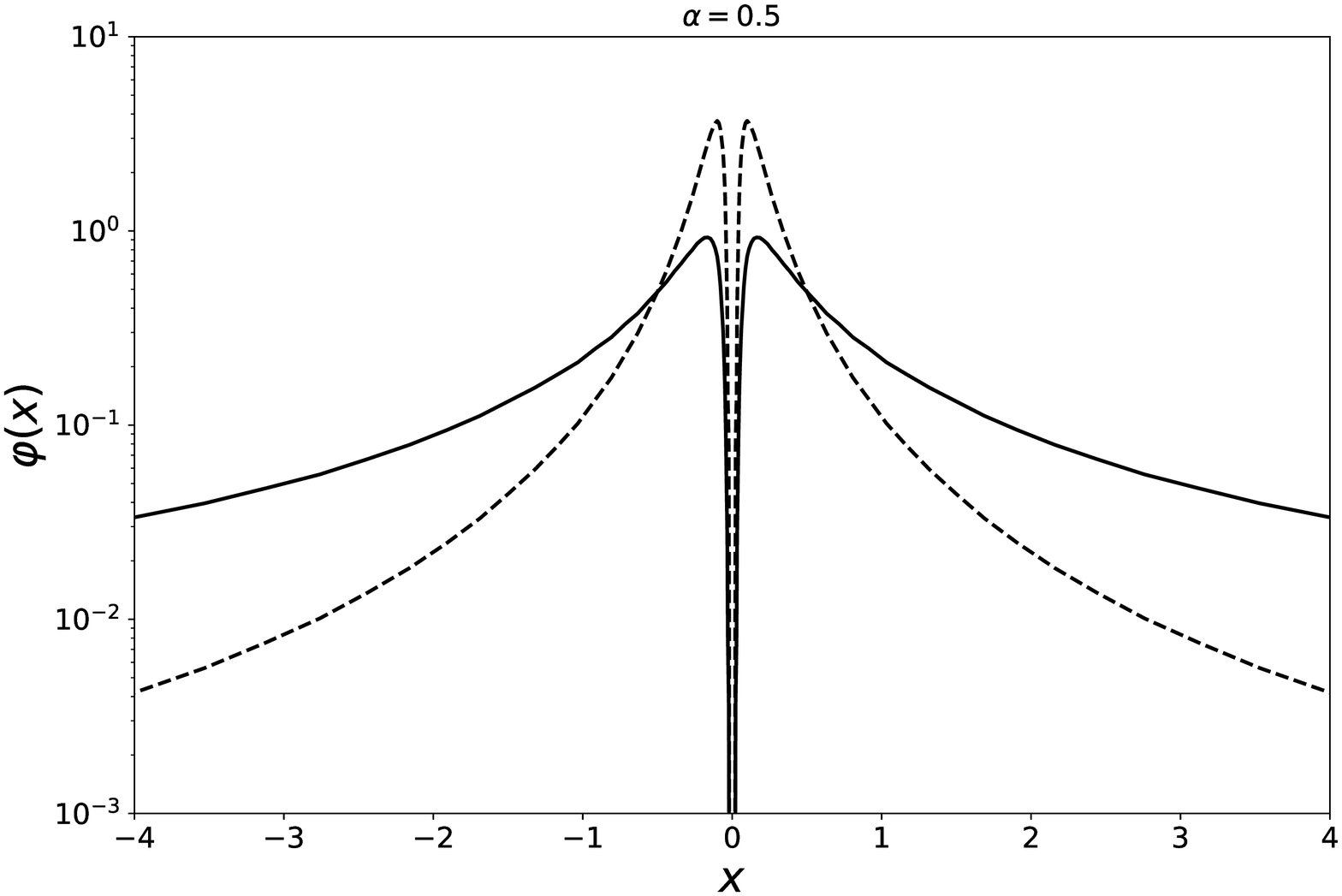} 
  \includegraphics[scale=0.14]{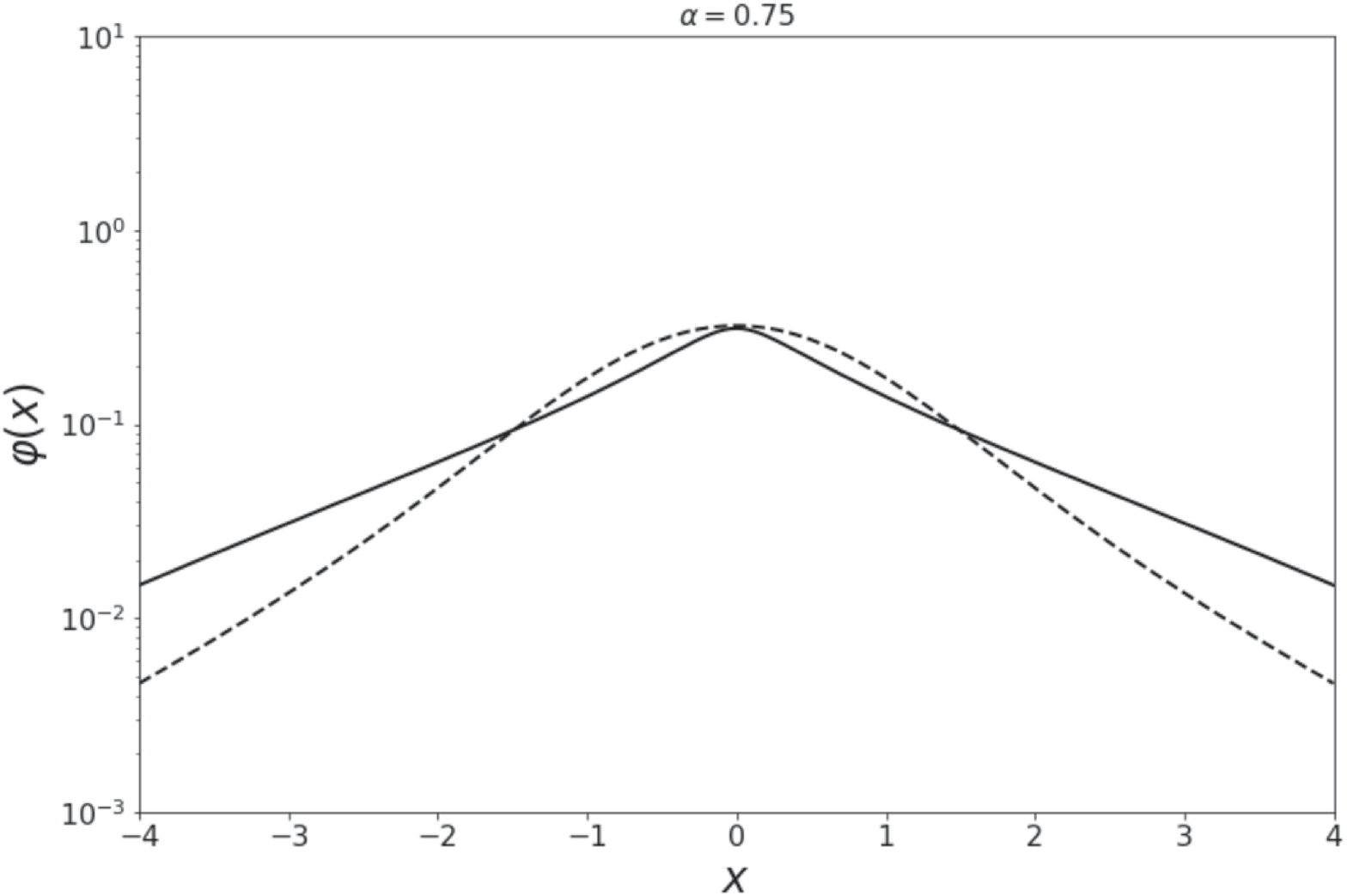}
  \includegraphics[scale=0.195]{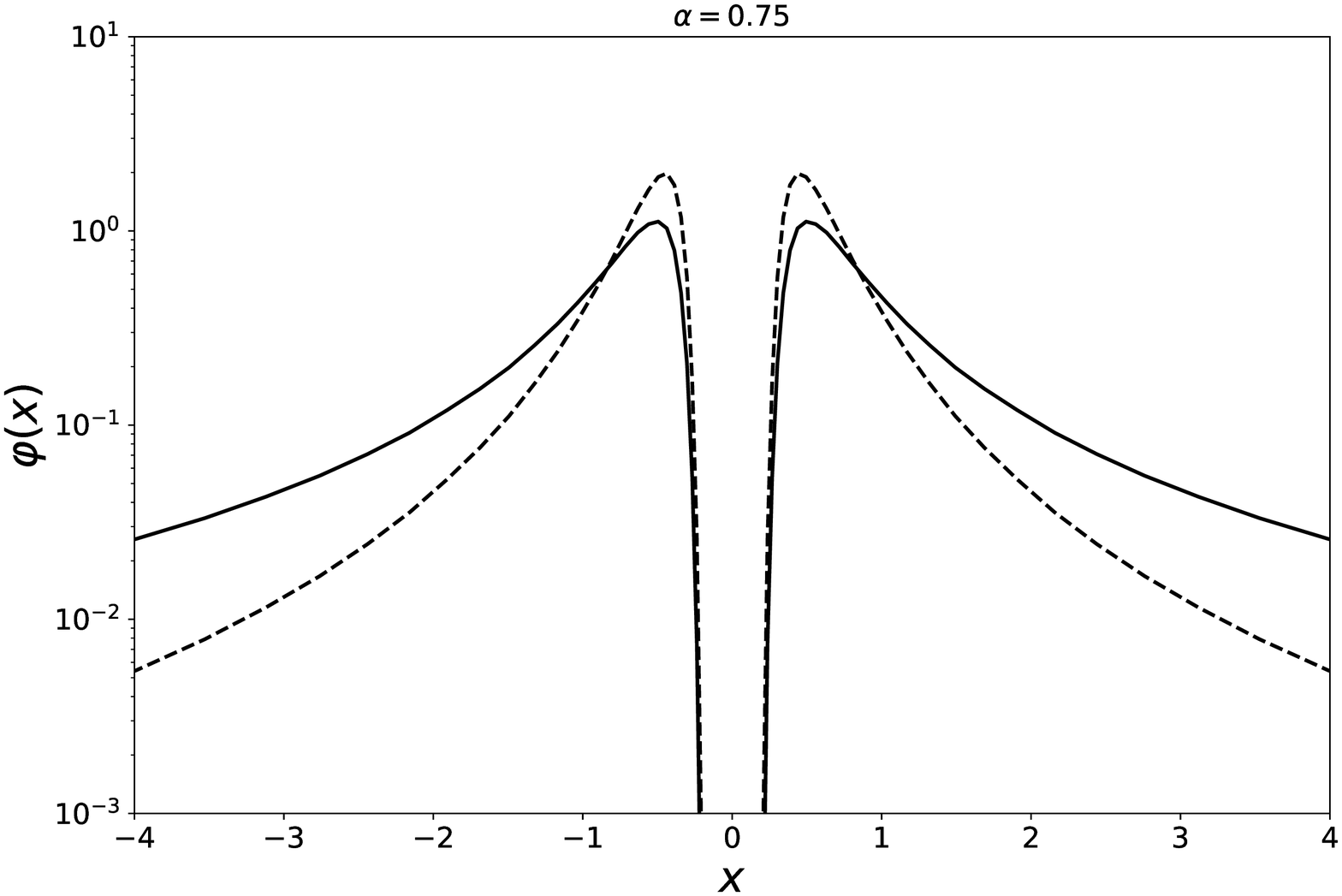}
  \includegraphics[scale=0.14]{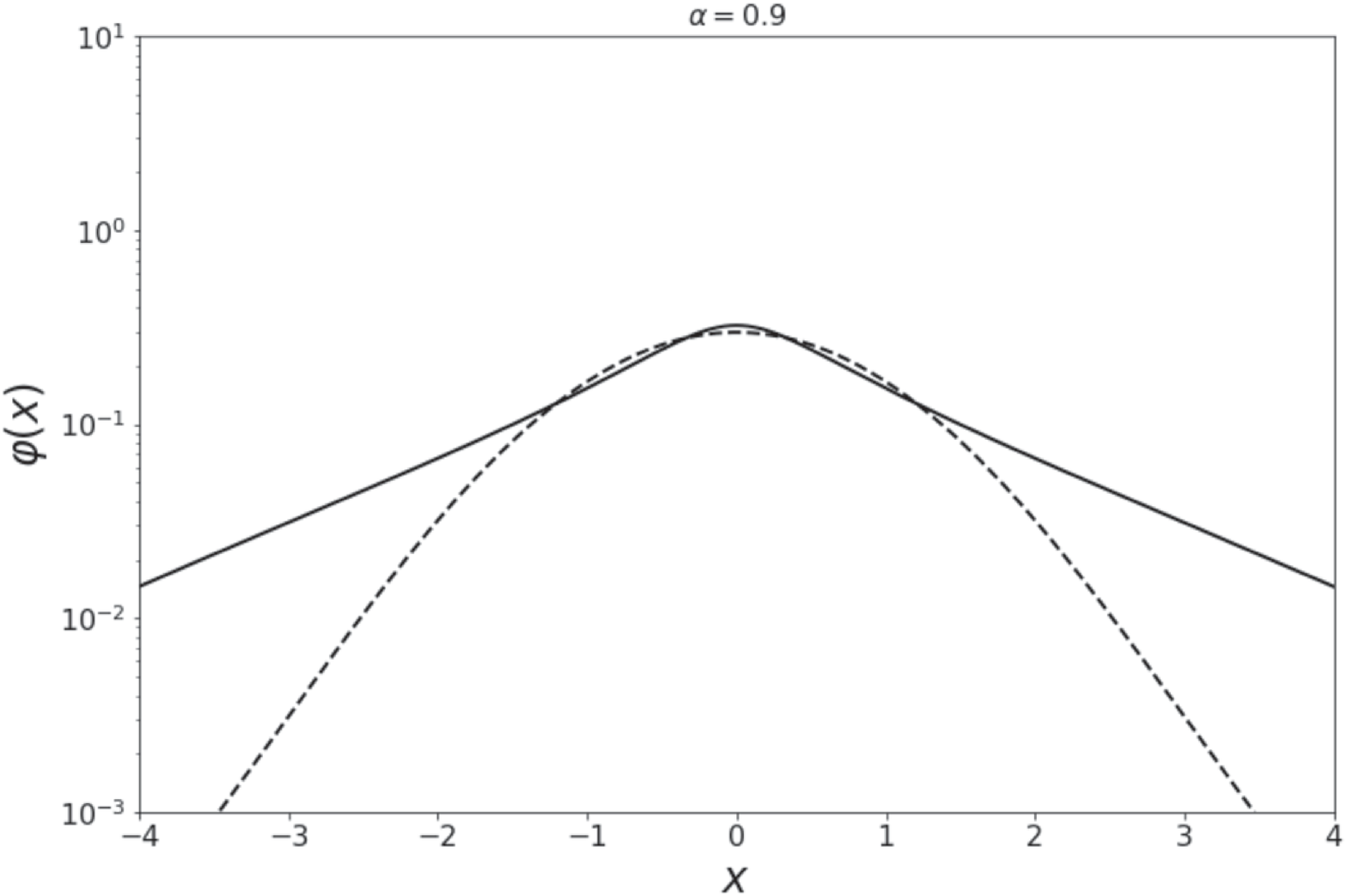}
  \includegraphics[scale=0.195]{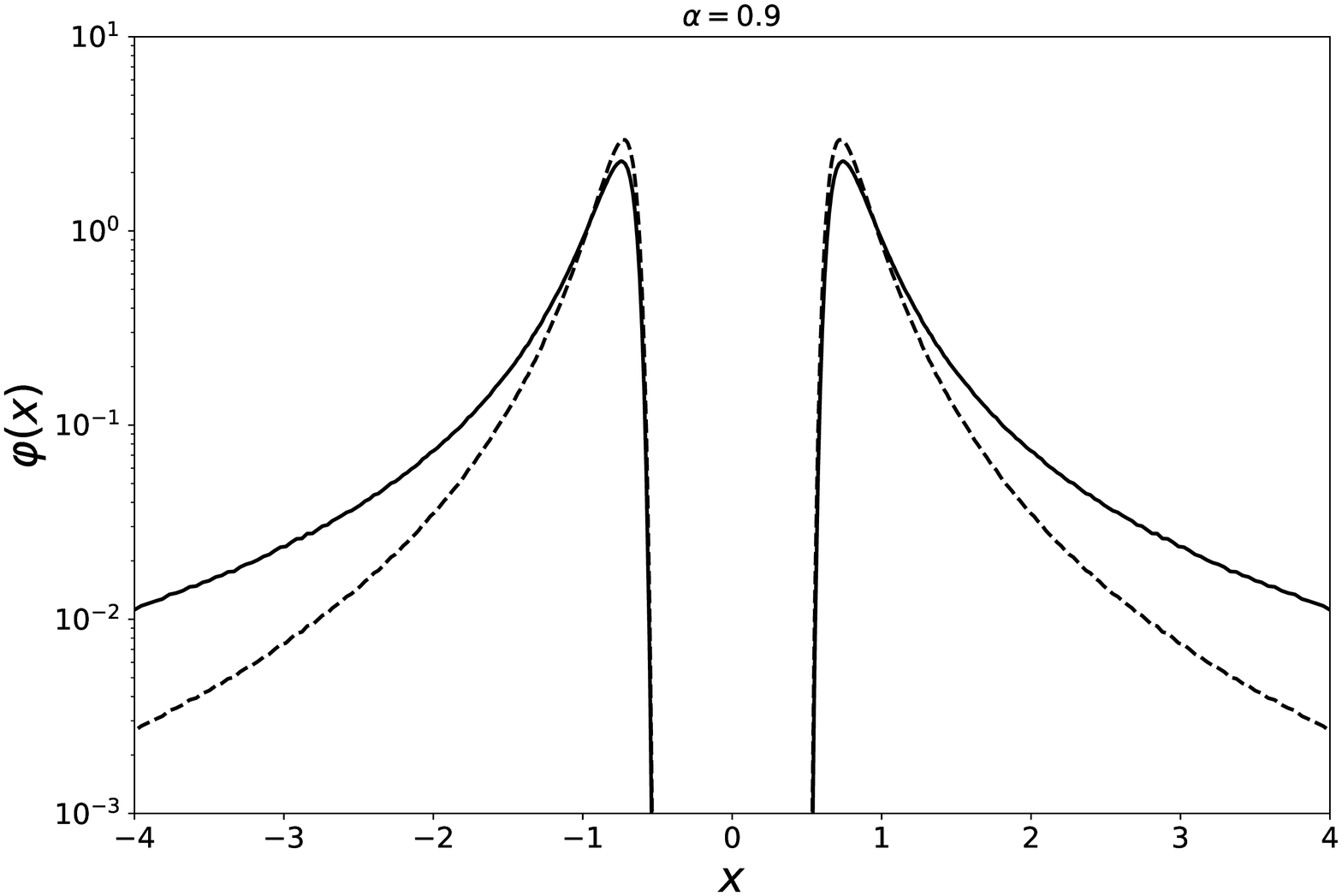}
 \bigskip  \bigskip
\caption{Plots of the one-dimensional ($N=1$)
jump $pdfs$ (\ref{Ljump}) - left column - and
(\ref{jumppdf}, \ref{jumppdf2}) - right column -
corresponding to the {\it Should I stay?} and {\it Should I go?} conditions,
respectively.
Left column: solid lines decrease as $|x|^{-(\alpha+1)}$ 
and dashed lines as $|x|^{-(2\alpha+1)}$ 
with $0 < \alpha < 1$.
Right column: solid lines decrease as $|x|^{-(\alpha + 1)}$
and dashed lines as $|x|^{-[(\alpha +1) + 1]}$ with
$0 < \alpha < 1$.}
\label{fig:levyjumps}
\end{figure}
  \end{center} 

See plots of the one-dimensional ($N=1$) jump $pdfs$ (\ref{Ljump}) and (\ref{jumppdf}, \ref{jumppdf2})
in Figure \ref{fig:levyjumps}.

 \medskip

\begin{center} 
\begin{figure}
  \includegraphics[scale=0.14]{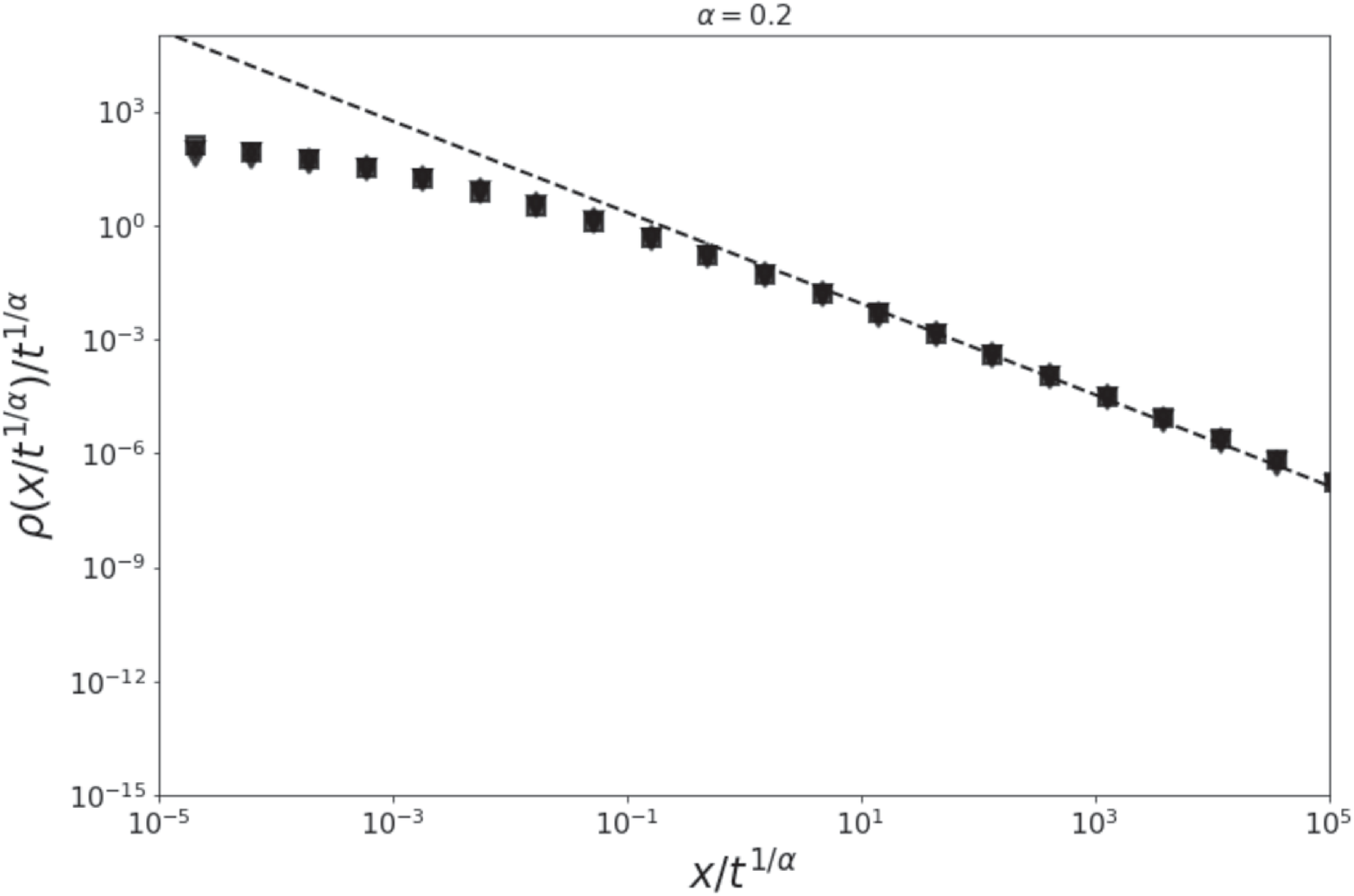}
  \includegraphics[scale=0.14]{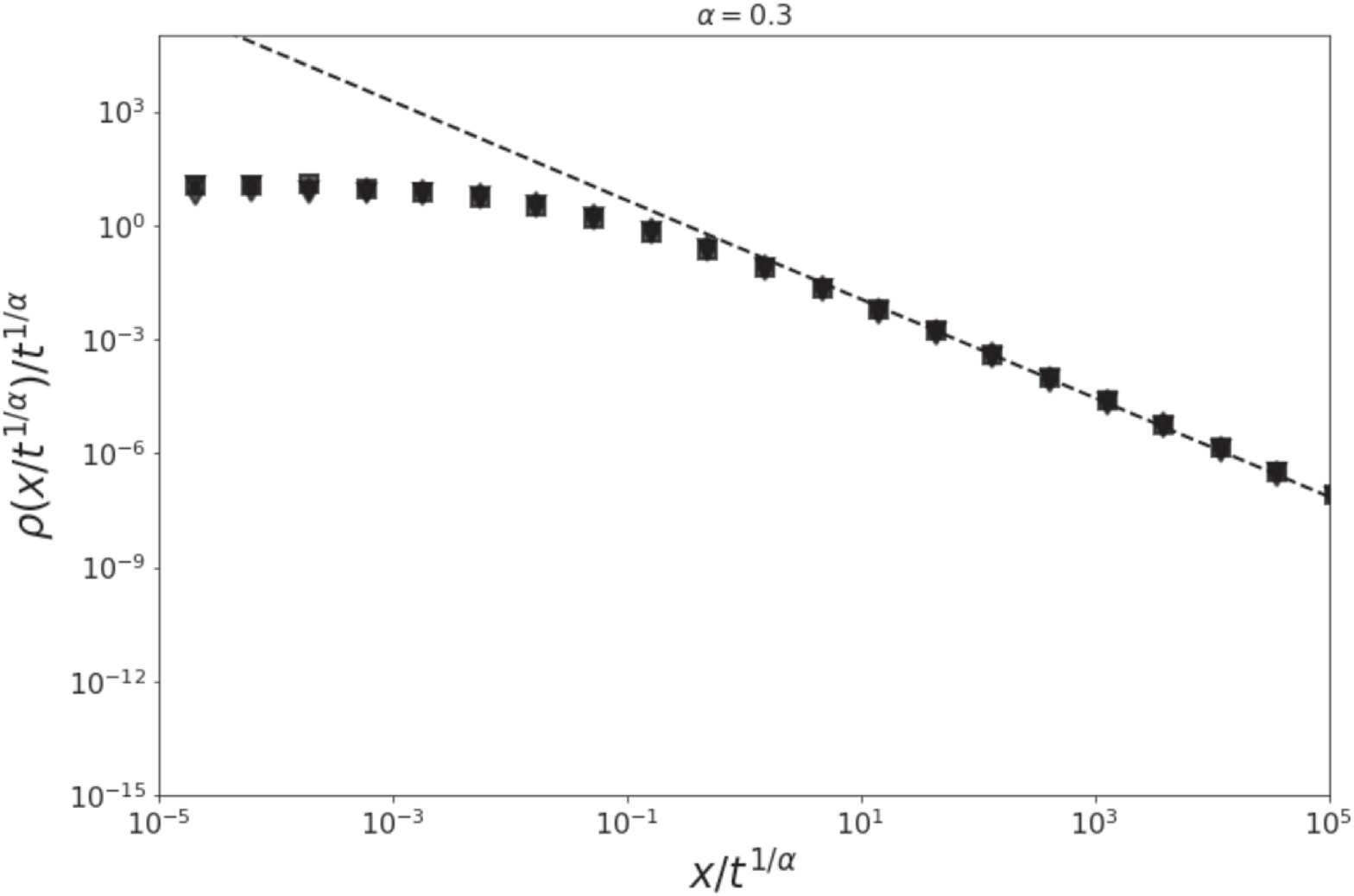}
  \includegraphics[scale=0.14]{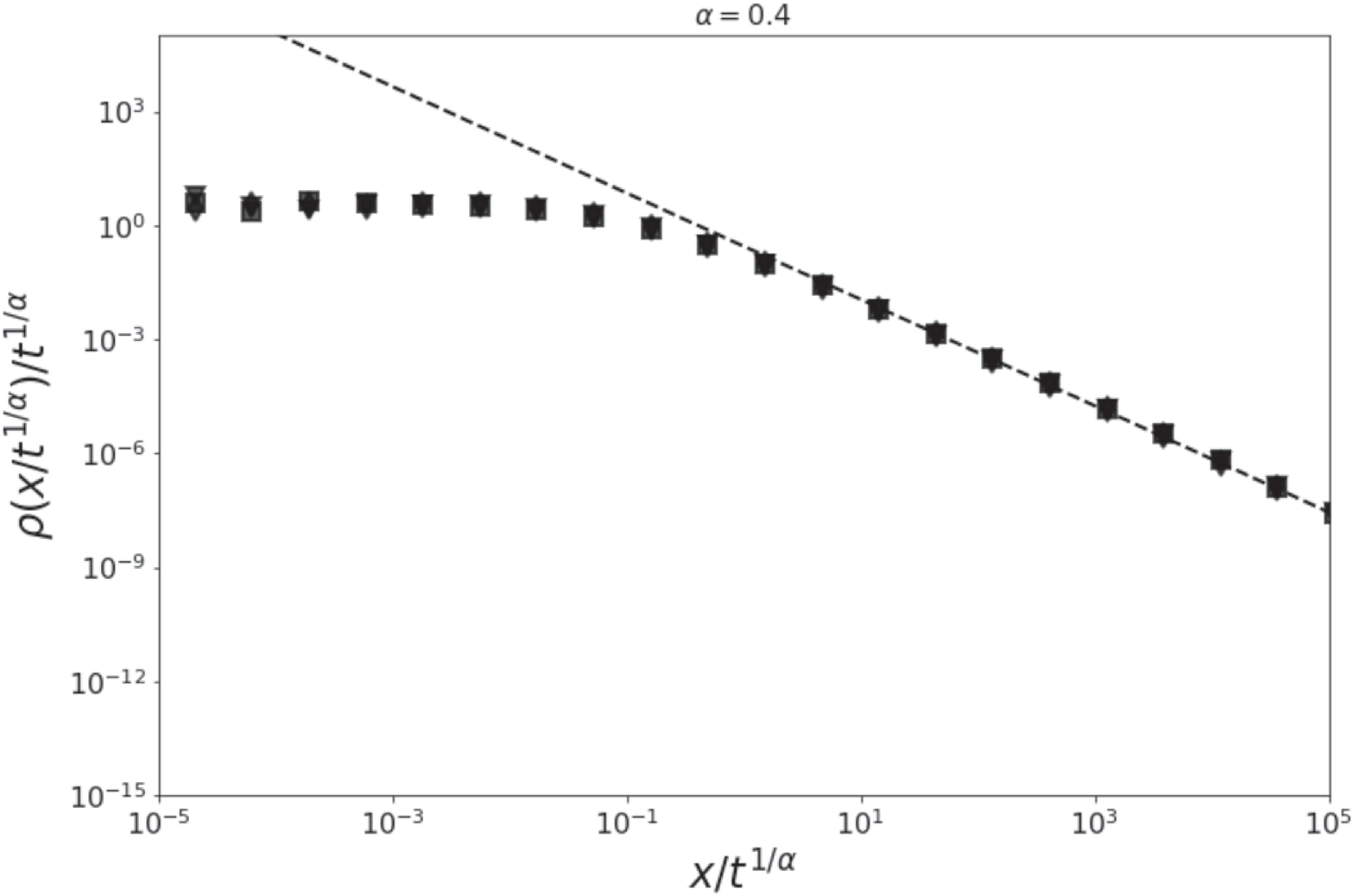}
  \includegraphics[scale=0.14]{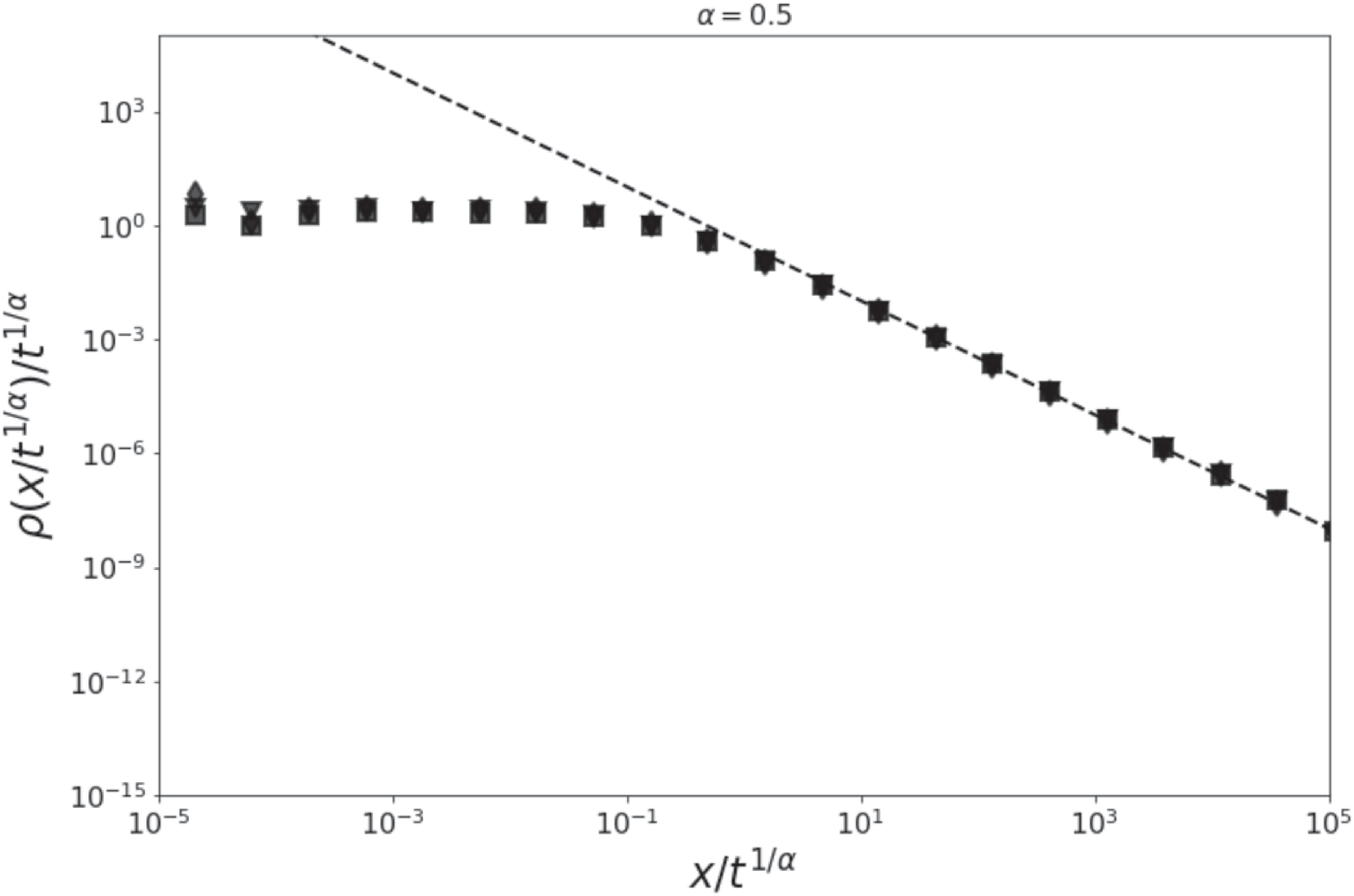}
  \includegraphics[scale=0.14]{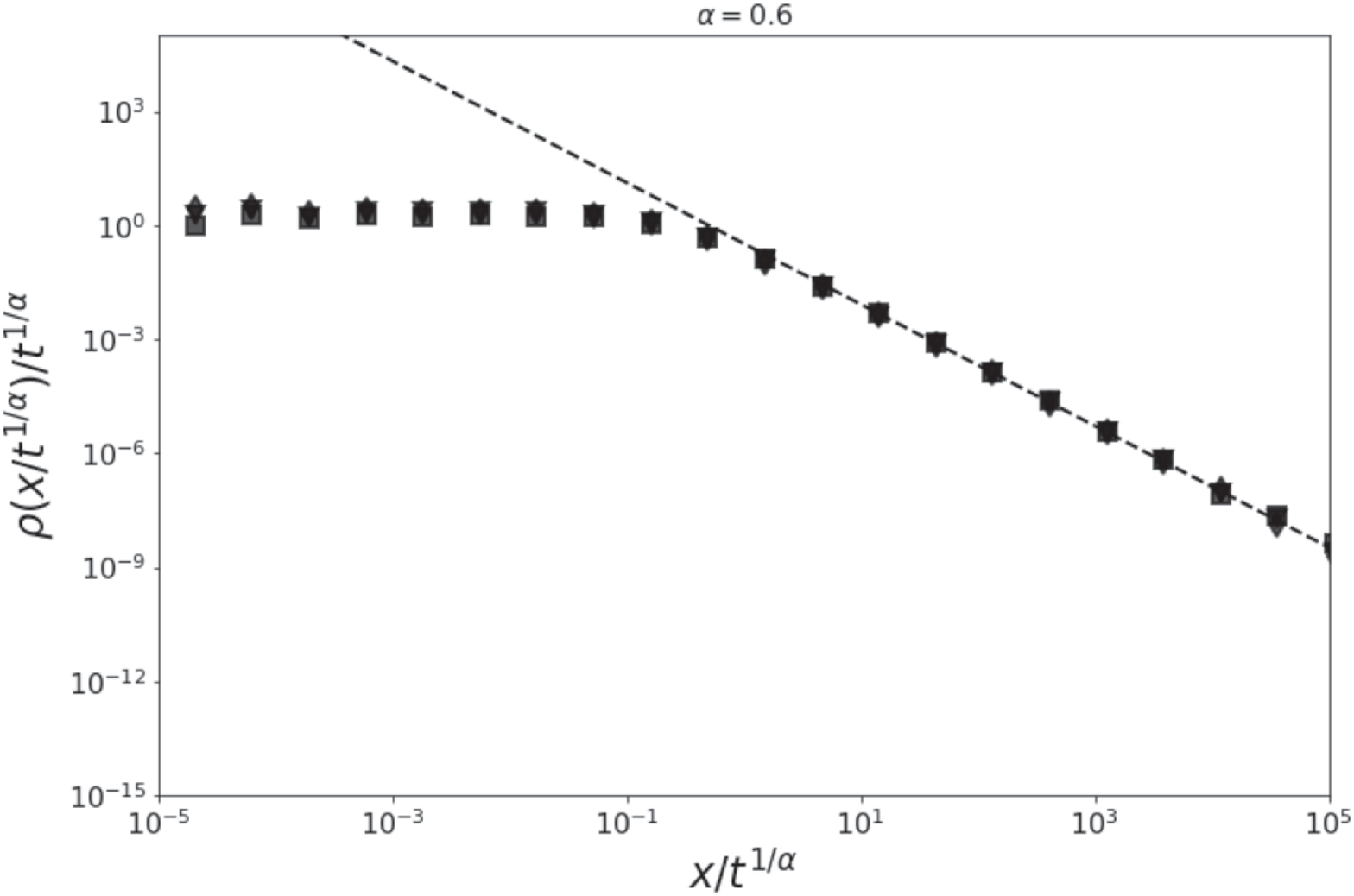}
  \includegraphics[scale=0.14]{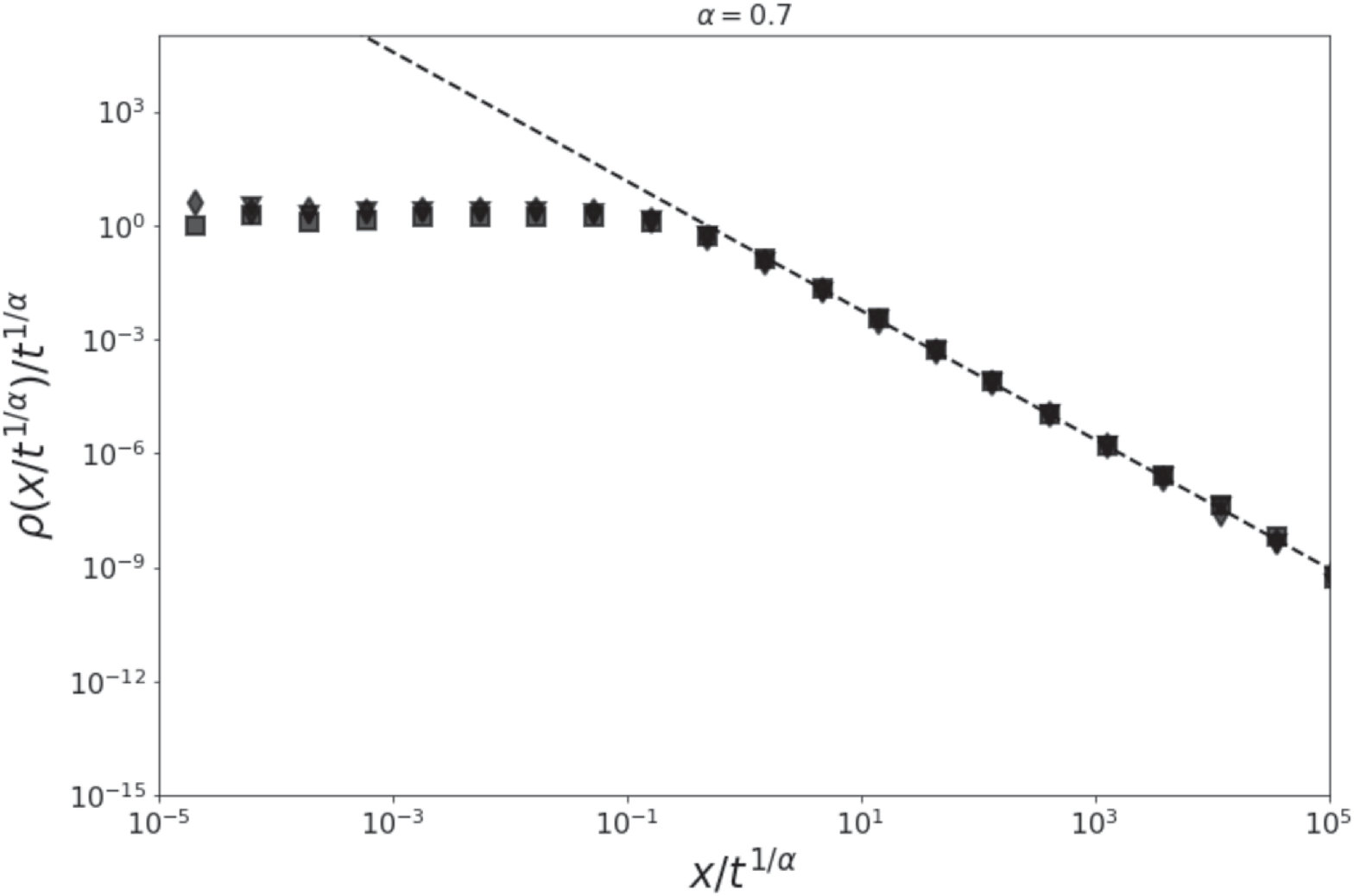}
  \includegraphics[scale=0.14]{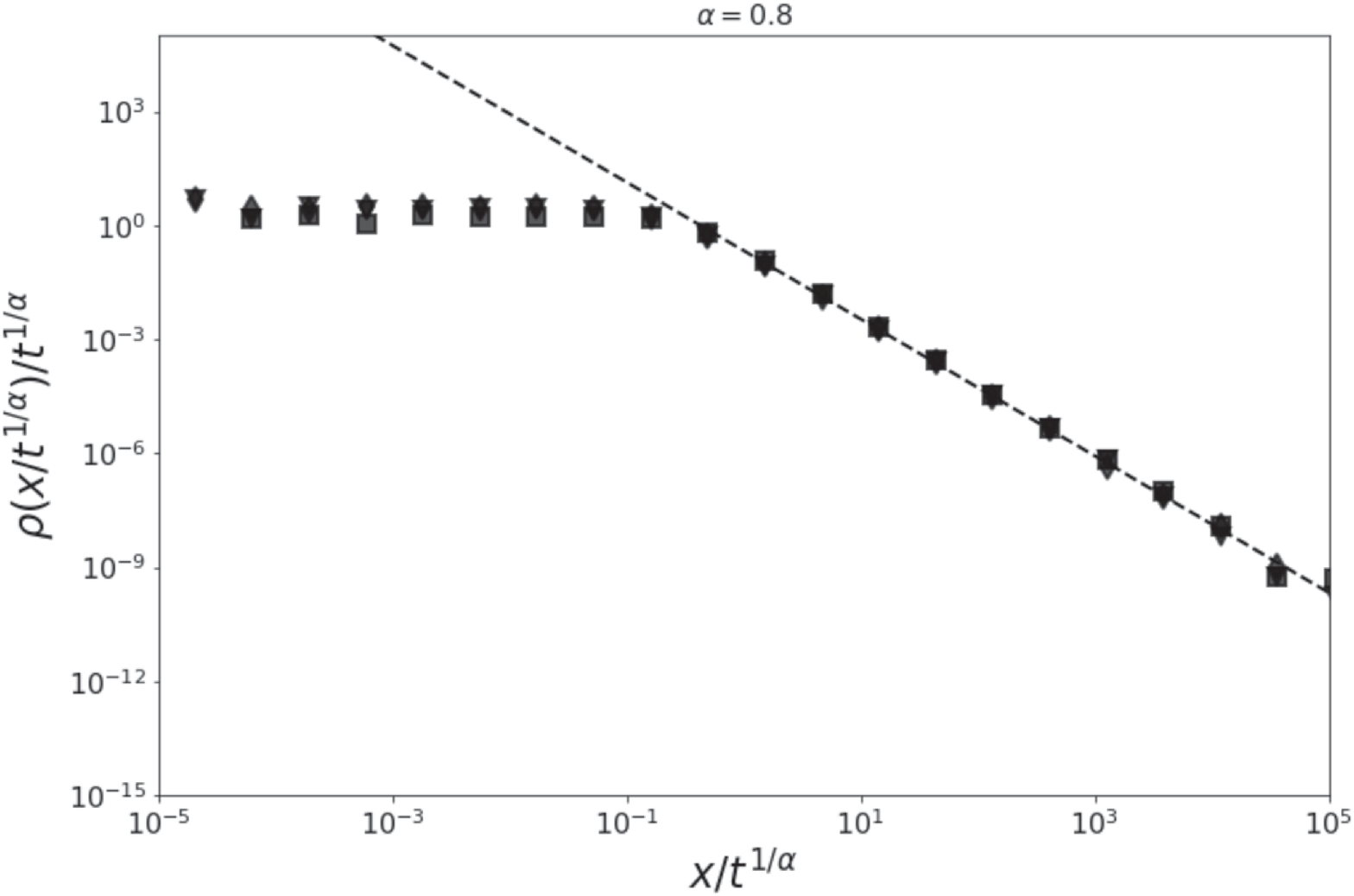}
  \includegraphics[scale=0.14]{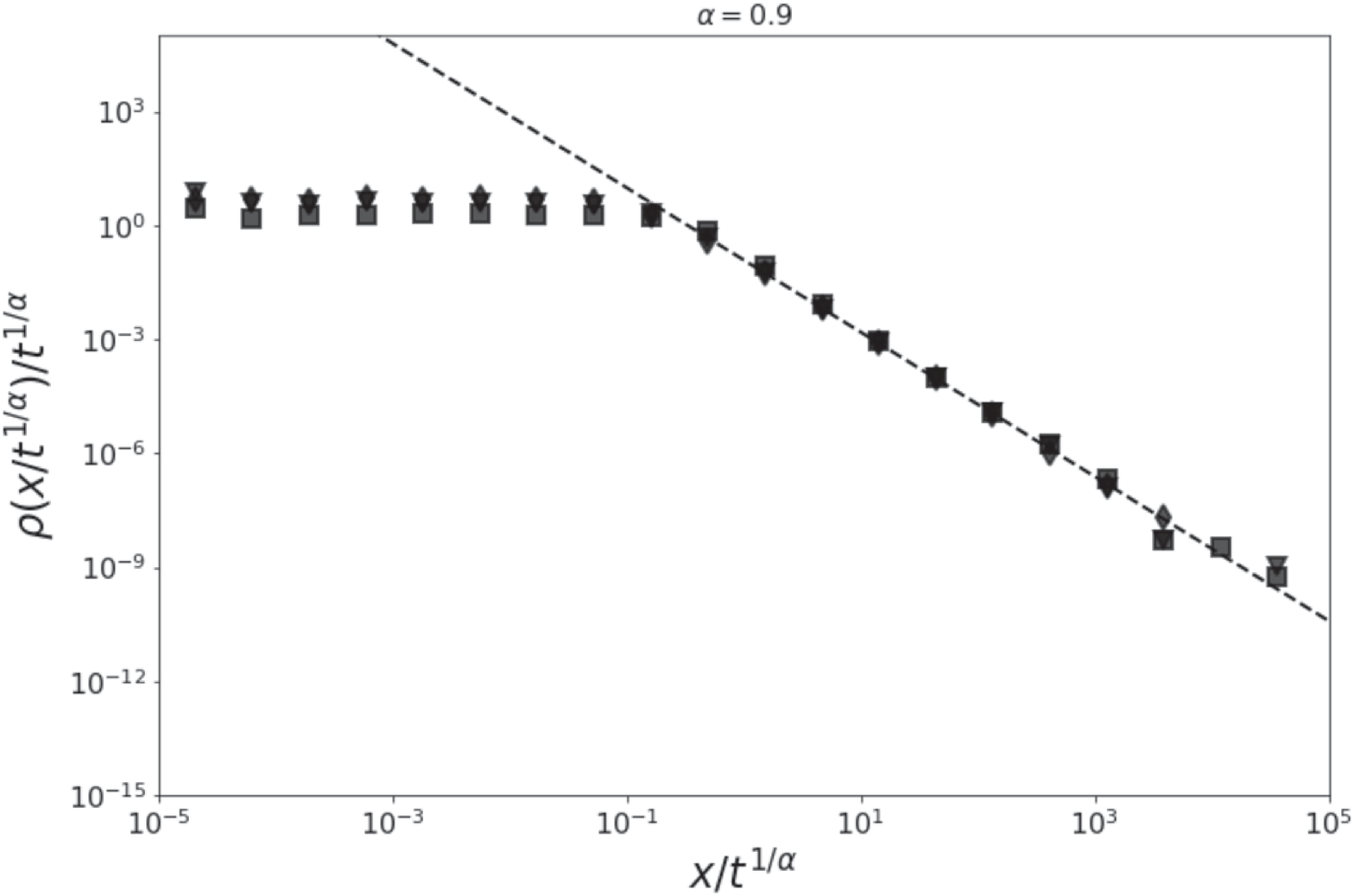}
\caption{Plots of the tails of the walker's distribution $\rho(x;t)$
obtained with jump $pdf$ (\ref{jumppdf})
at times $t=10 \tau, 100\tau, 1000\tau$ marked
by squares, triangles and diamonds, respectively.
The dashed lines represent the power-law
decaying $|x|^{-(\alpha + 1)}$.
}
\label{fig:tails1a}
\end{figure}
\end{center}

\begin{center} 
\begin{figure}
  \includegraphics[scale=0.2]{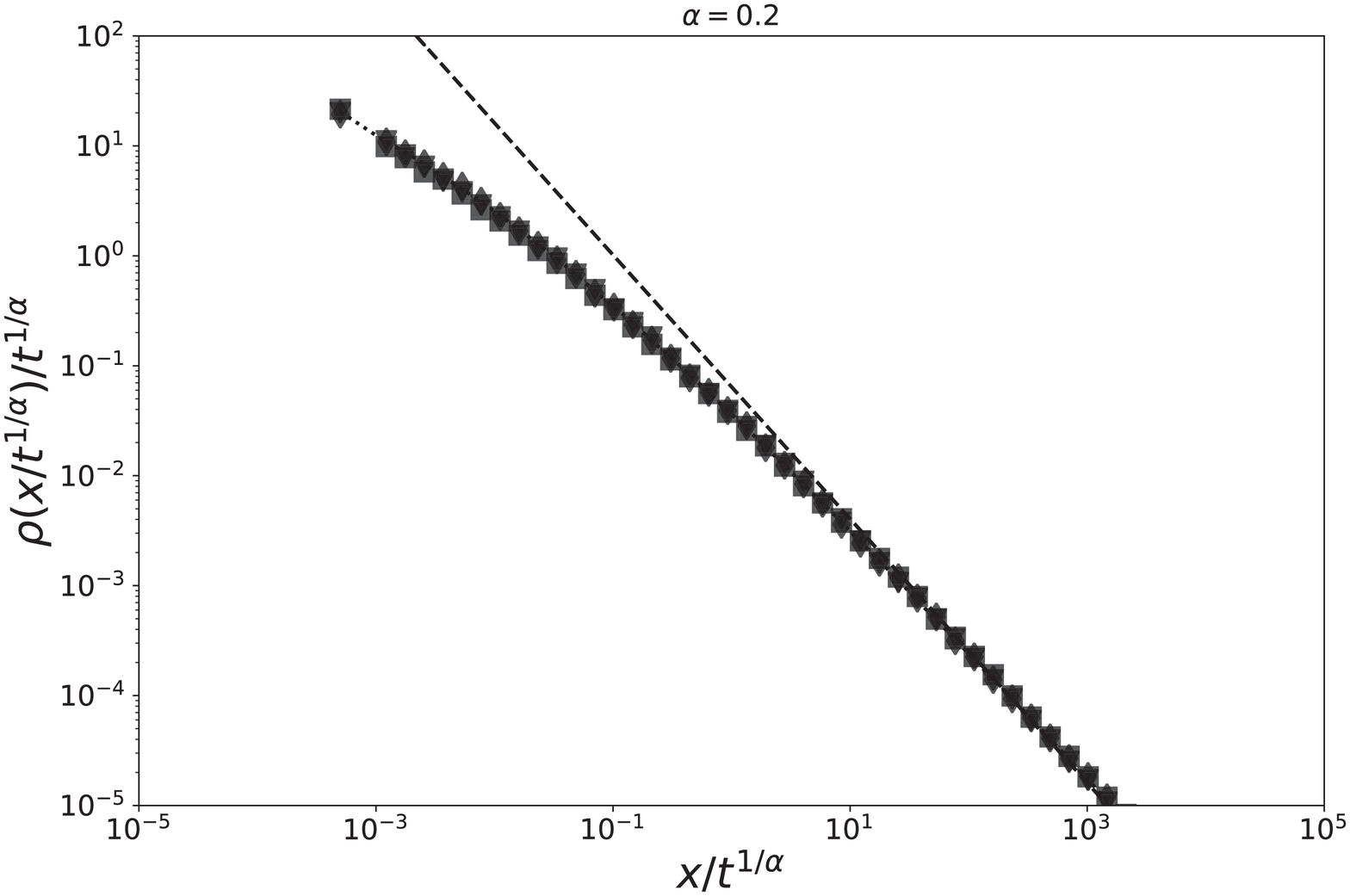}
  \includegraphics[scale=0.2]{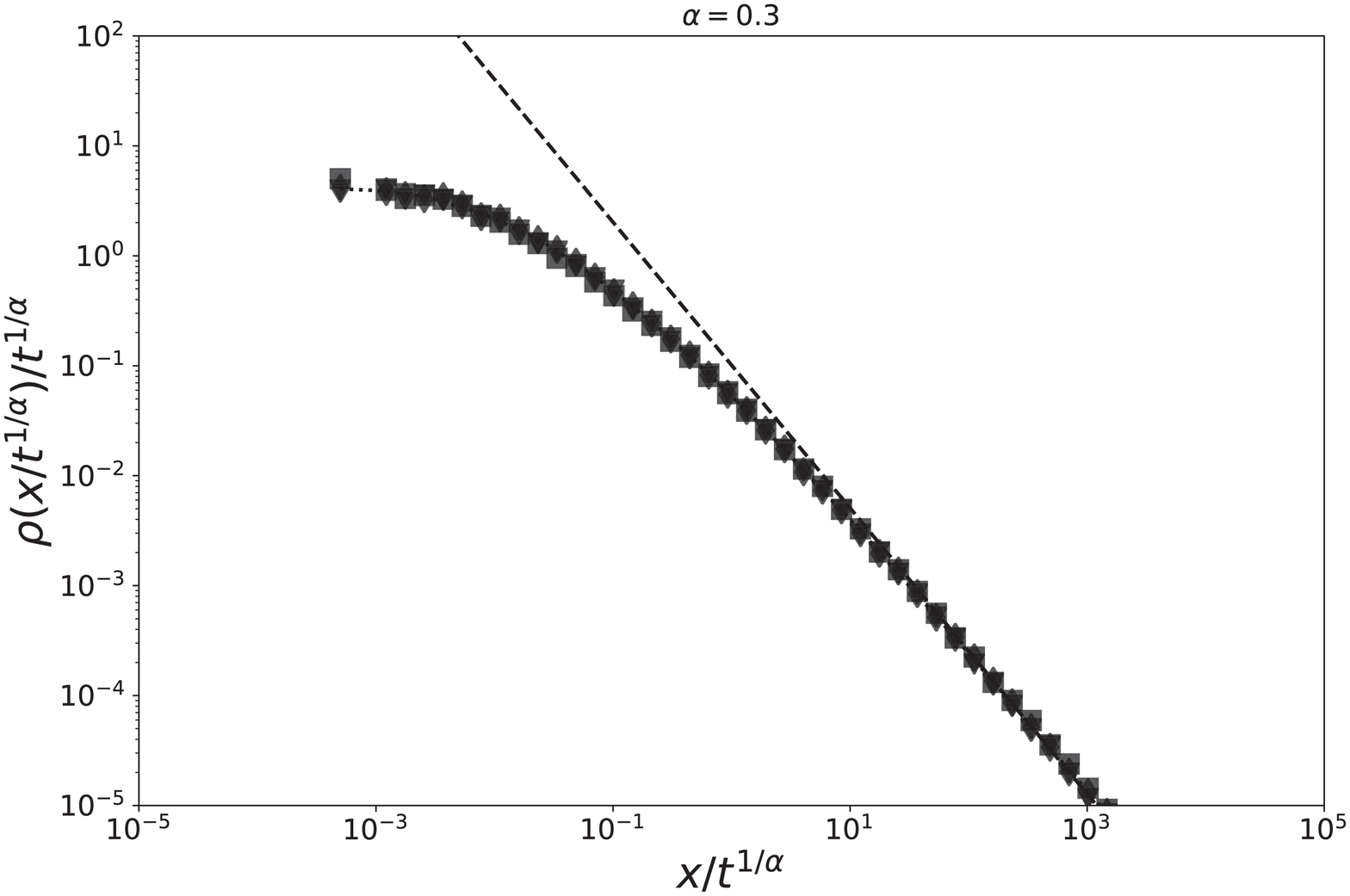}
  \includegraphics[scale=0.2]{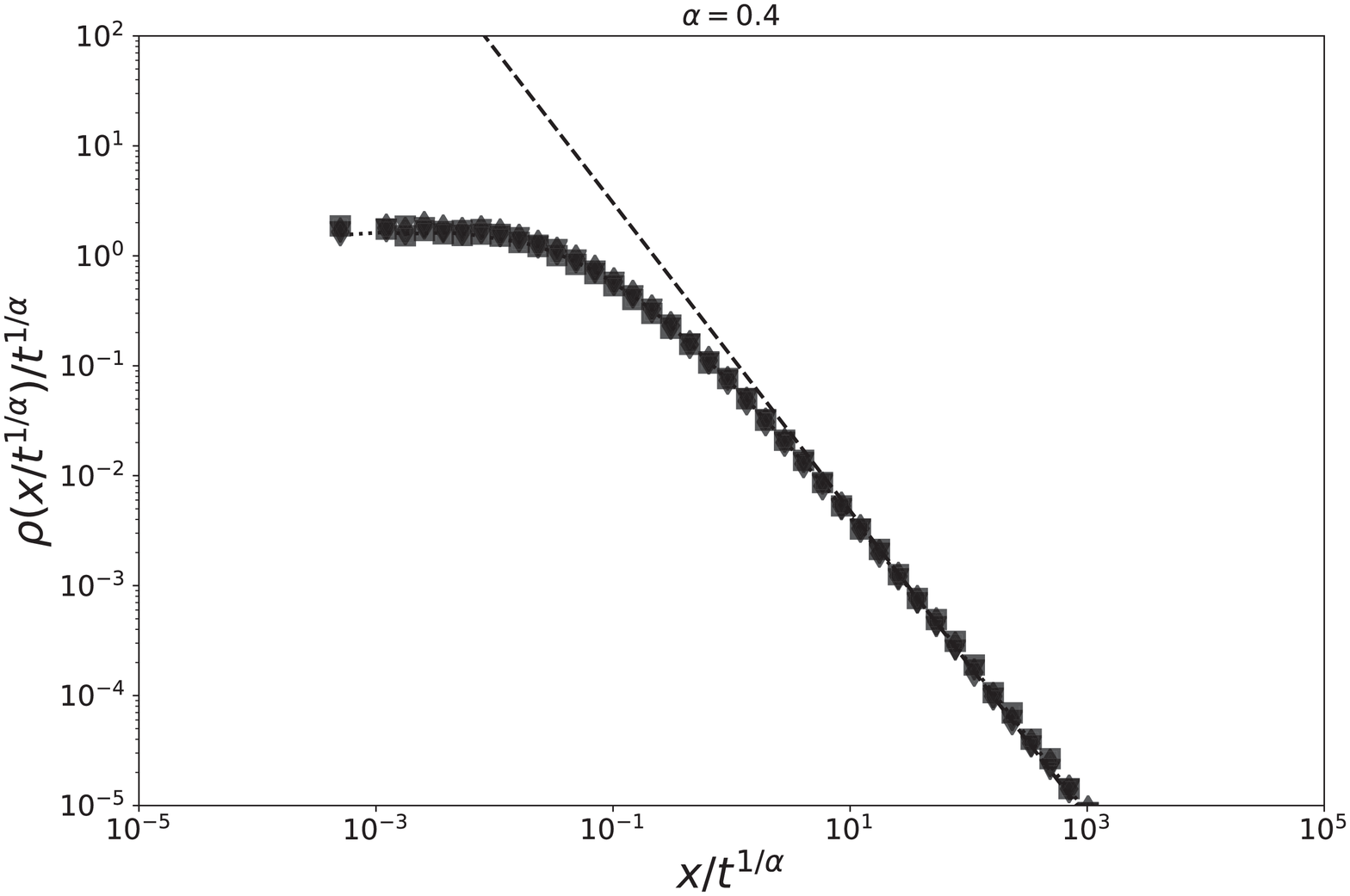}
  \includegraphics[scale=0.2]{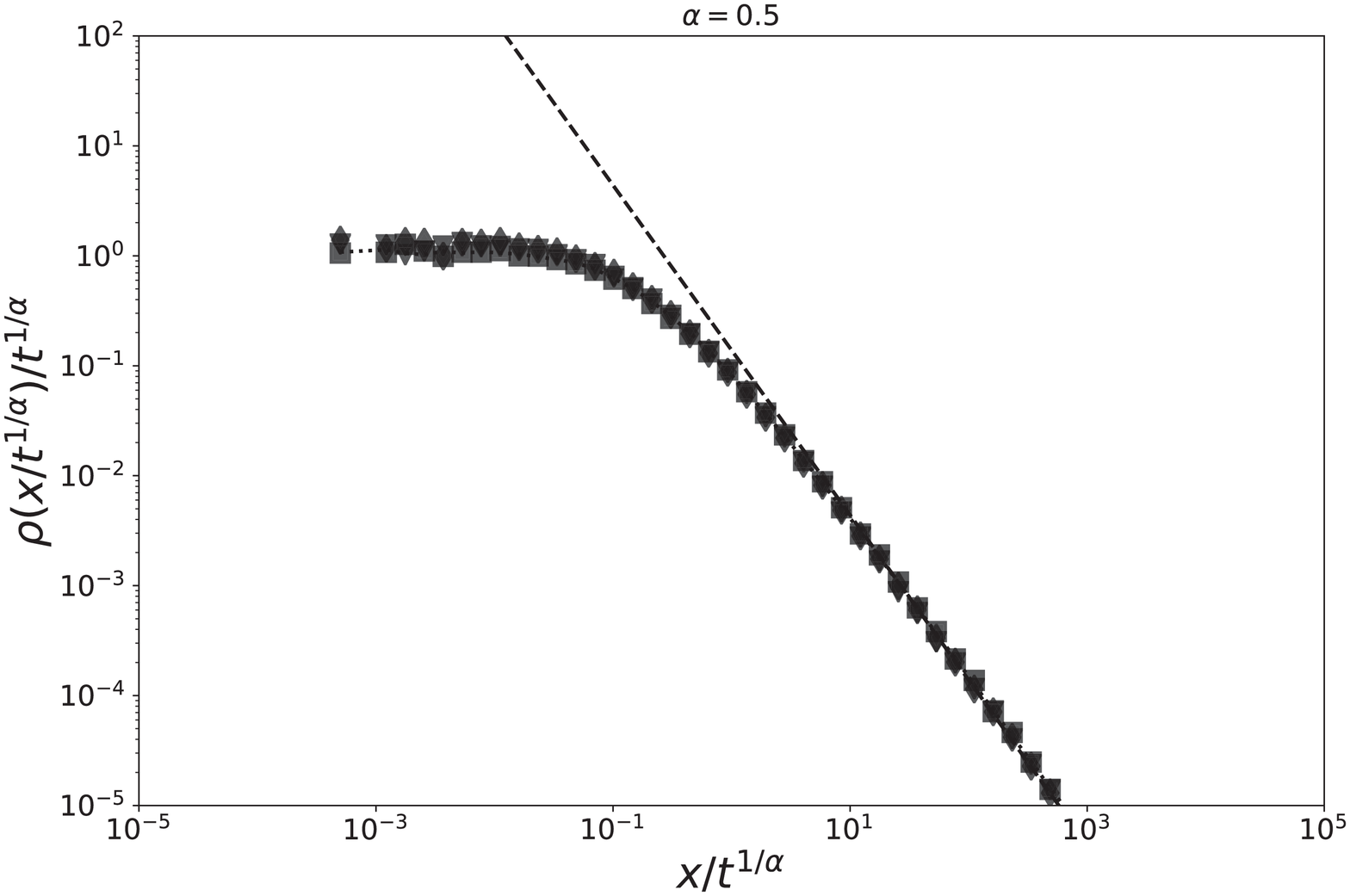}
  \includegraphics[scale=0.2]{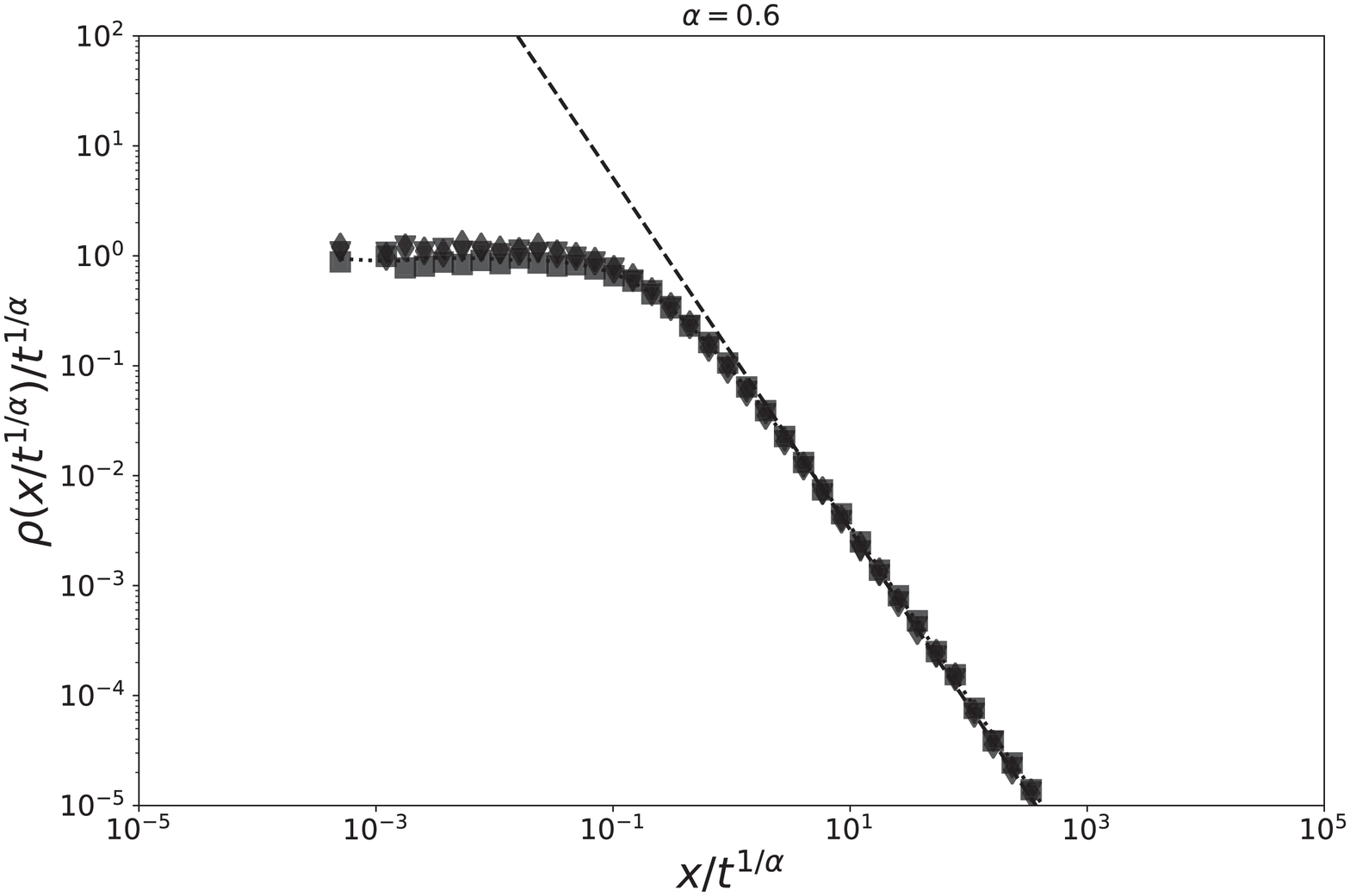}
  \includegraphics[scale=0.2]{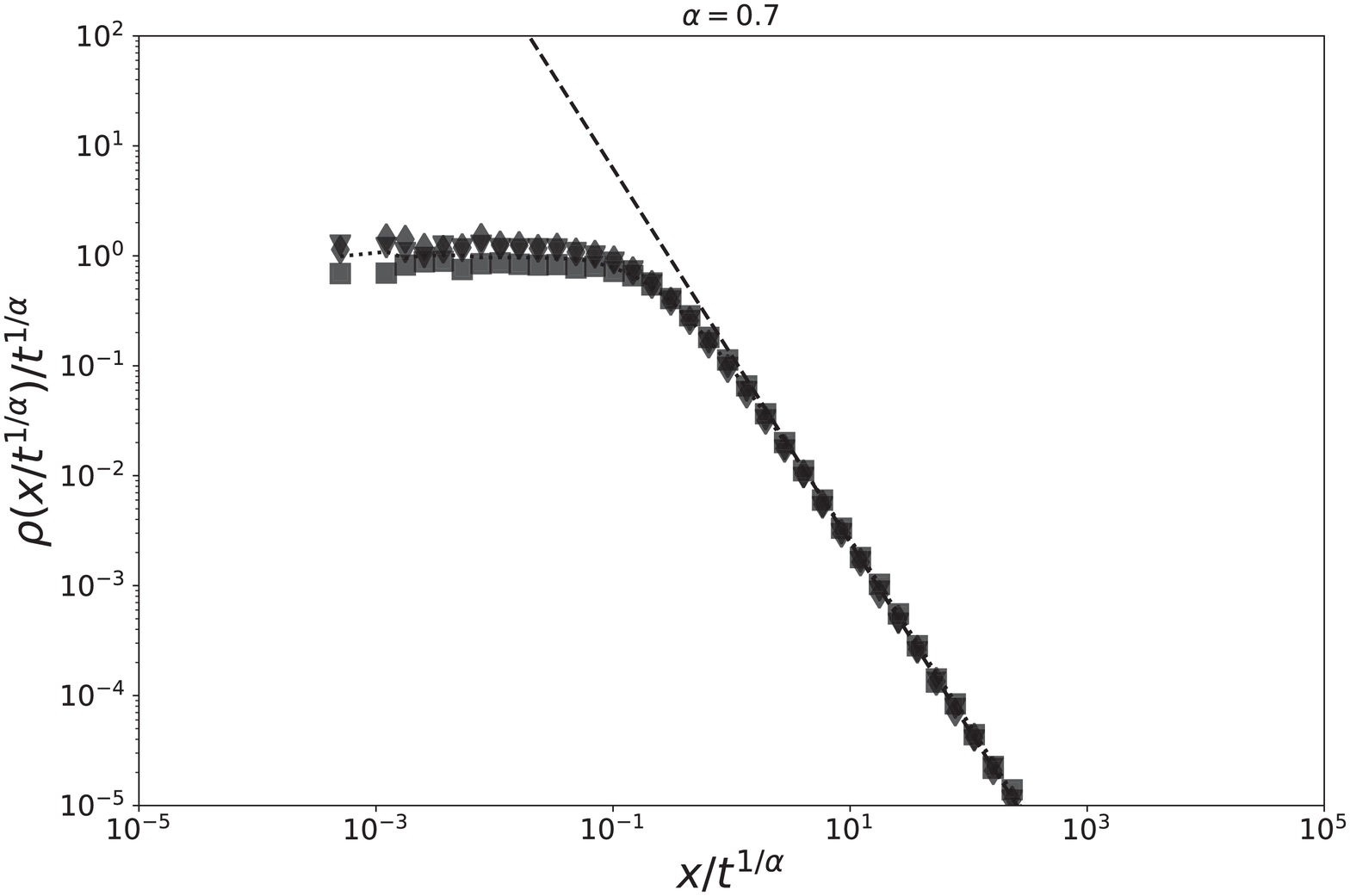}
  \includegraphics[scale=0.2]{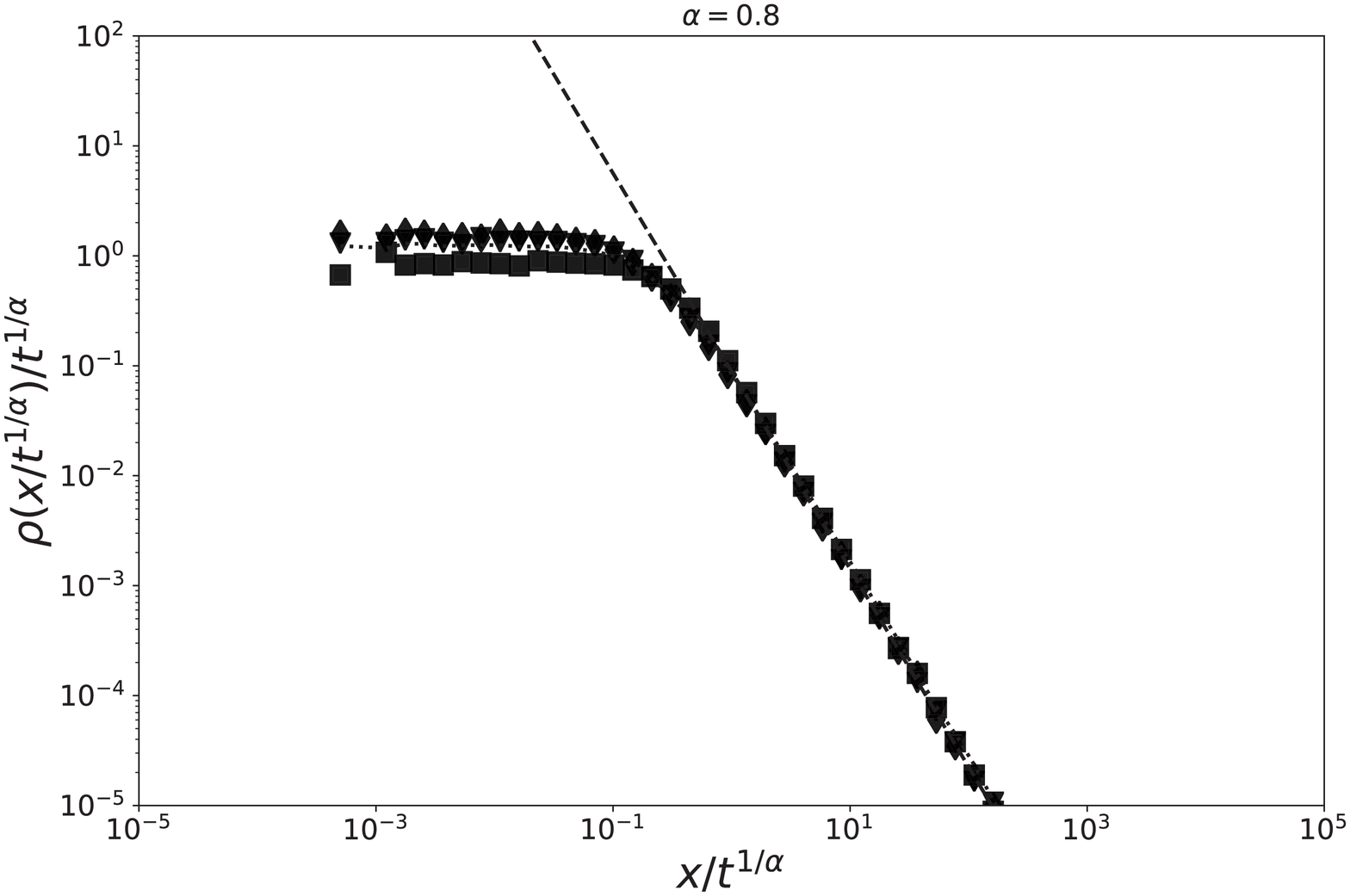}
  \includegraphics[scale=0.2]{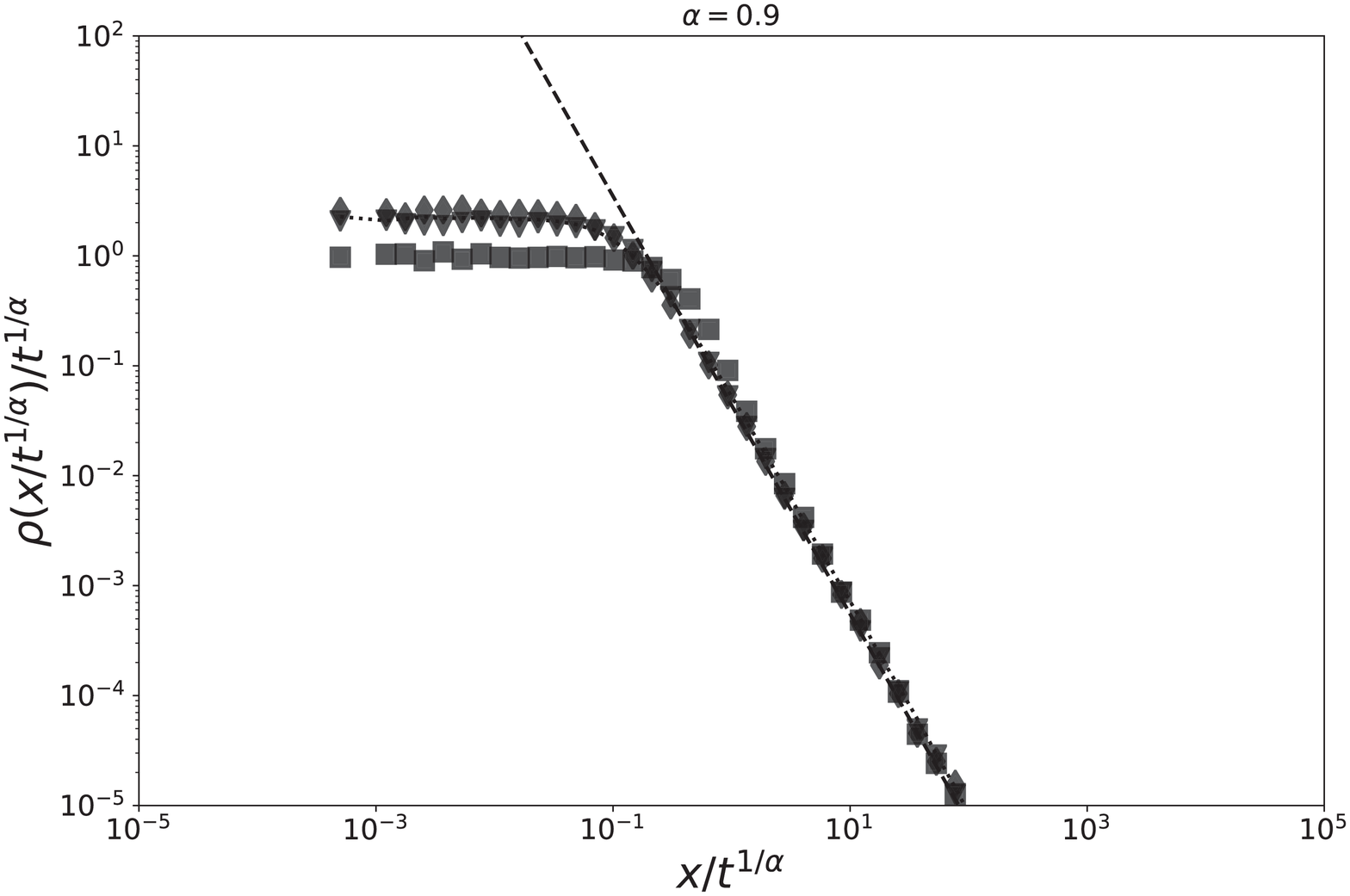}
\caption{Plots of the central part of the walker's distribution $\rho(x;t)$
obtained with jump $pdf$ (\ref{jumppdf})
at times $t=10 \tau, 100\tau, 1000\tau$ marked
by squares, triangles and diamonds, respectively.
The dotted lines represent L\'evy stable densities of
index $\alpha$ and the dashed lines the power-law
decaying $|x|^{-(\alpha + 1)}$.
The loss of self-similarity in the interval $1/2 < \alpha < 1$
is evident.
}
\label{fig:tails1b}
\end{figure}
\end{center}

From the L\'evy coin-flipping rule (\ref{jumppdf}),
we have that for $\kappa \in \R$ the characteristic function is
\vskip -10pt
\begin{eqnarray}
\widehat{\varphi}(\kappa)
&=& \sum_{n=0}^\infty \frac{(-1)^n}{n!}
\sin\left[\frac{\pi}{2}(1 + n \alpha)\right]
(\sqrt{2}\ell \, |\kappa|)^{n \alpha} \\
&\simeq& 1
- \sin\left[\frac{\pi}{2}(1 + \alpha)\right] (\sqrt{2}\ell \, |\kappa|)^\alpha
\nonumber \\
& & \quad + \frac{1}{2}\sin\left[\frac{\pi}{2}(1 + 2 \alpha)\right]
(\sqrt{2}\ell \, |\kappa|)^{2 \alpha}
+ o(|\kappa|^{2\alpha})
\,, \quad \ell |\kappa| \ll 1 \,, \nonumber
\label{jumppdfseries}
\end{eqnarray}
hence expansion (\ref{jumppdfseries}) is an alternating series
if $0 < \alpha \le 1/2$
such that $\rho(x;t)$ solves (\ref{SFDE}),
but if $1/2 < \alpha < 1$ then we have
\be
\sin\left[\frac{\pi}{2}(1 + \alpha)\right] > 0
\quad {\rm and} \quad
\sin\left[\frac{\pi}{2}(1 + 2 \alpha)\right] < 0 \,,
\ee
and expansion (\ref{jumppdfseries}) is not a series
with alternating signs.
In this case, from equation (\ref{MWeq}) it follows that
\be
\displaystyle{
\widehat{\rho}(\kappa;t)=\e^{- (\ell_{\!\alpha}|\kappa|^\alpha +
\frac{1}{2}\ell_{\!2\alpha}|\kappa|^{2\alpha}) \, t/\tau}
}\,,
\label{convolutionF}
\ee
\vskip -4pt \noindent
with
\vskip -12pt
\be
\displaystyle{
\ell_{\!\alpha}=(\sqrt{2}\ell)^\alpha
\left|\sin\left[\frac{\pi}{2}(1+\alpha)\right]
\right| \,,
}
\quad
\displaystyle{
\ell_{\!2\alpha}=(\sqrt{2}\ell)^{2\alpha}
\left|\sin\left[\frac{\pi}{2}(1+2\alpha)\right]\right| \,,
}
\ee
and therefore $\rho(x;t)$ solves the fractional evolution problem
\vskip -10pt
\be
\left\{
\begin{array}{l}
\displaystyle{
\frac{\partial \rho}{\partial t}
+ \mK_{\!\alpha} \, (-\Delta)^{\frac{\alpha}{2}} \rho
+ \frac{1}{2}\mK_{\!2\alpha} \, (-\Delta)^{\alpha} \rho = 0} \,,
\quad {\rm in} \,\, \R \times (0,+\infty) \,, \\ 
\rho(x;0)=\delta(x) \,,\\ 
\displaystyle{
\frac{1}{2} < \alpha < 1 \,,}
\end{array}
\right.
\label{convolutionproblem}
\ee
\vskip -2pt \noindent
where
\vskip -13pt
\be
\mK_{\!\alpha}
=\frac{\ell_\alpha}{\tau}
=\mD_{\!\alpha} \, 2^{\alpha/2}
\left|\sin\left[\frac{\pi}{2}(1+\alpha)\right]\right| \,,
\quad
\mK_1=0 \,,
\ee
\vskip -4pt \noindent
and
\vskip -13pt
\be
\mK_{\!2\alpha}
=\frac{\ell_{2\alpha}}{\tau}
=\mD_{\!2\alpha} \, 2^{2\alpha/2}
\left|\sin\left[\frac{\pi}{2}(1+2\alpha)\right]\right| \,,
\quad
\frac{1}{2}\mK_2=\mD_2=\mD \,,
\ee
such that $\rho(x;t)$ is a convolution of L\'evy stable densities:
\begin{eqnarray}
\rho(x;t)
&=&
\int_{\R^n} \mathcal{L}_\alpha(x - \xi;t)
\mathcal{L}_{2\alpha}(\xi;t) \, d\xi \\
&=&
\frac{1}{(\mK_{\!\alpha}\sqrt{\mK_{\!2\alpha}/2} \,t^{3/2})^{N/\alpha}}
\int_{\R^n}
\mathcal{L}_\alpha\!\left(\frac{x-\xi}{(\mK_{\!\alpha} t)^{1/\alpha}};1\right)
\nonumber \\
& & \hspace{4.0truecm}
\times \mathcal{L}_{2\alpha}\!
\left(\frac{\xi}{(\mK_{\!2\alpha} t/2)^{1/(2\alpha)}};1\right) \, d\xi
\,, \nonumber
\label{convolution}
\end{eqnarray}
with fractional absolute moments
\cite{chechkin_etal-pre-2002,pagnini_etal-jcam-2010}
\be
\sigma^q 
\propto
\left\{
\begin{array}{ll}
(\mK_{\!2\alpha} t)^{q/(2\alpha)} \,, & t \to 0 \,, \\
\\
(\mK_{\!\alpha} t)^{q/\alpha} \,, & t \to + \infty \,, \\
\end{array}
\right.
\quad 0 < q < \alpha \,.
\ee
From the characteristic function (\ref{convolutionF}),
we have that the tails of the distribution
$\rho(x;t)$ (\ref{convolution})
follow the same power-law of the tails of a stable density of
stability parameter $\alpha$, see Figures
\ref{fig:tails1a} and \ref{fig:tails1b},
namely
\be
\rho(x;t) \simeq \mathcal{L}_\alpha(x;t) \,, \quad
0 < \alpha < 1 \,,
\quad |x| \gg \ell \,.
\ee
Convolution integral (\ref{convolution}) has been studied
in a number of papers as fundamental solution of
double-order space-fractional diffusion equation
\cite{chechkin_etal-pre-2002}, as generalised Voigt function
\cite{mainardi_etal-fcaa-2008,pagnini_etal-jcam-2010},
or as a sum of two independent stable random variables
\cite{otiniano_etal-jcam-2013,nadarajah_etal-jcam-2019}.

\begin{center}  
\begin{figure}
  \includegraphics[scale=0.2]{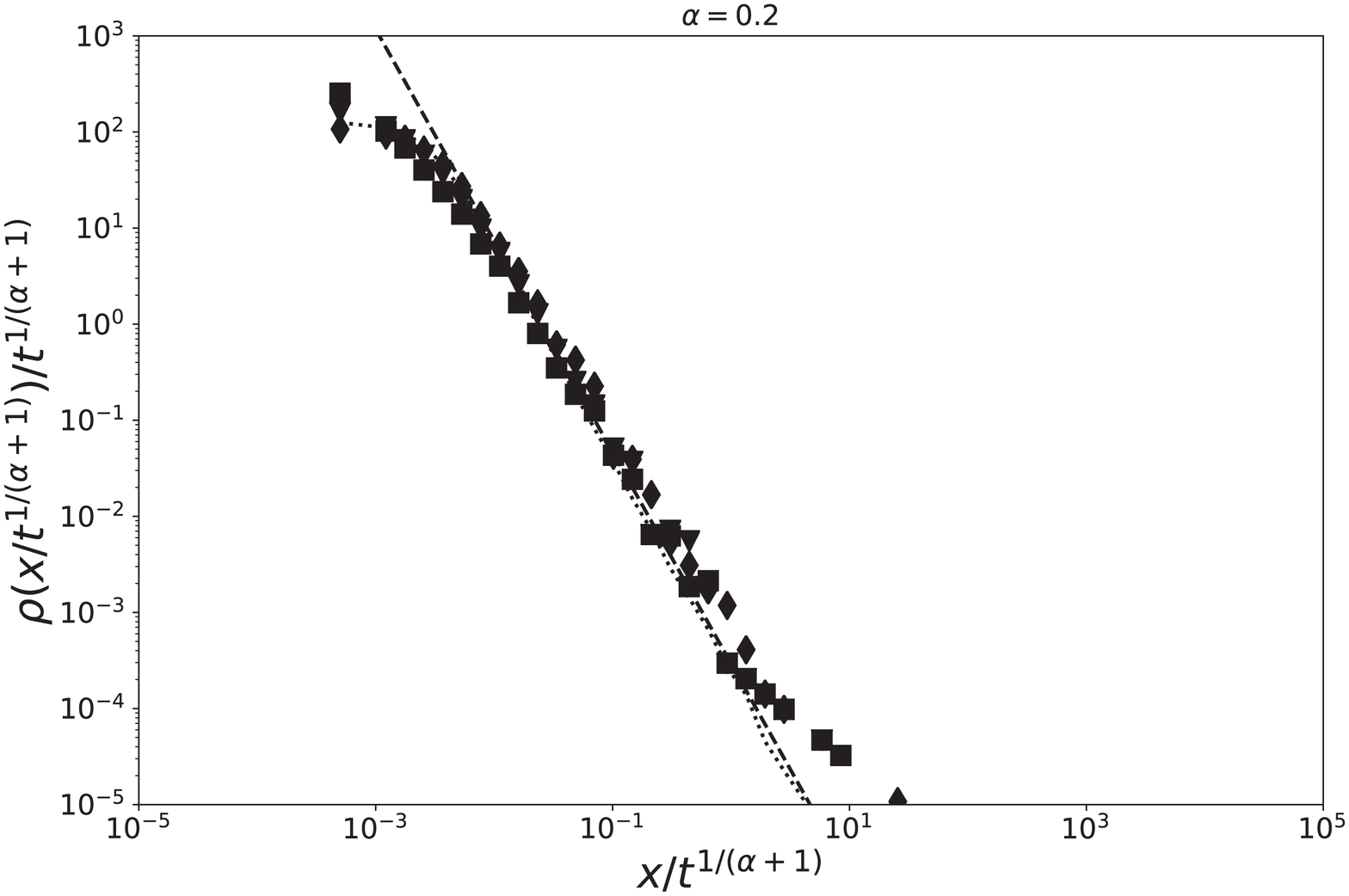}
  \includegraphics[scale=0.2]{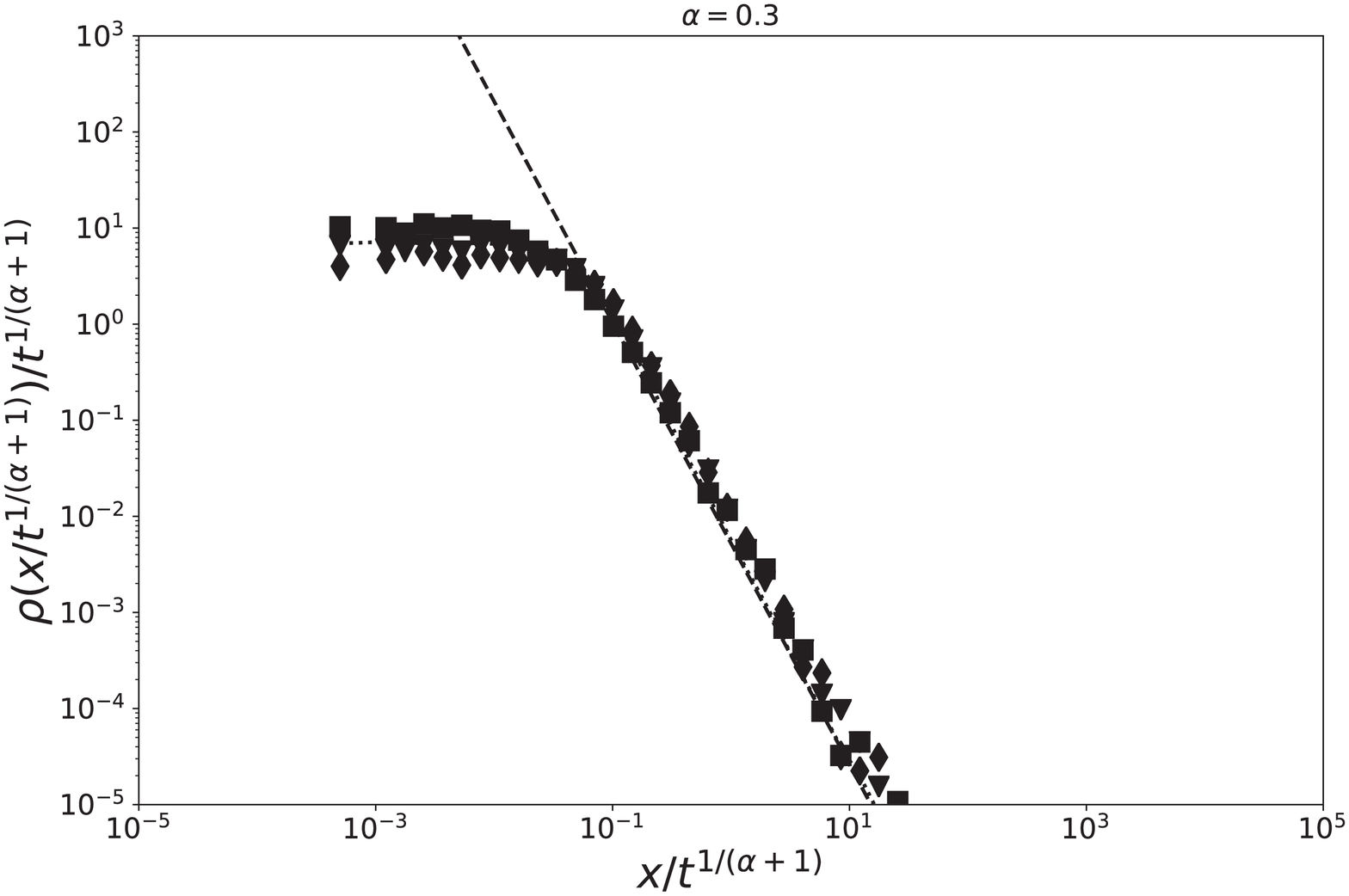}
  \includegraphics[scale=0.2]{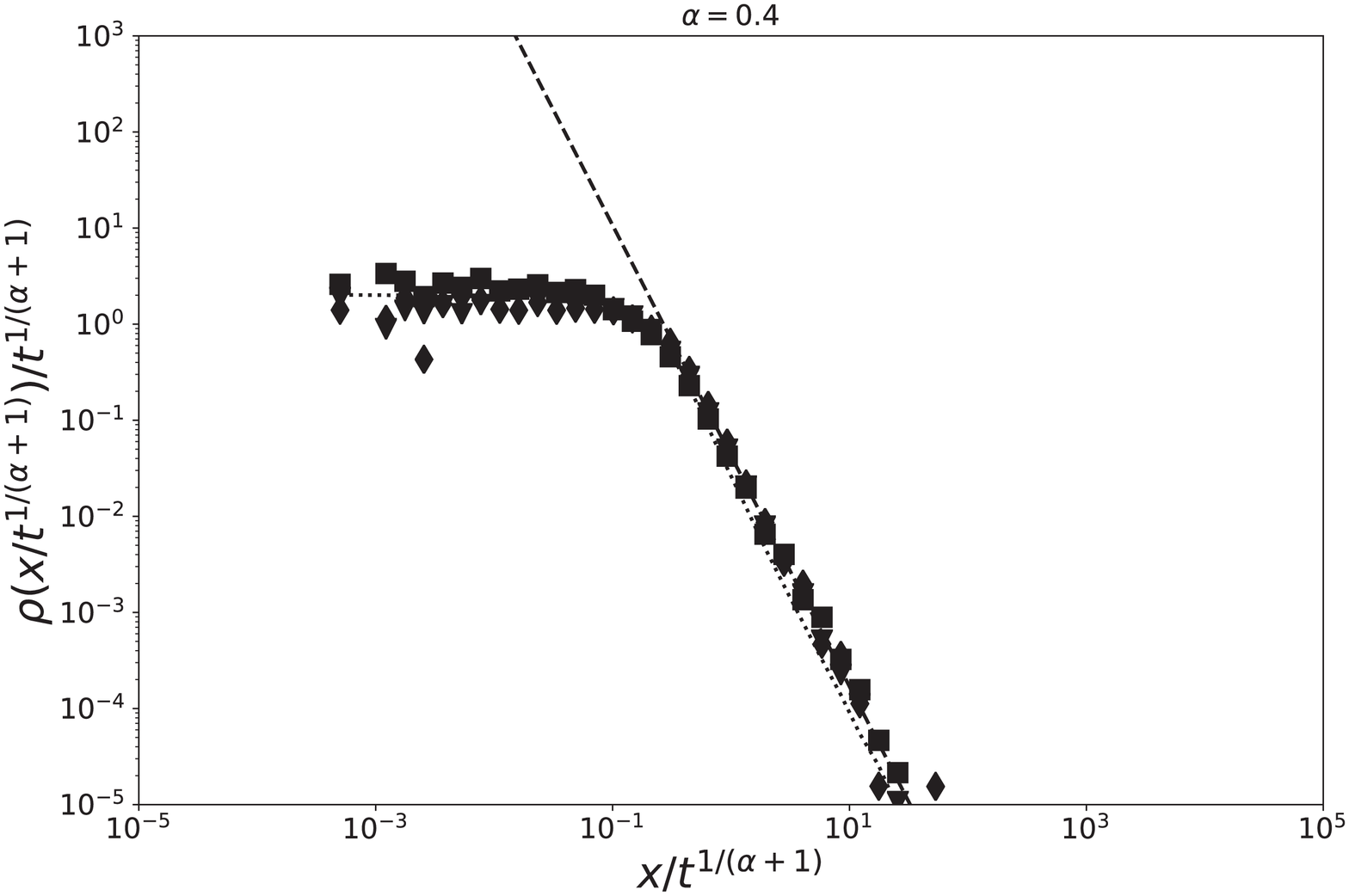}
  \includegraphics[scale=0.2]{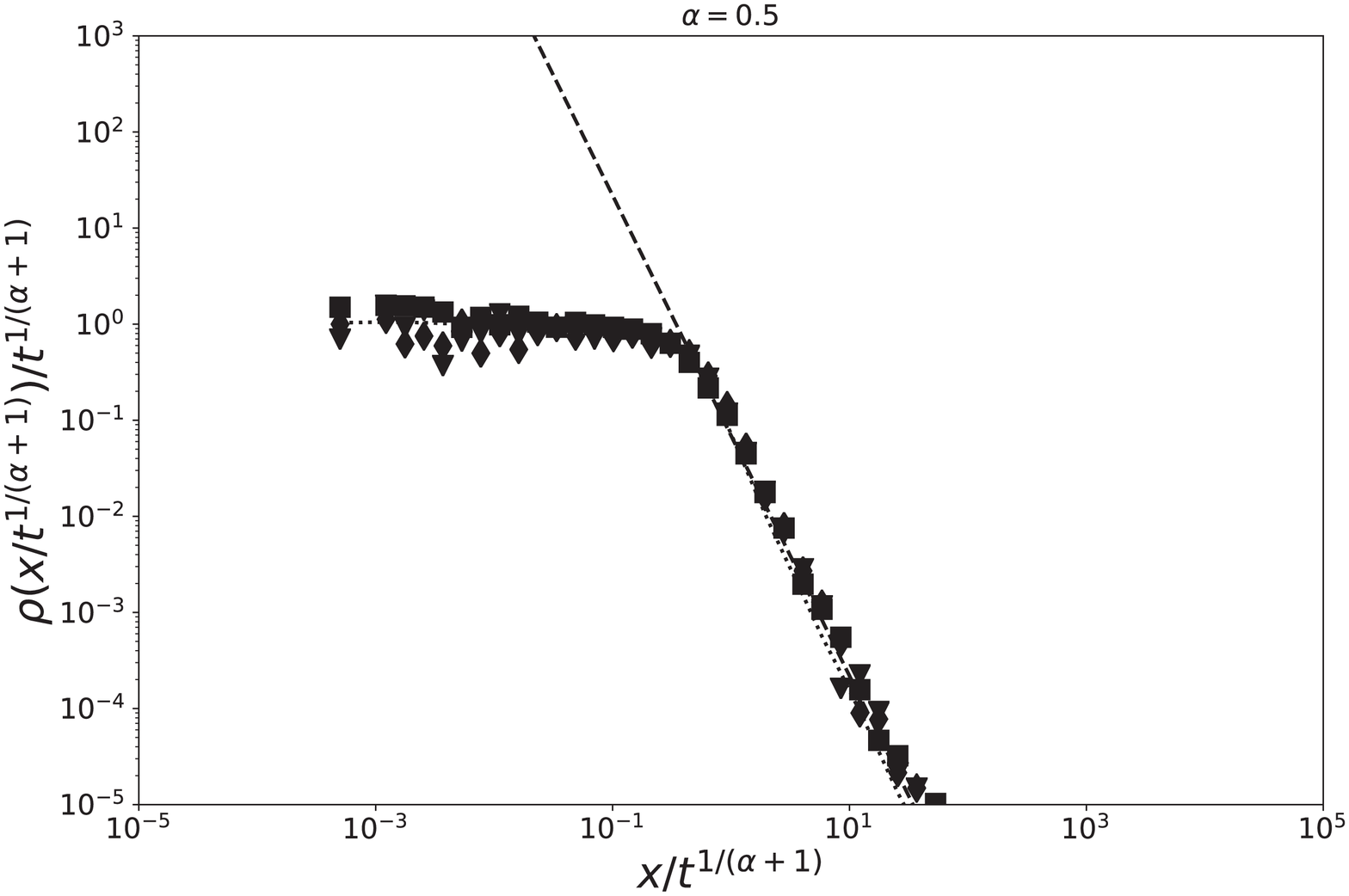}
  \includegraphics[scale=0.2]{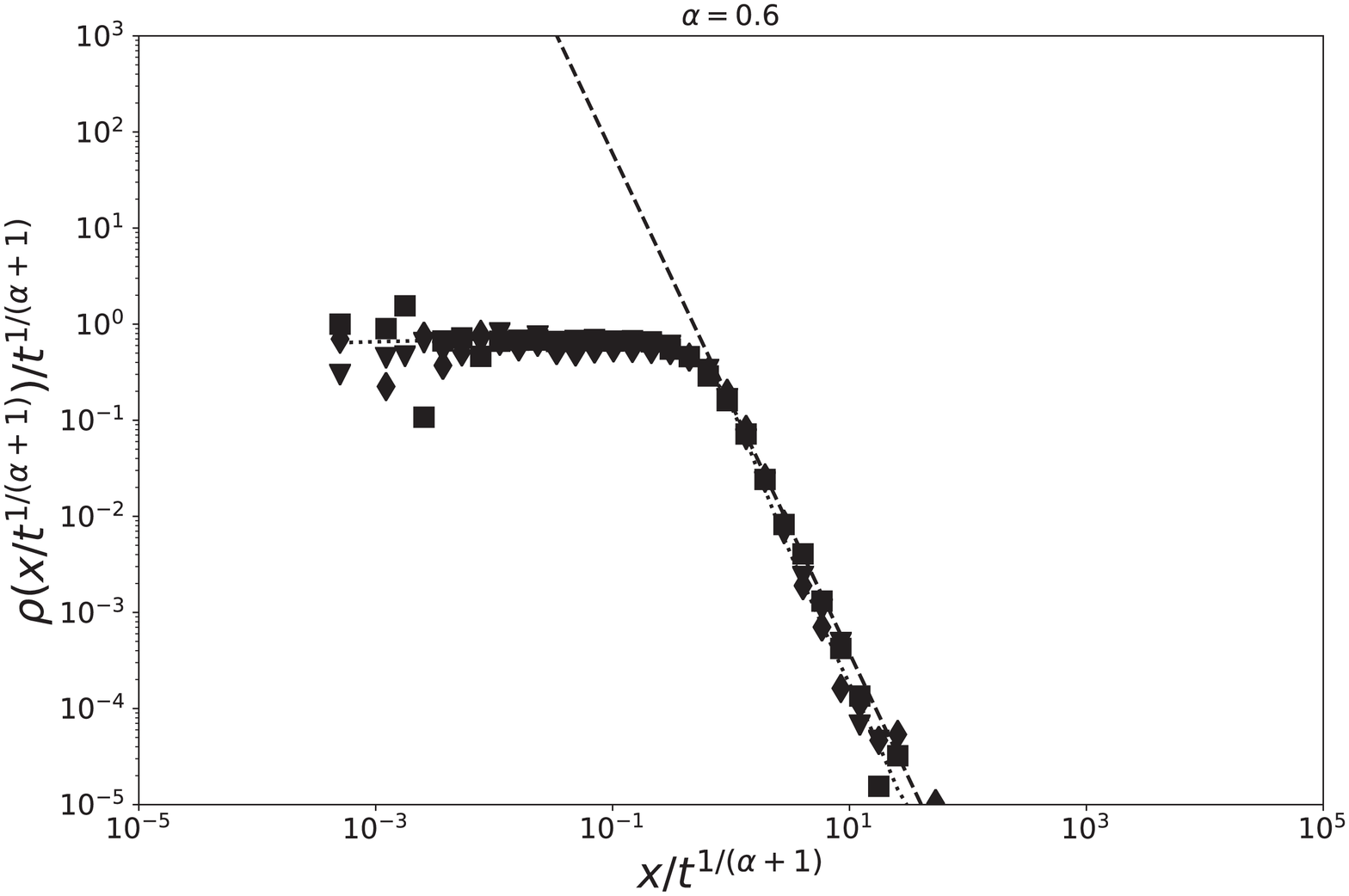}
  \includegraphics[scale=0.2]{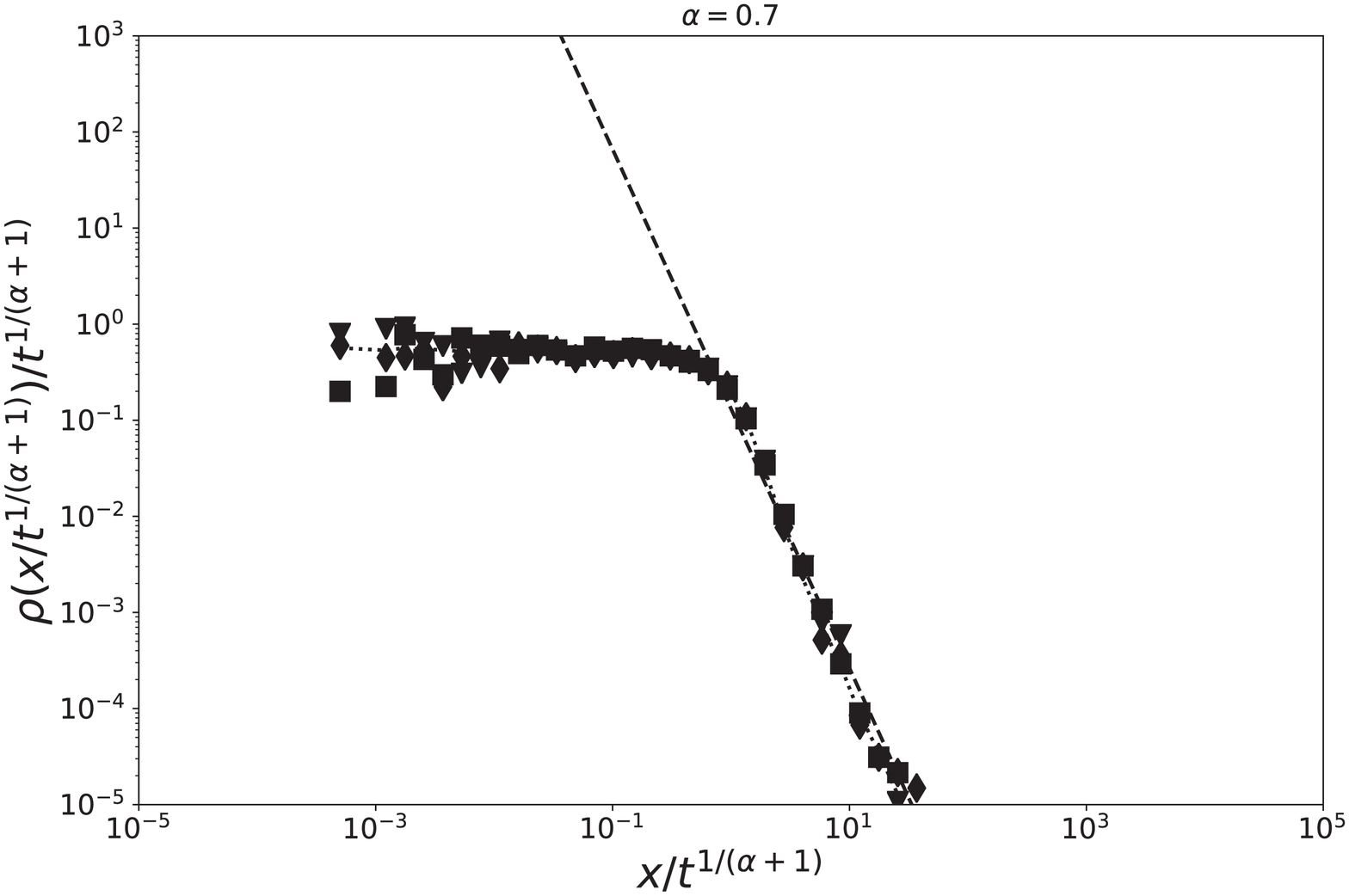}
  \includegraphics[scale=0.2]{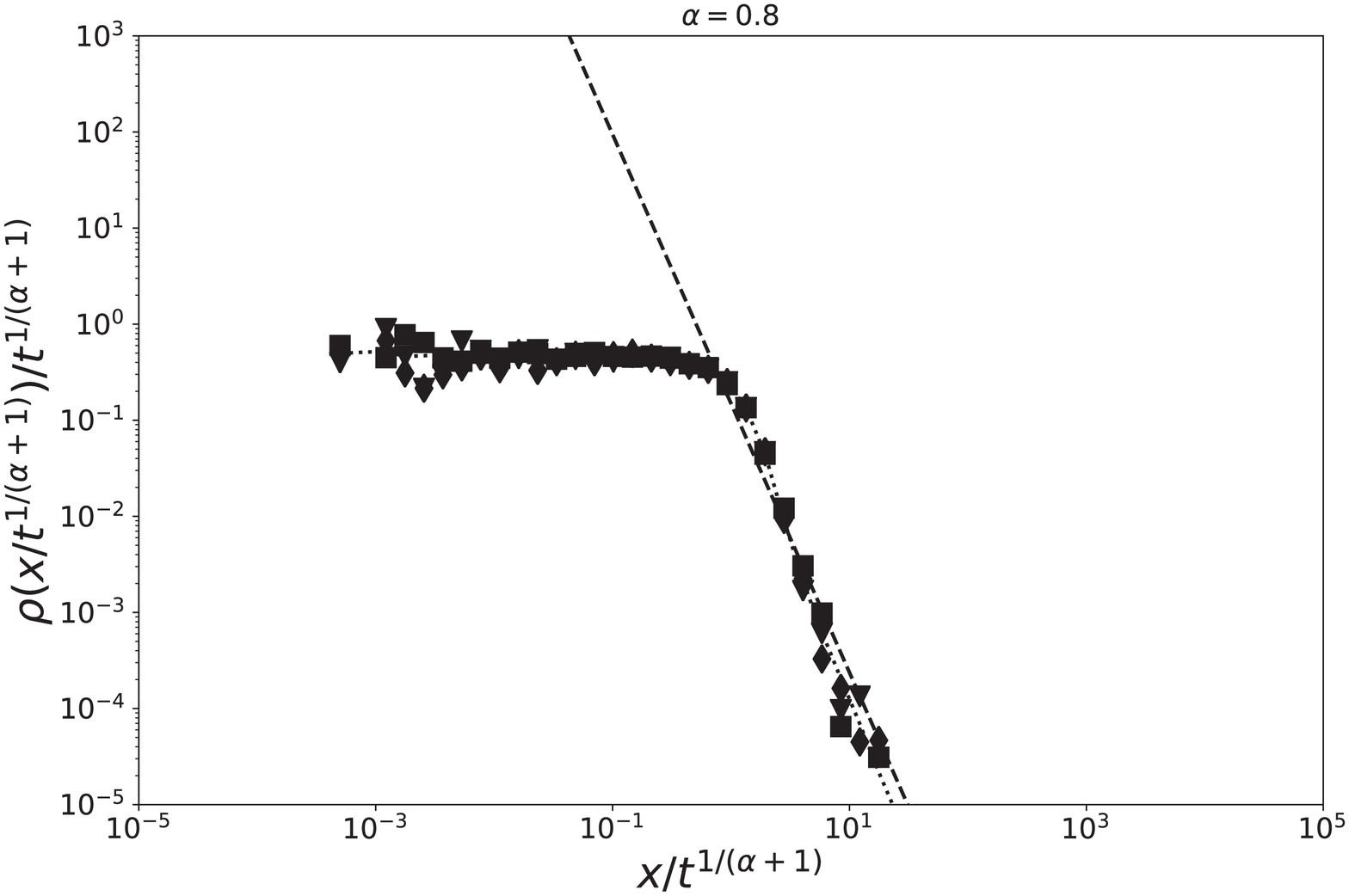}
  \includegraphics[scale=0.2]{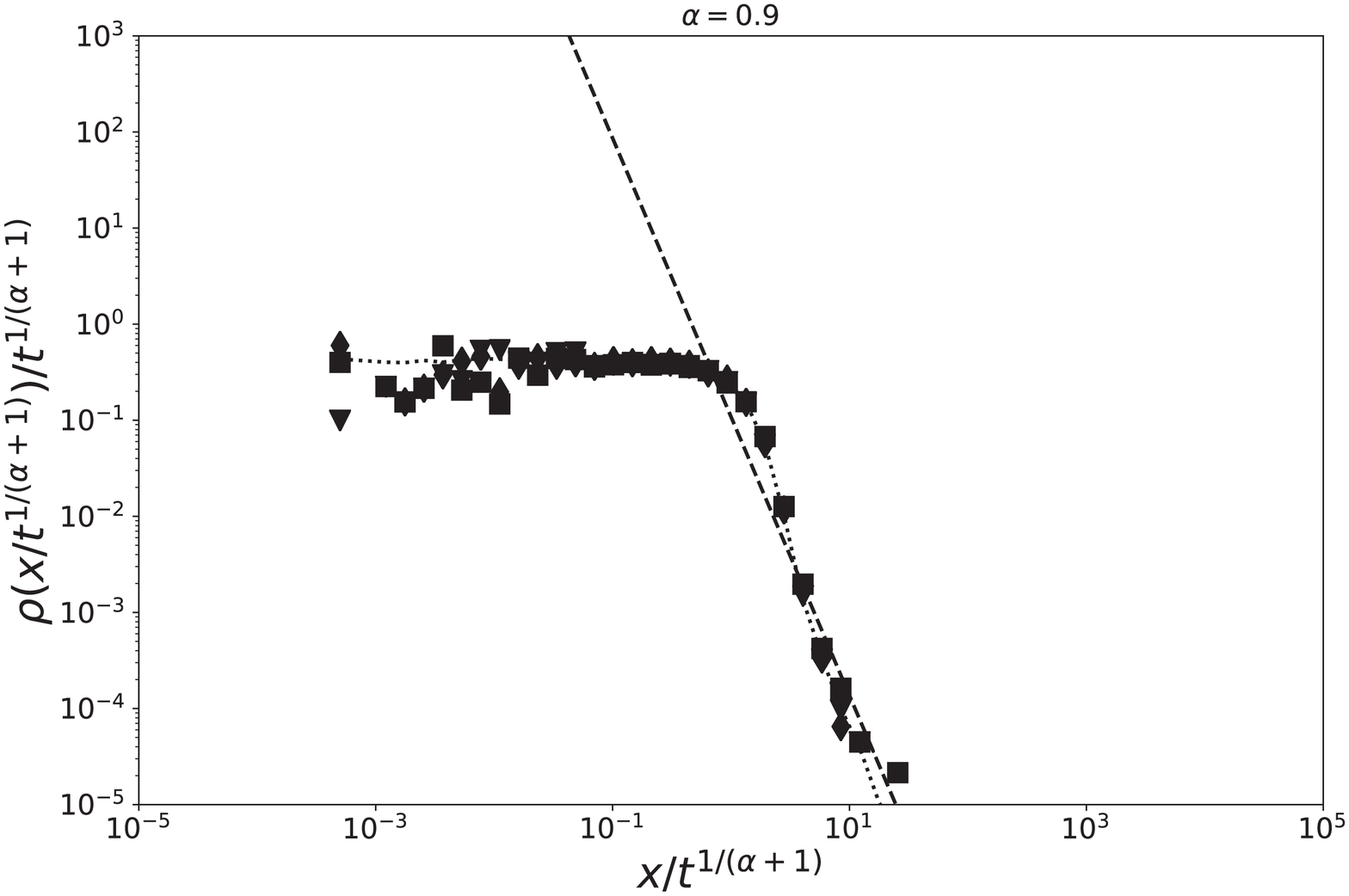}
\caption{Plots of the walker's distribution $\rho(x;t)$
obtained with jump $pdf$ (\ref{jumppdf2})
at times $t=10 \tau, 100\tau, 1000\tau$ marked
by squares, triangles and diamonds, respectively.
The dotted lines represent L\'evy stable densities of
index $(\alpha +1)$ and the dashed lines the power-law decaying
$|x|^{-[(\alpha + 1) + 1]}$.
}
\label{fig:tails2}
\end{figure}
\end{center}

From the L\'evy coin-flipping rule (\ref{jumppdf2}), we have that
\begin{eqnarray}
\widehat{\varphi}(\kappa)
&=&
\frac{1}{\Gamma(1/\alpha)}
\sum_{n=0}^\infty \frac{(-1)^n}{n!}
\frac{\Gamma\left(\frac{1}{\alpha}-\frac{n}{\alpha}\right)}
{\Gamma(1-n)}
\sin\left[\frac{\pi}{2}(1 + n)\right]
(\sqrt{2}\ell \, |\kappa|)^n \\
& & 
+\,
\frac{\alpha \sqrt{2}\ell \kappa}{\Gamma(1/\alpha)}
\sum_{n=0}^\infty \frac{(-1)^n}{n!}
\frac{\Gamma(-1-\alpha n)}{\Gamma(-\alpha n)}
\sin\left[\frac{\pi}{2}(2 + \alpha n)\right]
(\sqrt{2}\ell \, |\kappa|)^{\alpha n} \nonumber \\
&\simeq&
1 - \frac{\alpha}{\Gamma(1/\alpha)}
\frac{\sin(\pi\alpha/2)}{1+\alpha}
(\sqrt{2}\ell \, |\kappa|)^{\alpha + 1}
\nonumber \\
& & + \frac{\alpha}{\Gamma(1/\alpha)}
\frac{\sin(\pi\alpha)}{1+2\alpha}
(\sqrt{2}\ell \, |\kappa|)^{2\alpha + 1}
\!+ o(|\kappa|^{2\alpha+1})
\,, \quad \ell |\kappa| \ll 1 \,, \nonumber
\label{jumppdfseries2}
\end{eqnarray}
since $0 < \alpha < 1$,
expansion (\ref{jumppdfseries2}) is an alternating series
and $\rho(x;t)$ solves (\ref{SFDE}) by replacing
$\alpha \to (\alpha + 1)$ and $\displaystyle{
\mD_\alpha \to \mD_\alpha=
\frac{2^{(\alpha+1)/2}}{\Gamma(1/\alpha)}\frac{\alpha}{1+\alpha}
\sin\left[\frac{\pi}{2}\alpha\right]
\frac{\ell^{\alpha + 1}}{\tau}}$,
see Figure \ref{fig:tails2}.

\smallskip

Before ending this section, 
we want to highlight that both jump $pdf$s (\ref{jumppdf})
and (\ref{jumppdf2}) tend to the coin-flipping rule (\ref{coinflip})
when $\alpha \to 1$ and so both the resulting processes tend to
the Brownian motion.
However, from series expansions (\ref{jumppdfseries})
and (\ref{jumppdfseries2}) it emerges that they tend
to the Brownian motion in a very different way.
In fact, we observe that series expansion (\ref{jumppdfseries})
reduces to (\ref{coinflipseries})
through the third term $\propto |\kappa|^{2\alpha}$
because the coefficient of the second term goes to $0$,
while series expansion (\ref{jumppdfseries2})
reduces to (\ref{coinflipseries})
through the second term $\propto |\kappa|^{\alpha +1}$
because the coefficient of the third term goes to $0$.

\begin{center} 
\begin{figure}
  \includegraphics[scale=0.28]{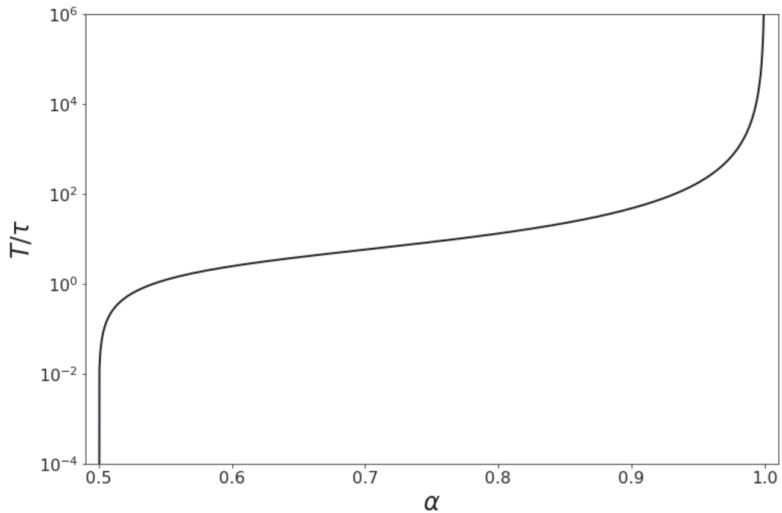}
\caption{Plot of the time-scale $T$ as defined in
formula (\ref{def:T}).}
\label{fig:T}
\end{figure}
\end{center} 

\vspace*{-5pt} 
\section{Discussion and significance} 
\label{sec:discussion}
\subsection{Jump $pdf$ and fractional Laplacian singularity} 

Our first remark on the above results concerns
the comparison with the probabilistic derivation
of the fractional diffusion equation (\ref{SFDE})
provided by Valdinoci \cite{valdinoci-bsema-2009},
see Appendix B for a short reminder.
In fact, from a generic probabilistic framework where
the walker's distribution function $\rho(\bx;t)$
is updated at each fixed time-step $\Delta t$
through a symmetric jump $pdf$ $\varphi(\Delta x)$,
i.e.,
\be
\rho(\bx;t + \Delta t) = \int_{\R^N}
\varphi(\Delta \bx) \rho(\bx - \Delta \bx;t) \,
d \Delta \bx \,,
\label{integrale}
\ee
fractional diffusion equation (\ref{SFDE}) is obtained
only if $\varphi(\Delta \bx) \propto |\Delta \bx|^{-N-\alpha}$
with $\varphi(0)=0$,
namely only if the distribution of the jumps follows a rule
{\it \`a la} coin-flipping.
As a matter of fact, condition $\varphi(0)=0$ turns out to be straightforwardly related
to the singularity of the fractional Laplacian at $0$
and for this reason the {\it ``Should I go?"} condition emerges to be
of paramount importance for fractional diffusion modelling,
see Appendix B.

\smallskip

Fractional diffusion equation (\ref{SFDE})
can be derived on the basis of (\ref{integrale})
also by considering formula (\ref{integrale}) in the
domain of the characteristic functions and by applying
the small wavelength expansion \cite{klafter_sokolov-2011}.
However, the previous discussion about the small wavelength expansion
of the characteristic function of jumps
and the consequence due to alternating and non-alternating series holds
also for this derivation method.
This procedure does not catch the peculiar role
of the {\it ``Should I go?"} condition that is mapped
through the jump rule {\it \`a la} coin-flipping
- namely $\varphi(0)=0$ - into the distinctive singularity of the fractional Laplacian, see Appendix B.

\smallskip

As it follows from the previous analysis,
a Markovian CTRW converges always to a density function that solves
the fractional diffusion equation (\ref{SFDE})
only if the jump $pdf$ has its maximum in $0$,
i.e., in the {\it ``Should I stay?"} condition,
and indeed when the jumps follow a rule {\it \`a la} coin-flipping,
i.e., the {\it ``Should I go?"} condition,
that is the one compatible with the probabilistic derivation
\cite{valdinoci-bsema-2009}, it is not guaranteed that
the resulting density function solves (\ref{SFDE}).

\smallskip

Hence, concerning the significance of the
previous analysis, we state that
when the small wavelength expansion of the characteristic function
of the jump $pdf$ is not an alternating series
then the process is not self-similar and it is not governed
by the fractional diffusion equation (\ref{SFDE}).
More concretely, in the studied example,
the jump rule {\it \`a la} coin-flipping (\ref{jumppdf}),
when $1/2 < \alpha < 1$,
generates a process that converges to a Voigt-like distribution
\cite{mainardi_etal-fcaa-2008,pagnini_etal-jcam-2010},
it is not self-similar and is governed by the double-order
fractional diffusion equation (\ref{convolutionproblem}).
Besides the many fields where the Voigt profile emerges,
we report here that recently it has been highlighted that
the Voigt profile is a good descriptor of the processes occurring
in protein folding and in the native state
\cite{maisuradze_etal-pj-2021}.

\vspace*{-3pt}

\subsection{Indetermined homecoming: the effect of the
L\'evy coin-flipping rule
on transience (and recurrence) of anomalous diffusion processes}
\label{sec:recurrence} 

Our second remark on the above results concerns
the problem of transience (and recurrence) for
power-law processes and its
relation with the concept of site fidelity in animal behaviour
\cite{greenwood-ab-1980,benhamou-e-2007,
boyer_etal-prl-2014,berthelot_etal-bioRxiv-2020,
giuggioli_etal-jmb-2012,gautestad_etal-me-2013}
and with the L\'evy flights foraging hypothesis
\cite{fritz_etal-prslb-2003,humphries_etal-pnas-2012,
palyulin_etal-pnas-2014,zeng_etal-fcaa-2014,
benhamou_etal-jtb-2015,pyke-mee-2015,klages-2018}.
Affili, Dipierro \& Valdinoci \cite{affili_etal-matrix-2020}
developed an approach for
deriving the conditions for transience and recurrence
of Markovian random processes that is based on the decaying in time
of the walker's distribution in the starting site,
i.e., $\rho(0;t)$.
We briefly report that approach \cite{affili_etal-matrix-2020}
in Appendix C, where we have re-arranged it
accordingly to the present aims.

\smallskip

Actually, diffusive processes, whose walker's distribution converges
to a stable density with stable parameter $0 < \alpha < 2$,
are always transient except in the one-dimensional ($N=1$)
case when $1 \le \alpha < 2$
\cite{affili_etal-matrix-2020,michelitsch_etal-jpa-2017},
see Appendix C.

\smallskip

We observe that distribution $\rho(x;t)$ (\ref{convolution})
in the {\it Should I go?} condition (\ref{jumppdf})
with $1/2 < \alpha < 1$ is {\it not self-similar},
on the contrary $\rho(x;t)$ is self-similar in the same
{\it Should I go?} condition (\ref{jumppdf})
but with $0 < \alpha \le 1/2$ or in
the {\it Should I go?} condition (\ref{jumppdf2})
with $0 < \alpha < 1$,
and  in the {\it Should I stay?} condition
(\ref{Ljump}) with $0 < \alpha < 2$.
As a consequence, in the considered case study,
the loss of self-similarity introduces a time-scale $T$
necessary for defining the large-time limit and
so for attaining the scaling law in time of $\rho(0;t)$
and determining transience and recurrence. This time-scale $T$
emerges to be dependent on $\alpha$ and tending to infinity
when $\alpha \to 1$ and this makes the large-time limit
unattainable in real systems. In fact, by starting from the formula
\be
\rho(0;t)
=\frac{1}{\pi}\int_0^\infty \widehat{\rho}(\kappa;t) \, d\kappa
=\frac{1}{\pi}\int_0^\infty
\e^{-(1-\widehat{\varphi}(\kappa))t/\tau} \, d\kappa \,,
\label{rho0}
\ee
we have that when $\ell |\kappa| \ll 1$
the expansion (\ref{convolutionF}) holds and then
(\ref{rho0}) can be approximated by
\be
\rho(0;t)
\simeq \frac{1}{\pi}\left\{
\int_0^{1/\ell}
\e^{- (\ell_\alpha \kappa^\alpha +
\frac{1}{2}\ell_{2\alpha} \kappa^{2\alpha})t/\tau}
\, d\kappa
+ \e^{-t/\tau} \int_{1/\ell}^\infty
\e^{\widehat{\varphi}(\kappa)t/\tau} \, d\kappa
\right\} \,,
\ee
and, after the change of variable $\ell_\alpha k^\alpha t/\tau=\xi$,
it becomes
\begin{eqnarray}
\rho(0;t)
&\simeq& \frac{t^{-1/\alpha}}{\alpha\pi\mathcal{K}_\alpha}
\left\{
\int_0^{\frac{\ell_\alpha t}{\ell^\alpha \tau}}
\e^{- \xi -
\frac{1}{2}\frac{\ell_{2\alpha}}{\ell_\alpha^2}\frac{\tau}{t}
\xi^{2\alpha}}
\, \xi^{1/\alpha -1} d\xi \right. \\
& & \quad \left.
+ \, \e^{-t/\tau}
\int_{\frac{\ell_\alpha t}{\ell^\alpha \tau}}^\infty
\exp\left\{
\widehat{\varphi}\left(\xi^{1/\alpha}\right)
\frac{t}{\tau} \right\}
\, \xi^{1/\alpha -1} d\xi
\right\} \,, \nonumber
\end{eqnarray}
that in the limit $t/\tau \to \infty$ reduces to
\begin{eqnarray}
\rho(0;t)
&\simeq&
\frac{t^{-1/\alpha}}{\alpha\pi \mathcal{K}_\alpha^{1/\alpha}}
\int_0^\infty
\e^{- \xi -
\frac{1}{2}\frac{\ell_{2\alpha}}{\ell_\alpha^2}\frac{\tau}{t}
\xi^{2\alpha}}
\, \xi^{1/\alpha -1} d\xi \\
&=&
\displaystyle{
\frac{\Gamma(1/\alpha) \, t^{-1/\alpha}}
{\alpha\pi\mathcal{K}_\alpha^{1/\alpha}}
\left(\frac{\ell_{2\alpha}}{\ell_\alpha}
\frac{\tau}{t}\right)^{-\frac{1}{2\alpha}}
\e^{\frac{\ell_\alpha^2 t}{4 \ell_{2\alpha} \tau}}
\, D_{\!\!-\frac{1}{\alpha}}\left(
\sqrt{\frac{\ell_\alpha^2 t}{\ell_{2\alpha}\tau}}\right)} \,,
\nonumber
\end{eqnarray}
where $D_p(z)$, with ${\rm Re} \, p<0$, is the parabolic cylinder function,
see \cite[3.462(1) p. 365 and 9.24-9.25 (9.246) p. 1028]{gradshteyn_ryzhik-1980}
with asymptotic behaviour
\be
D_p(z) \sim \e^{z^2/4} \, z^p \,,
\quad |z| \gg 1 \quad {\rm and} \quad |z| \gg |p| \,.
\ee
To conclude, we obtain that,
when the following large-time limit is reached
\be
t \gg T=\frac{\tau}{\alpha^2}
\frac{\left|\sin\left[\frac{\pi}{2}(1+2\alpha)\right]\right|}
{\left[\sin\left[\frac{\pi}{2}(1+\alpha)\right]\right]^2}
\,,
\label{def:T}
\ee
it holds
\vskip -10pt
\be
\rho(0;t) \sim
\frac{\Gamma(1/\alpha)}{\alpha \pi \mK_\alpha^{1/\alpha}} \,
t^{-1/\alpha} \,.
\label{decayinglaw}
\ee
However $T \to \infty$ when $\alpha \to 1$,
see Figure \ref{fig:T}, that poses an issue on the attainability
of such large-time limit in real systems.


\medskip 	

\begin{center} 
\begin{figure}
  \includegraphics[scale=0.14]{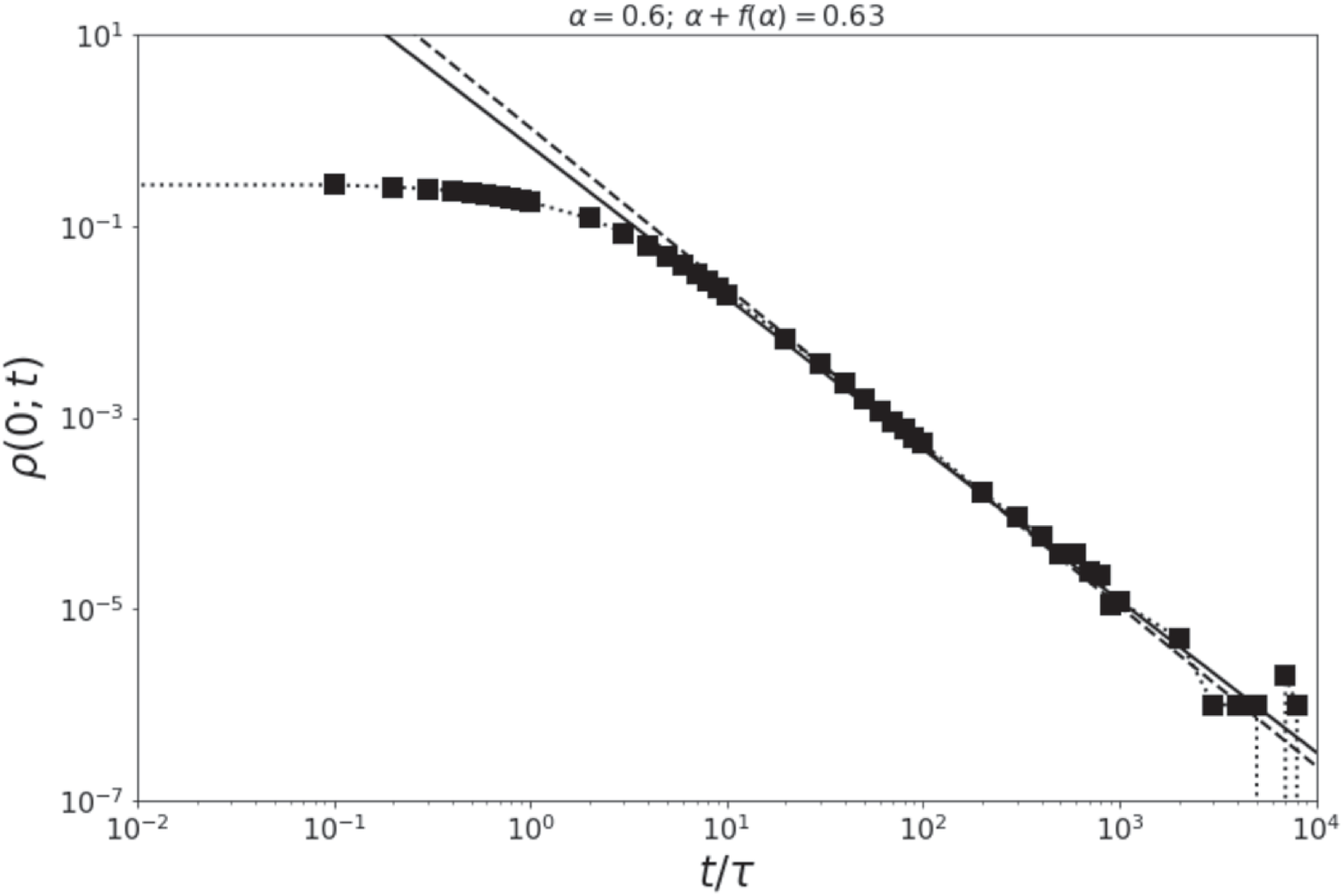}
  \includegraphics[scale=0.14]{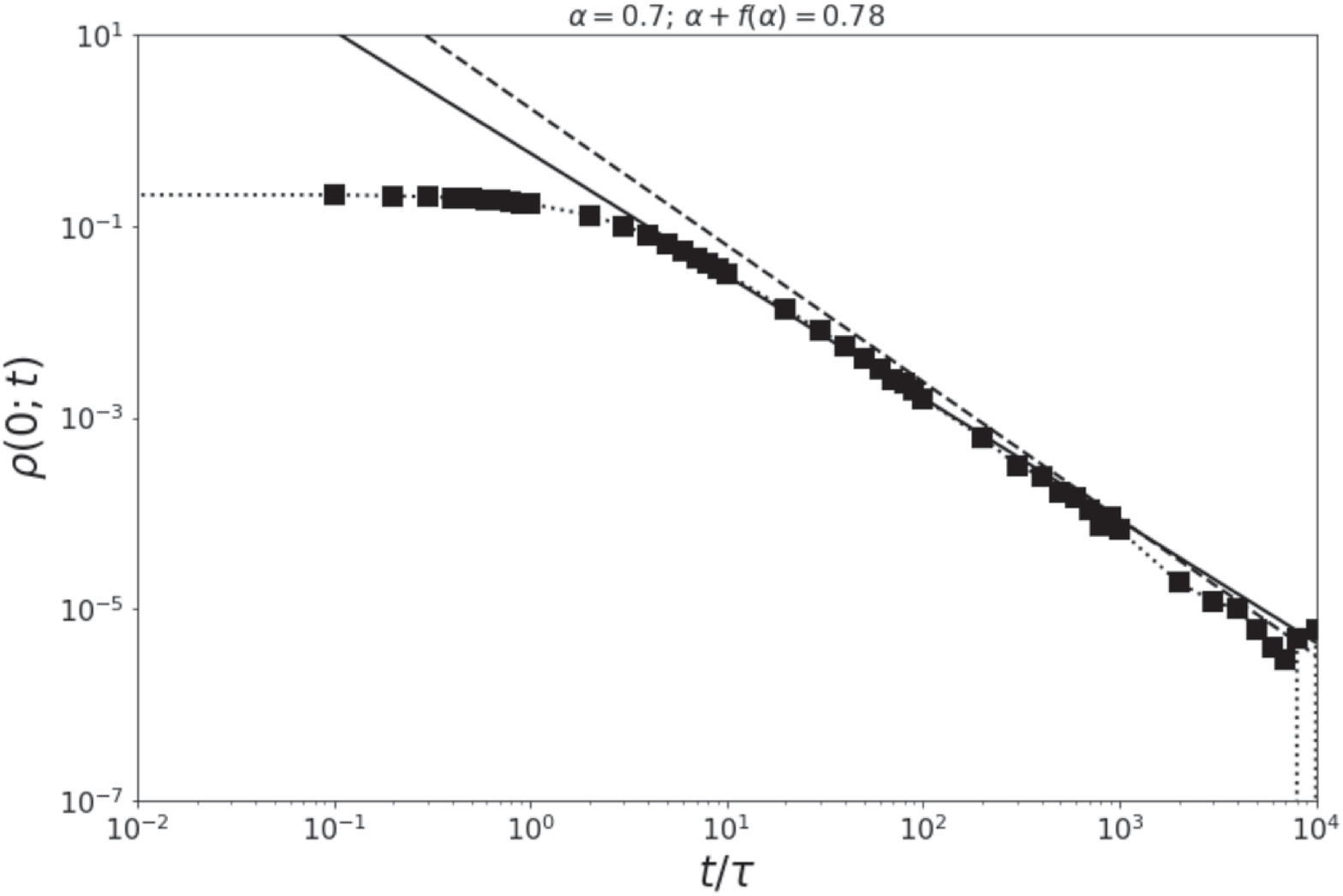}
  \includegraphics[scale=0.14]{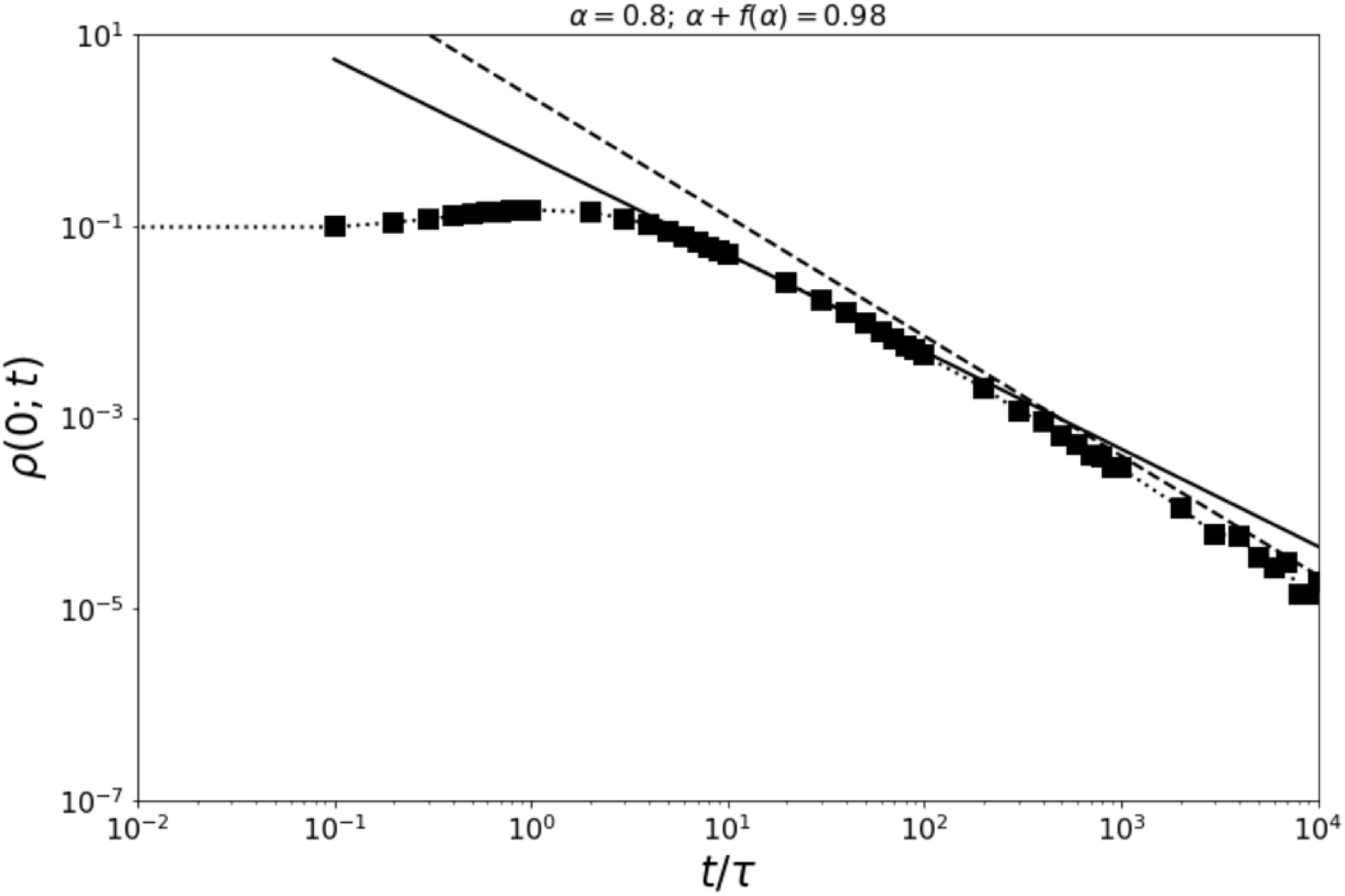}
  \includegraphics[scale=0.14]{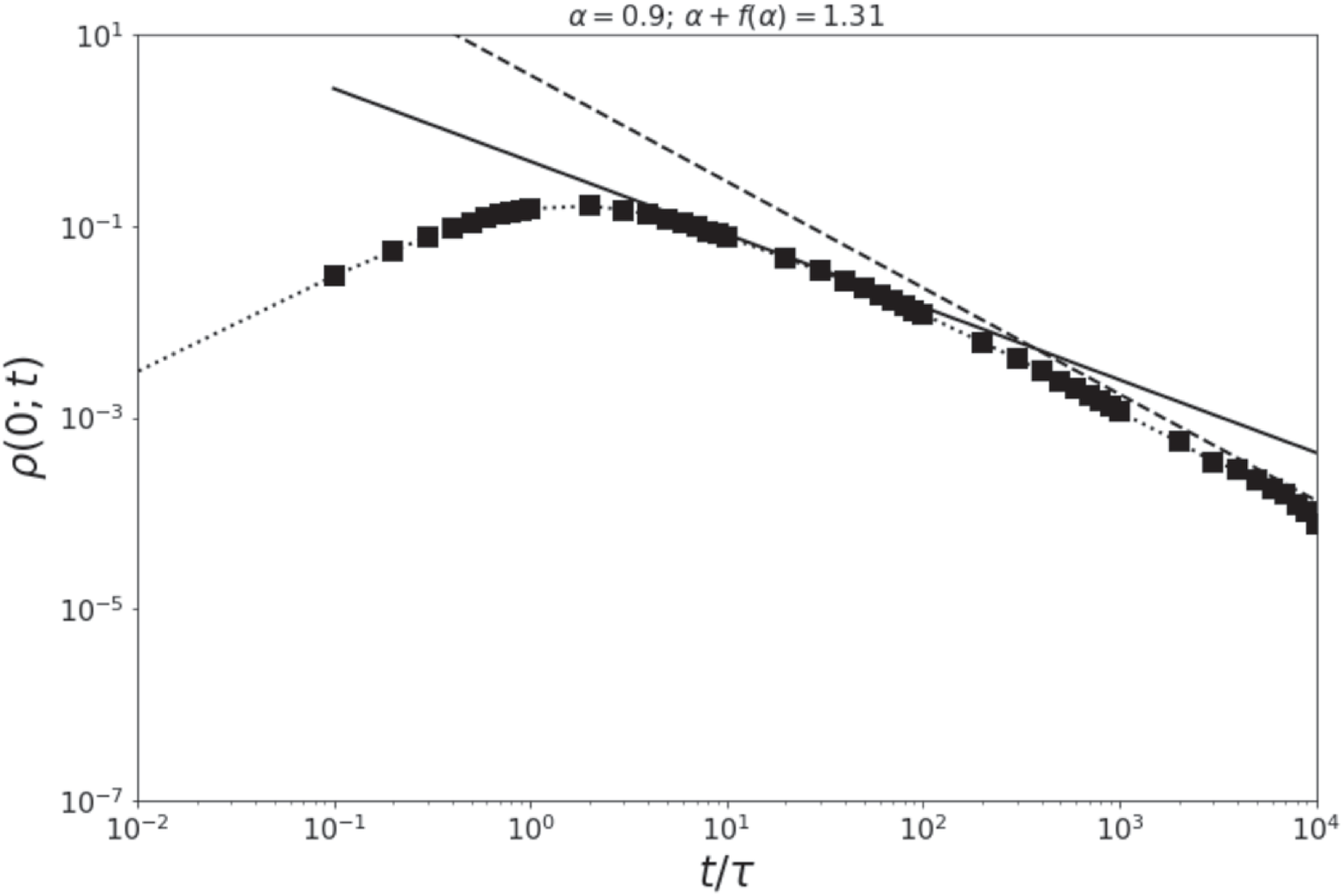}
\caption{Plots of the decreasing in time of the
maximum of the walker's distribution $\rho(0;t)$ generated
through the jump $pdf$ (\ref{jumppdf}) with $\alpha= 0.6 \,, 0.7 \,, 0.8 \,, 0.9$.
The solid line represent the decaying-law $t^{-[\alpha + f(\alpha)]}$ (\ref{scaling0})
and the dashed line is the large-time decaying-law $t^{-\alpha}$ (\ref{decayinglaw}).
The plots show the duration of the intermediate regime $\tau \ll t \ll T$
and its enlarging as $\alpha \to 1$.
}
\label{fig:scaling1}
\end{figure}
\end{center}

\begin{center} 
\begin{figure}
  \includegraphics[scale=0.14]{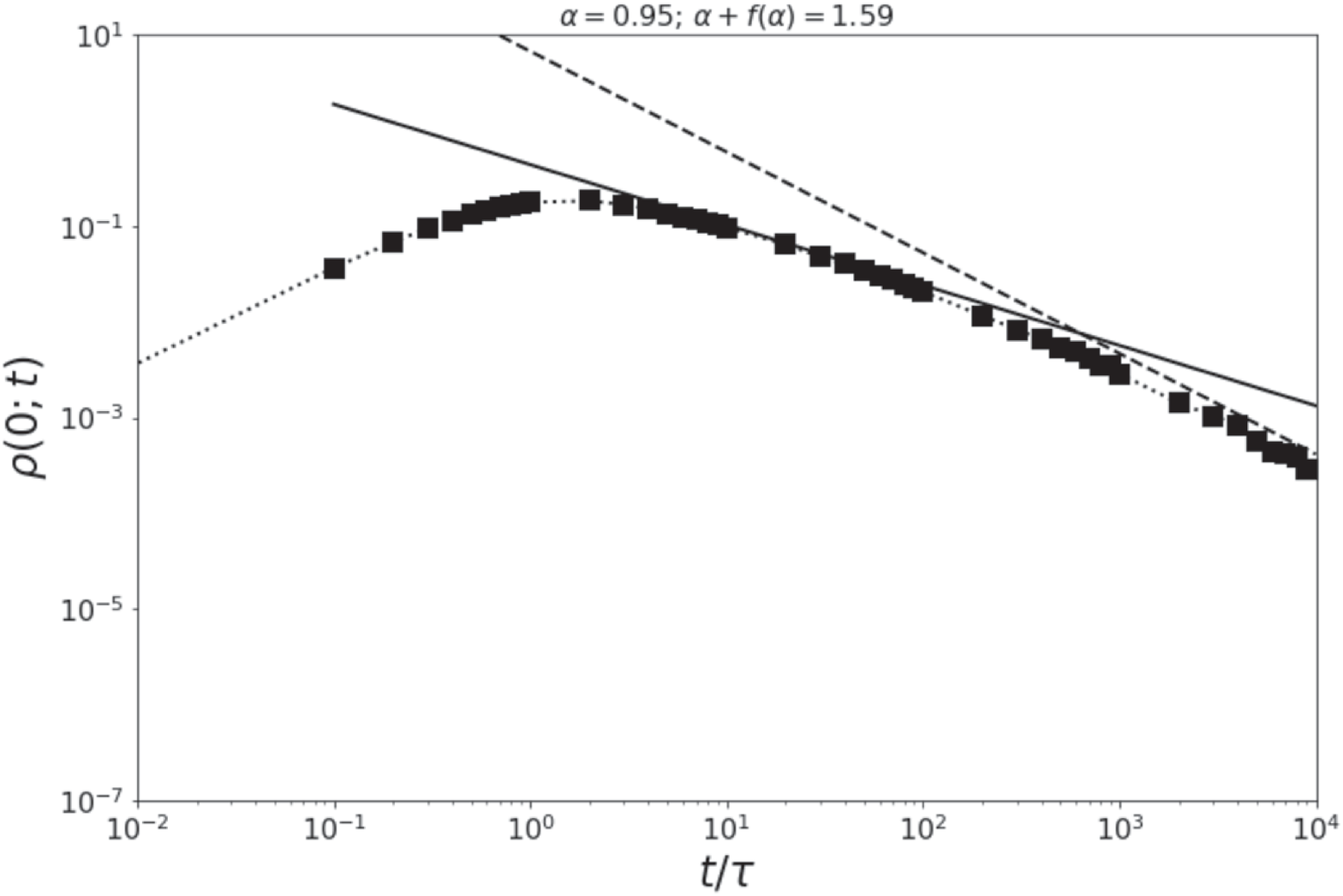}
  \includegraphics[scale=0.14]{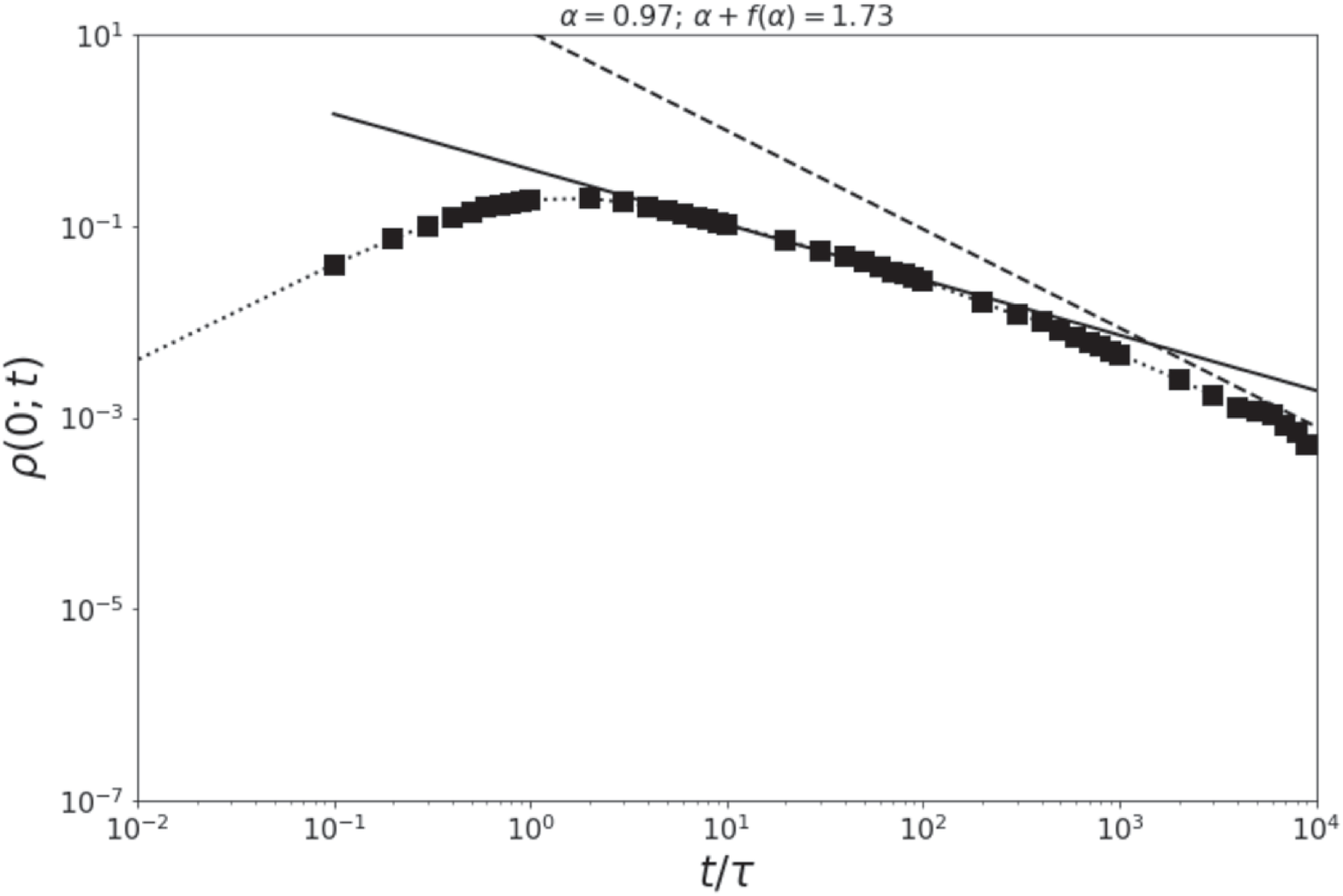}
  \includegraphics[scale=0.14]{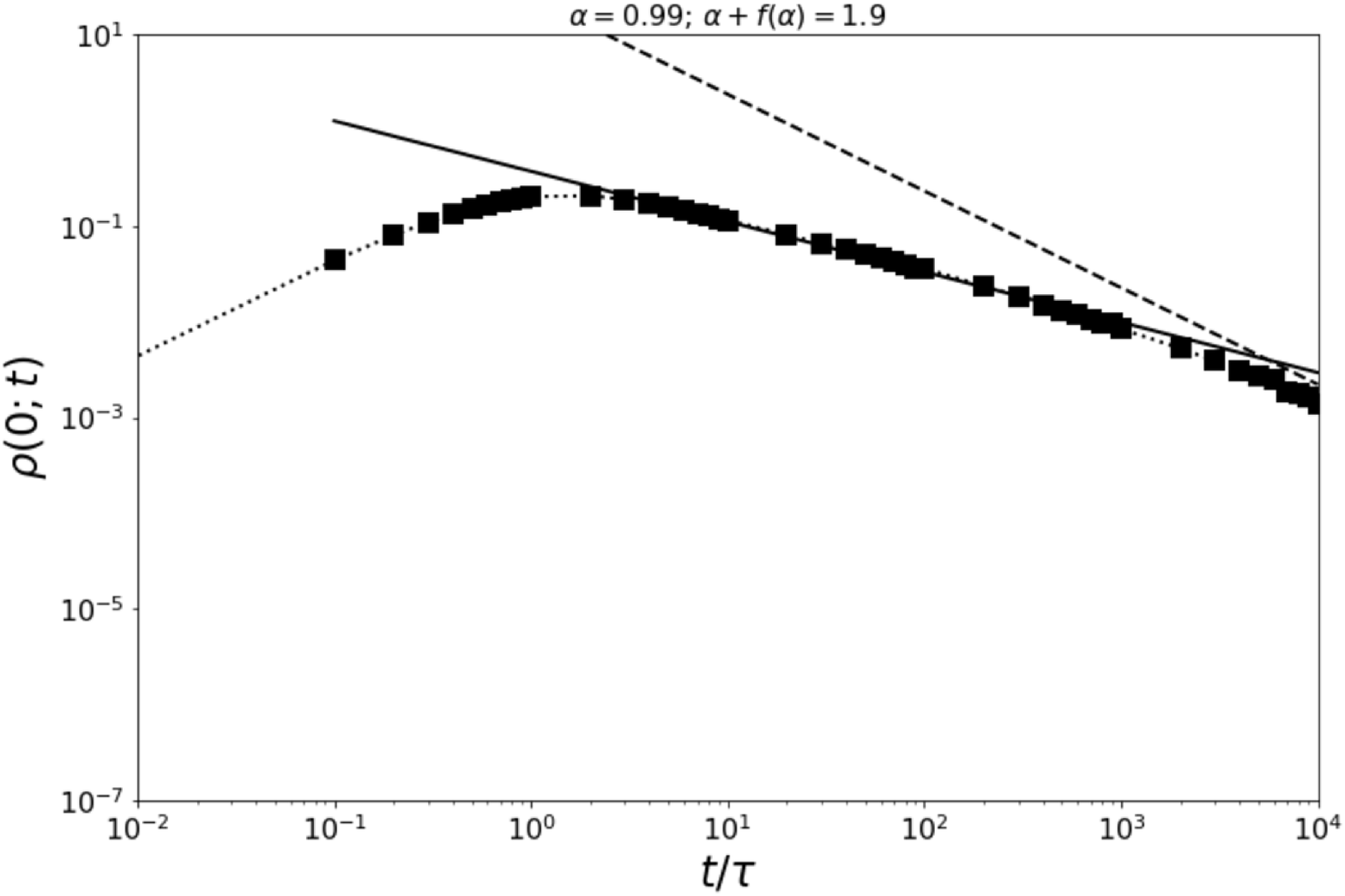}
  \includegraphics[scale=0.14]{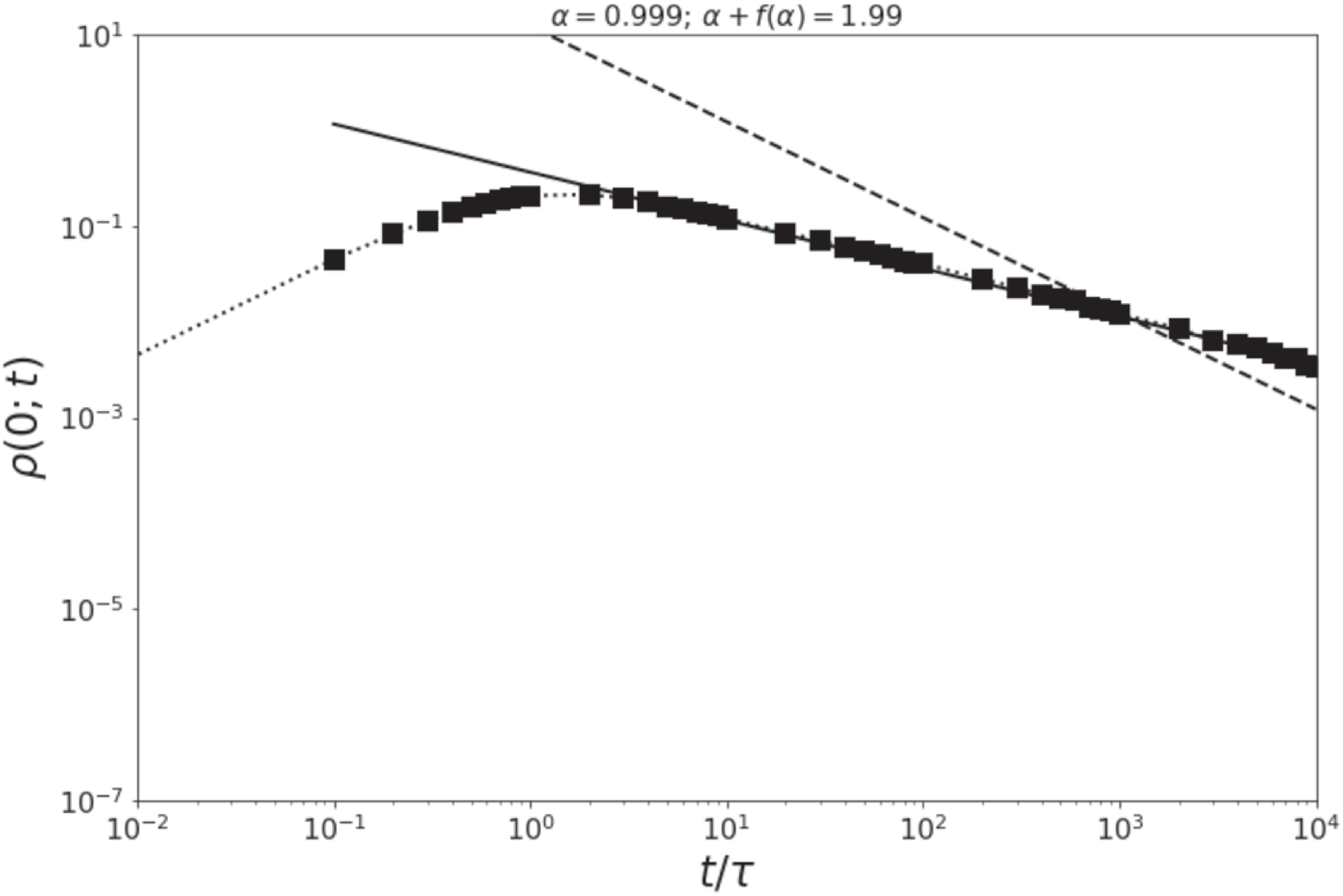}
\caption{The same as in Figure \ref{fig:scaling1}
but with $\alpha= 0.95 \,, 0.97 \,, 0.99 \,, 0.999$
for highlithing the delay in attaining the large-time limit $t \gg T$.}
\label{fig:scaling2}
\end{figure}
\end{center}

\begin{center} 
\begin{figure}
  \includegraphics[scale=0.28]{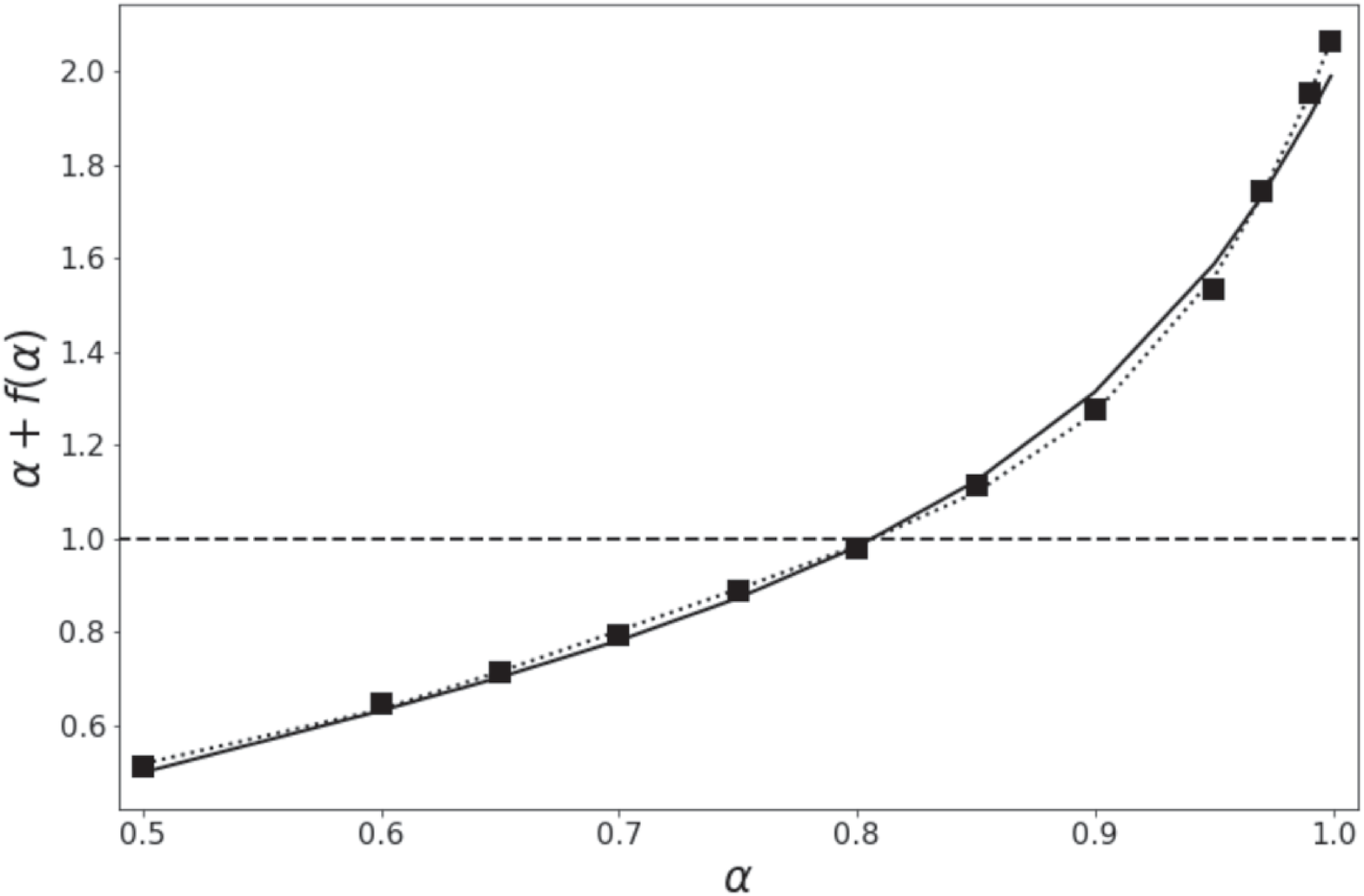}
\caption{Plot of the decaying-law of $\rho(0;t)$
as estimated by simulations (black squares).
The dotted line corresponds to
the formula $\alpha + c_2\alpha^2 + \dots + c_6\alpha^6$
as provided by the fitting routine {\tt scipy.optimize.curve\_fit}
while the solid line corresponds to the formula
$\alpha + f(\alpha)$ (\ref{scaling0})
and the dashed line is the reference-line
indicating the transient-to-recurrence conversion
at $\alpha+f(\alpha)=1$.}
\label{fig:fitalpha}
\end{figure}
\end{center}

We have simulated the process that converges to $\rho(x;t)$
(\ref{convolution}) and the decaying in time of $\rho(0;t)$
is shown in Figures \ref{fig:scaling1} and \ref{fig:scaling2}.
By using a fitting procedure, we have estimated
the scaling-law in the transient regime $\tau \ll t \ll T$
and we found that it is a not power-law, see Figure \ref{fig:fitalpha}.
In particular, we have approximated it
with the easy-to-read formula
\be
\rho(0;t) \sim t^{-1/[\alpha + f(\alpha)]} \,, \quad
f(\alpha)=\frac{1}{\alpha^2}
\frac{\Gamma(2\pi\alpha)-\Gamma(\pi)}
{\Gamma(2\pi)-\Gamma(\pi)} \,, \quad \tau \ll t \ll T \,,
\label{scaling0}
\ee
that meets the constraints $f(1/2)=0$ and $f(1)=1$
in order to recover the limit scaling-laws
$\rho(0;t) \sim t^{-2}$, when $\alpha = 1/2$,
and $\rho(0;t) \sim t^{-1/2}$, when $\alpha = 1$,
as expected from (\ref{jumppdfseries}).

\medskip

Simulations show that, during the intermediate regime $\tau \ll t \ll T$,
when $\alpha \to 1$ it holds $\alpha + f(\alpha) > 1$
(see also Figures \ref{fig:scaling1}, \ref{fig:scaling2} and \ref{fig:fitalpha}).
Moreover, if $\alpha \to 1$ then $T \to \infty$
and this makes unattainable the large-time limit $t \gg T$
whenever the studied Markovian CTRW model with
jumps following a rule {\it \`a la} coin-flipping corresponds to
a real system. In particular, this means that,
a recurrence-like scaling, i.e., $\rho(0;t) \sim t^{-\beta}$
with $1 \le \beta < 2$, could be observed for a very extended temporal interval
because $T \to \infty$, in spite of the transient
theoretical scaling $\rho(0;t)\sim t^{-1/\alpha}$, with $1/2 < \alpha < 1$.

 \medskip

Therefore, since recurrence can be understood as homecoming probability
and power-law distributions are used for explaining animal behaviour,
the significance of this apparent recurrence
- in spite of the actual transience - lays into an indetermined homecoming.
This indetermination provides a further weakness
of the hypothesis of L\'evy-like motions for animal behaviour
that can be overcome, for example, in the framework of truncated
L\'evy flights \cite{mantegna_etal-prl-1994,koponen-pre-1995}.

\section{Summary and conclusions} 
\label{sec:conclusion}
\setcounter{section}{4}
\setcounter{equation}{0}\setcounter{theorem}{0}

In this paper we have analysed random walks
and we have discussed the role of a jump rule {\it \`a la} coin-flipping,
namely jumps with a bi-modal distribution that is equal to zero in zero.
In particular, we have studied an example of jump process
that displays tails decaying with a power-law and we found that,
within the framework of Markovian CTRW models for L\'evy flights, i.e.,
the walkers's distribution converges to a stable density,
the self-similarity of the diffusive process is lost
for a certain interval of the stability parameter
$0 < \alpha < 2$: in the particular case of our example
the self-similarity is lost when $1/2 < \alpha < 1$.

\smallskip

In the derivation of L\'evy flights from the CTRW,
the key role is played by the asymptotic limit for small wavelength of the
characteristic function of jumps and this leaves open
the specific form of the full characteristic function
of the jump $pdf$. This asymptotic limit re-phrases
in a single step the Kramers--Moyal expansion
and the Pawula theorem by reducing the so-called Montroll--Weiss
equation, that governs the walker's distribution
in the CTRW approach, to the fractional diffusion equation in
the domain of the wavelength. Actually, this procedure is an example
for showing the Central Limit Theorem in the sense of L\'evy.
As a matter of fact, when the jump-sizes follow a rule {\it \`a la} coin-flipping,
the small wavelength expansion of their characteristic function is not always a series with
alternating signs and this fact causes the loss of self-similarity.
In the framework of the studied example,
the resulting diffusive process converges to a generalised
Voigt profile that is given by the convolution of
two stable densities.

\smallskip

We have highlighted that this loss of self-similarity has a
double significance.
At the mathematical level, the use of a jump $pdf$
corresponding to a rule {\it \`a la} coin-flipping
makes the model for L\'evy flights consistent with the
probabilistic derivation of the fractional diffusion equation
where the distinctive singularity of the fractional Laplacian
is a consequence of the jump-rule {\it \`a la} coin-flipping,
but the resulting evolution equation is indeed
a double fractional-order equation
in the stability interval $1/2 < \alpha < 1$.
At the application level,
the loss of self-similarity generates an intermediate temporal regime
which defines a time-scale for large-time limit.
This time-scale results to be depended on $\alpha$ and it tends
to infinite when $\alpha$ tends to $1$:
this means that, whenever the studied process is a reliable model for
a physical system, the large-time limit could not be observed in real measurements.
This unattainability of the large-time limit has an effect
on the transience and recurrence of the process: actually,
in spite of the expected transience of the process,
the long-extended intermediate regime could display a recurrence-like
scaling that leads to an indetermined situation in real cases.
This apparent recurrence because of the unattainability
of the large-time limit is a property of the studied CTRW model
that deserves attention in the future.
In fact, if animal movement is modelled
through L\'evy-like motions then the serching for food,
and also the searching for home,
can be affected by the adopted jump rule: the searching for food
could lead to a double-order equation and the searching for home to an indetermined
homecoming in real systems.

\smallskip

To conclude, we state that the research on the derivation of
random walks models for L\'evy flights and fractional diffusion
is not concluded yet, that a further deep investigation on the
role of jump-rules {\it \`a la} coin-flipping is necessary.
This calls for a generalisation of the present results
both in terms of the choice of the jump $pdf$
and in terms of the considered random walk model characterised by
power-law tails as, for example, L\'evy walks.
Moreover, this distinguishible effect due to
the jump-rule, i.e., it is
{\it \`a la} coin-flipping or not,
turns into a distinguishible feature of
the motion of animals: namely if they stand in the majority of the
iterations or if they always move.

\smallskip

Definitively, the difference between the
{\it ``Should I stay?"} and {\it ``Should I go?"} conditions
cannot be disregarded.

\vspace*{-3pt} 
\section*{Appendix A} 
\setcounter{equation}{0}\setcounter{theorem}{0}
\renewcommand{\theequation}{A.\arabic{equation}}

We report here the main steps related to the calculations
concerning the jump $pdf$s (\ref{jumppdf}) and (\ref{jumppdf2})
providing the L\'evy coin-flipping rules
for the {\it ``Should I go?"} condition.

\smallskip

Since the considered jump $pdf$s are symmetric,
the corresponding characteristic functions are defined by
\be
\widehat{\varphi}(\kappa)=2 \int_0^\infty
\cos(\kappa x) \varphi(x) dx \,,
\ee
that are symmetric as well, i.e.,
$\widehat{\varphi}(\kappa)=\widehat{\varphi}(-\kappa)$,
and they can be expressed through their Mellin transform
\cite[see from (2.26) to (2.31)]{mainardi_etal-fcaa-2001}, i.e.,
\be
\widehat{\varphi}(\kappa)=
\frac{2}{\kappa}
\frac{1}{2\pi i}\int_{L}
\varphi^*(s)\Gamma(1-s)
\sin\left(\frac{\pi s}{2}\right) \kappa^s ds \,,
\quad \kappa > 0 \,,
\ee
where $L$ is the integration path in the
sense of the Mellin--Barnes integrals
and $\varphi^*(s)$, with $s \in \C$,
is the Mellin transform of $\varphi(x)$, with $x > 0$:
\be
\varphi^*(s)=\int_0^\infty \varphi(x) \, x^{s-1} dx \,, \quad
\varphi(x)=\frac{1}{2\pi i} \int_L \varphi^*(s) \, x^{-s} ds \,,
\quad x > 0 \,.
\ee
For further details on the Mellin transform and
Mellin--Barnes integrals we refer the reader to, for example,
the textbook by Marichev \cite{marichev-1983}.

\smallskip

By reminding the Mellin--Barnes integral representation of
extremal L\'evy densities \cite{mainardi_etal-fcaa-2001, mainardi_etal-fcaa-2008}
\vskip -10pt
\be
\mL(x)
= \frac{1}{\alpha} \frac{1}{2 \pi i} \int_L
\frac{\Gamma\left(\frac{1}{\alpha}-\frac{s}{\alpha}\right)}
{\Gamma(1-s)} \, x^{-s} ds \,,
\ee
and then the Mellin transform
\be
\int_0^\infty \mL(x) \, x^{s-1} dx
= \frac{1}{\alpha}
\frac{\Gamma\left(\frac{1}{\alpha}-\frac{s}{\alpha}\right)}
{\Gamma(1-s)}
=
\frac{\Gamma\left(1+\frac{1}{\alpha}-\frac{s}{\alpha}\right)}
{\Gamma(2-s)} \,,
\ee
where the two formulae
are related by the property $\Gamma(1+\xi)=\xi\Gamma(\xi)$
such that the normalization condition when $s=1$ is
straightforwardly checked, for the jump $pdf$ (\ref{jumppdf}) it holds
\be
\widehat{\varphi}(\kappa)=
\frac{1}{\alpha \kappa}
\frac{1}{2 \pi i} \int_L
\Gamma\left(\frac{1}{\alpha} - \frac{s}{\alpha}\right)
\sin\left(\frac{\pi s}{2}\right) \kappa^s ds \,,
\quad \kappa > 0 \,,
\ee
and by applying the residue theorem for $\kappa \to 0$
formula (\ref{jumppdfseries}) is obtained,
and analogously for the jump $pdf$ (\ref{jumppdf2}) it holds
\be
\widehat{\varphi}(\kappa)=
\frac{1}{\Gamma(1/\alpha) \kappa}
\frac{1}{2 \pi i}\int_L
\Gamma\left(\frac{2}{\alpha} - \frac{s}{\alpha}\right)
\frac{\Gamma(1-s)}{\Gamma(2-s)}
\sin\left(\frac{\pi s}{2}\right) \kappa^s ds \,,
\,\, \kappa > 0 \,,
\hspace{-0.8truecm}
\ee
and by applying the residue theorem for $\kappa \to 0$
formula (\ref{jumppdfseries2}) is obtained.

\section*{Appendix B} 
\label{App:Valdinoci1}
\setcounter{equation}{0}\setcounter{theorem}{0}
\renewcommand{\theequation}{B.\arabic{equation}}

We briefly report the probabilistic derivation
of the fractional diffusion equation (\ref{SFDE})
due to Valdinoci \cite{valdinoci-bsema-2009}.
If $\rho(\bx;t)$ is the walker's distribution function
and $\varphi(\Delta \bx)$ is the symmetric jump $pdf$, then
the generic update of $\rho(\bx;t)$ at any
constant time-step $\Delta t$ is given by
\be
\rho(\bx;t + \Delta t) = \int_{\R^N}
\varphi(\Delta \bx) \rho(\bx - \Delta \bx;t) \,
d \Delta \bx \,.
\label{uPu}
\ee
We discretise the jump $pdf$ in a lattice
$h\Z^N$, where $\Z^N$ is a regular lattice
with unitary grid-size and $h > 0$,
such that $\Delta \bx = h \bz$ and $\bz \in \Z^N$,
then (\ref{uPu}) reads
\vskip -10pt
\be
\rho(\bx;t + \Delta t) = \sum_{\bz \in \Z^N}
\varphi(h\bz) \rho(\bx - h\bz;t) h^N \,.
\ee
If we assume a power-law jump $pdf$
up to a normalizing constant, i.e.,
\be
\varphi(\Delta \bx)=|\Delta \bx|^{-N-\alpha} \,, \quad
{\rm with} \quad \varphi(0)=0 \,,
\ee
then, by using the normalization condition
\be
\int_{\R^N} \varphi(\Delta \bx) \, d\Delta \bx =
\sum_{\bz \in \Z^N} \varphi(h\bz) h^N=
\sum_{\bz \in \Z^N} \varphi(\bz) =1 \,,
\ee
the evolution in time of $\rho(\bx;t)$ results to be governed by
\begin{eqnarray}
\frac{\rho(\bx;t+\Delta t)-\rho(\bx;t)}{\Delta t}
&=&\sum_{\bz \in \Z^N} \frac{\varphi(\bz)}{\Delta t}
[\rho(\bx - h \bz;t) - \rho(\bx;t)] \nonumber \\
&=&\mD_\alpha \sum_{\bz \in \Z^N}
\frac{\rho(\bx - h \bz;t) - \rho(\bx;t)}{|h\bz|^{N+\alpha}} h^{N} \,,
\label{quasifatta1}
\end{eqnarray}
where $\mD_\alpha=h^\alpha/\Delta t$ and
$\varphi(\bz) = |\bz|^{-N-\alpha}$.
By applying the change of variable $\by=h\bz$,
the rhs of (\ref{quasifatta1}) is the sum approximation of a Riemann integral,
and in the limits $h \to 0$ and $\Delta t \to 0$, it holds
\be
\frac{\partial \rho}{\partial t}
=\mD_\alpha \int_{\R^N}
\frac{\rho(\bx - \by;t) - \rho(\bx;t)}{|\by|^{N+\alpha}} d\by \,.
\label{quasifatta2}
\ee

\smallskip

To conclude, by applying in (\ref{quasifatta2}) the shift $\bx - \by \to \by$,
we finally obtain
\be
\frac{\partial \rho}{\partial t}
=- \mD_\alpha (-\Delta)^{\frac{\alpha}{2}} \rho \,,
\ee
where we used the following definition,
up to a normalizing constant, of the fractional Laplacian \cite{valdinoci-bsema-2009}
\vskip -8pt
\be
(-\Delta)^{\frac{\alpha}{2}} g
= \int_{\R^N}
\frac{g(\bx) - g(\by)}{|\bx-\by|^{N+\alpha}} \, d\by \,,
\quad 0 < \alpha < 2 \,.
\ee

\section*{Appendix C} 
\label{App:Valdinoci2}
\setcounter{equation}{0}\setcounter{theorem}{0}
\renewcommand{\theequation}{C.\arabic{equation}}

We briefly report here an analytical approach
due to Affili, Dipierro \& Valdinoci \cite{affili_etal-matrix-2020}
for the determination of recurrence and transience
of random processes. That approach \cite{affili_etal-matrix-2020}
is based on the partial differential equation that governs the evolution
in time of the walker's distribution $\rho(\bx;t)$,
and we re-arrange it according to the present aim
by remembering that we assume as initial datum $\rho(\bx;0)=\delta(\bx)$.

\smallskip

We introduce a ball of radius $r > 0$ that we denote by $B_r$
and we center it in the starting point $\bx=0$,
then we consider the probability for a walker to be outside of the ball $B_r$
at time $t$, i.e.,
\be
Q(r,t) = \int_{\R^N \setminus B_r} \rho(\bx;t) \, d\bx \,,
\ee
\vskip -3pt \noindent
or equivalently
\vskip -12pt
\be
Q(r,t) = 1 - \int_{B_r} \rho(\bx;t) \, d\bx \,,
\label{defQ}
\ee
\vskip -3pt \noindent
and then it holds
\vskip -10pt
\be
0 \le Q(r,t) \le \int_{\R^N} \rho(\bx;t) \, d\bx = 1 \,,
\ee
where the normalization condition (\ref{defrho}) is used.

\smallskip

At any instant $t$, the probability for the walker to step from some position
$X_t$ into the ball $B_r$ is the probability to make a
jump of the necessary size: $\mP(X_B \in B_r|X_t)=\mP(\Delta X=X_B - X_t)$.
If the jumps are statistically independent, at each instant $t$,
the probability to step into $B_r$ is the probability of an independent drawing, so
the probability to step into $B_r$ during the whole random walk is given by
the product of the probabilities of the necessary jumps at all instants,
that is, at any fixed time-step in a discrete time framework \cite{affili_etal-matrix-2020}.
Since we are considering a Markovian CTRW with exponentially distributed waiting-times
with mean value $\tau$, we replace the time-step with $\tau$
and we consider the probabilities at any integer multiples of $\tau$:
\vskip -13pt
\be
Q(r)=\prod_{h=1}^\infty Q(r,t=h\tau) \in [0,1] \,,
\ee
and the recurrence or the transience of the process
in the starting point $\bx=0$ is determined by the limit
\vskip -11pt
\begin{subnumcases}{
\lim_{r \to 0} Q(r) =}
\displaystyle{0 \,, \quad {\rm recurrent} \,,}
\\
\displaystyle{1 \,, \quad {\rm transient} \,.}
\end{subnumcases}

\smallskip

We assume that, inside the ball $B_r$,
the distribution $\rho(\bx;t)$ follows a self-similarity law of the form
\be
\rho(\bx;t)=\frac{1}{t^{N\beta}} \, 
\rho\!\left(\frac{\bx}{t^\beta};1\right) \,, \quad
\bx \in B_r  \,, \quad \beta > 0 \,,
\label{sslaw}
\ee
where the dimensional issues covered in the main text by the diffusion
coefficients are now disregarded for lighting the notation,
and thus from (\ref{defQ}) and (\ref{sslaw}) we have
\vskip -10pt
\be
Q(r,t)
=1 - \frac{1}{t^{N\beta}} \int_{B_r} \rho\!\left(
\frac{\bx}{t^\beta};1\right) d\bx \,.
\ee

\smallskip

Since $\rho(\bx;t)$ is the distribution function of a diffusion process with
initial datum $\rho(\bx;0)=\delta(\bx)$, it holds
\be
\sup_{\bx \in \R^N} \rho(\bx;t)=\rho(0;t) \,, \quad t > 0 \,,
\ee
and therefore we have that
\be
Q(r,t) \in \left[
1-\frac{\mu |B_r|}{t^{N\beta}}, 1-\frac{\nu |B_r|}{t^{N \beta}}\right] \,,
\label{intervallo}
\ee
\vskip -3pt \noindent
where
\vskip -12pt
\be
|B_r|=\int_{B_r} dx = C \, r^N \,, \quad C > 0 \,,
\ee
\vskip -3pt \noindent
and
\vskip -10pt
\be
\nu=\inf_{\xi \in B_r} \rho(\xi;1) > 0 \,, \quad
\mu=\sup_{\xi \in \R^N} \rho(\xi;1) =
\sup_{\xi \in B_r} \rho(\xi;1) < + \infty \,.
\ee
By using properties of logarithmic function, it follows that
\be
\log Q(r)=\log \prod_{h=1}^{+\infty} Q(r,t=h\tau)
=\sum_{h=1}^{+\infty} \log Q(r,t=h\tau) \,,
\ee
and then (\ref{intervallo}) becomes
\be
\log Q(r) \in
\left[
\sum_{h=1}^{+\infty}\log\left(
1-\frac{\mu C r^N}{(h\tau)^{N\beta}}\right) ,
\sum_{h=1}^{+\infty}\log\left(1-\frac{\nu C r^N}{(h\tau)^{N \beta}}
\right)\right] \,.
\label{preapproxlog}
\ee
Since $Q(r,t) \in [0,1]$, from (\ref{intervallo}) it results that also
\be
1 - \frac{C_0 r^N}{(h\tau)^{N\beta}} \in [0,1] \,,
\quad {\rm with} \quad C_0=\nu C
\quad {\rm or } \quad C_0= \mu C \,,
\ee
and then from the approximation rule $\log(1+z)\simeq z$ when $|z|< 1$
we obtain
\be
\log\left(
1-\frac{C_0 r^N}{(h\tau)^{N \beta}} \right)
\simeq -\frac{C_0 r^N}{(h\tau)^{N \beta}} \,,
\ee
\vskip -3pt \noindent
and finally
\vskip -13pt
\be
\sum_{h=1}^{+\infty}\log\left(
1-\frac{C_0 r^N}{(h\tau)^{N\beta}}\right)
\simeq - C_0 \frac{r^N}{\tau^{N\beta}}
\sum_{h=1}^{+\infty}\frac{1}{h^{N\beta}} \,,
\label{approxlog}
\ee
that converges if $N \beta > 1$.
To conclude, from (\ref{preapproxlog}) and (\ref{approxlog})
it results that
\vskip -13pt
\be
\log Q(r) \in
\left[
- \mu C \frac{r^N}{\tau^{N\beta}}
\sum_{h=1}^{+\infty}\frac{1}{h^{N\beta}} ,
- \nu C \frac{r^N}{\tau^{N\beta}}
\sum_{h=1}^{+\infty}\frac{1}{h^{N\beta}} \right]
\,,
\ee
and then from the convergence rule of (\ref{approxlog})
it follows that
\begin{subnumcases}{}
\log Q(r) = - \infty \,, \quad N\beta \le 1 \,,  \\
\log Q(r) \in \left[
- \mu C_* \frac{r^N}{\tau^{N\beta}}, - \nu C_* \frac{r^N}{\tau^{N\beta}}
\right] \,, \quad N\beta > 1 \,,
\end{subnumcases}
which turns into
\begin{subnumcases}{}
Q(r) = 0 \,, \quad N\beta \le 1 \,, \\
Q(r) \in \left[
\e^{- \mu C_* \frac{r^N}{\tau^{N\beta}}},
\e^{- \nu C_* \frac{r^N}{\tau^{N\beta}}}
\right] \,, \quad N\beta > 1 \,.
\end{subnumcases}

\smallskip

In conclusion the recurrence/transience criterium is
\begin{subnumcases}
{\lim_{r \to 0} Q(r) =}
0 \,, \quad N\beta \le 1 \quad {\rm (recurrent)} \,, \\
1 \,, \quad N\beta > 1 \quad {\rm (transient)} \,,
\label{formulavaldinoci}
\end{subnumcases}
and, by applying the self-similarity law (\ref{sslaw}),
we recover the well-known result for the Brownian motion
\cite{polya-1921,novak-amm-2014}
(i.e., $\beta=1/2$)
\begin{subnumcases}{}
Q(0) = 0 \,, \quad N \le 2 \quad {\rm (recurrent)} \,,  \\
Q(0)= 1 \,, \quad N > 2 \quad {\rm (transient)} \,,
\label{formulavaldinociBM}
\end{subnumcases}
and for L\'evy-distributed processes
\cite{affili_etal-matrix-2020,michelitsch_etal-jpa-2017}
(i.e., $\beta=1/\alpha$)
\begin{subnumcases}{} 
Q(0) = 0 \,,
\quad N = 1 \,\, {\rm with} \,\, 1 \le \alpha < 2 \quad {\rm (recurrent)} \,,
\label{formulavaldinociLevya}  \\
Q(0)= 1 \,, \quad N \ge 2 \,, \,\, {\rm and} \,\,
 N=1 \,\, {\rm with} \,\, 0 < \alpha < 1
\quad {\rm (transient)} \,.
\label{formulavaldinociLevyb}
\end{subnumcases}



\end{document}